\documentclass[acmsmall,screen,nonacm]{acmart}

\setcopyright{cc}
\setcctype{by}

\usepackage{faktor}
\usepackage{tikz}
\usepackage{scalerel}
\usepackage{array}
\usepackage{dashbox}
\usepackage{tensor}
\usepackage{xparse}
\usepackage{xifthen}
\usepackage{mathtools}

\usetikzlibrary{shapes}
\usetikzlibrary{arrows}
\usetikzlibrary{calc}
\usetikzlibrary{arrows.meta}

\newcommand{\textdom}[1]{\textit{#1}} %
\newcommand{\textlog}[1]{\textsf{\upshape #1}} %

\newcommand{\integer}{\mathbb{Z}}

\DeclareMathOperator*{\Sep}{\scalerel*{\ast}{\sum}} %

\makeatletter%
\@ifundefined{dplus}{%
\newcommand\dplus{\mathbin{+\!\!+}}
}{}
\makeatother%
\newcommand\fmap{\mathrel{\langle\$\rangle}}

\newcommand{\spac}{\nobreak\hskip 0.2em plus 0.1em} %

\def\All #1.{\forall #1.\spac}%
\def\Exists #1.{\exists #1.\spac}%
\def\Ret #1.{#1.\spac}%

\newcommand\fpfn{\xrightharpoonup{\smash{\raisebox{-.3ex}{\ensuremath{\scriptstyle\kern-0.25ex\textlog{fin}\kern-0.1ex}}}}}
\newcommand{\la}{\leftarrow}

\newcommand{\Ra}{\Rightarrow}

\newcommand\monra[1][]{\xrightarrow{\smash{\raisebox{-.3ex}{\ensuremath{\scriptstyle\kern-0.15ex\textlog{mon}_{#1}\kern-0.05ex}}}}}
\newcommand\monnra{\xrightarrow{\smash{\raisebox{-.3ex}{\ensuremath{\scriptstyle\kern-0.15ex\textlog{mon,ne}\kern-0.05ex}}}}}
\newcommand\nfn{\xrightarrow{\smash{\raisebox{-.3ex}{\ensuremath{\scriptstyle\kern-0.15ex\textlog{ne}\kern-0.05ex}}}}}
\newcommand{\eqdef}{\triangleq}

\newcommand*\set[1]{\left\{#1\right\}}

\newcommand{\dom}{\textlog{dom}}

\renewcommand{\lim}{\textlog{lim}}

\newcommand{\subst}[3]{{#1}[{#3} / {#2}]}

\newcommand{\mapelem}[2]{#1\mathop{\la}#2}
\newcommand{\mapinsert}[3]{#3\!\left[\mapelem{#1}{#2}\right]}
\newcommand{\mapdelete}[2]{#2\setminus\set{#1}}
\newcommand{\mapsingleton}[2]{\mapinsert{#1}{#2}{\,}}
\newcommand{\mapinsertComp}[4]
  {\mapinsert{#1}{#2 \spac\middle|\spac #3}{#4}}
\newcommand{\mapComp}[3]
  {\mapinsertComp{#1}{#2}{#3}{}}

\newcommand{\nil}{\epsilon}

\newcommand{\dispdot}[2][.2ex]{\dot{\raisebox{0pt}[\dimexpr\height+#1][\depth]{$#2$}}}%

\newcommand{\latertinj}{\textlog{next}}

\newcommand{\iProp}{\textdom{iProp}}

\newcommand{\mincl}[1][]{%
  \ensuremath{\mathrel{\vbox{\offinterlineskip\ialign{%
    \hfil##\hfil\cr
    \ensuremath{\scriptstyle #1}\cr
    \noalign{\kern-0.25ex}
    $\preccurlyeq$\cr
}}}}}

\newcommand{\gname}{\gamma}

\newcommand{\mask}{\mathcal{E}}
\newcommand{\namesp}{\mathcal{N}}
\newcommand{\namecl}[1]{{#1^{\kern0.2ex\uparrow}}}

\def\MU #1.{\mu #1.\spac}%
\def\Lam #1.{\lambda #1.\spac}%

\newcommand{\proves}{\vdash}
\newcommand{\provesIff}{\mathrel{\dashv\vdash}}

\newcommand{\wand}{\mathrel{-\!\!\ast}}

\NewDocumentCommand\wpre{O{} m O{} m}%
  {\textlog{wp}^{#1}_{#3}\spac#2\spac{\left\{#4\right\}}}

\newcommand{\later}{\mathop{{\triangleright}}}

\newcommand{\always}{\mathop{\square}}

\tikzstyle{boxedassert_border} = [sharp corners,line width=0.2pt]
\NewDocumentCommand \boxedassert {O{} m o}{%
	\tikz[baseline=(m.base)]{
		\node[rectangle,inner sep=0.8pt,outer sep=0.2pt,anchor=base] (m) {${\,#2\,}\mathstrut$};
		\draw[#1,boxedassert_border] ($(m.south west) + (0,0.65pt)$) rectangle ($(m.north east) + (0, 0.7pt)$);
	}\IfNoValueF{#3}{^{\,#3}}%
}
\newcommand*{\knowInv}[2]{\boxedassert{#2}[#1]}

\newcommand*{\ownGhost}[2]{\boxedassert[densely dashed]{#2}[#1]}

\NewDocumentCommand \vsGen {O{} m O{} O{}}%
  {\mathrel{%
    \ifthenelse{\equal{#3}{}}{%
      {#2}_{#1}%
    }{%
      \ifthenelse{\equal{#4}{}}{%
        \tensor*[_{#1}]{#2}{_{#3}}%
      }{%
        \ifthenelse{\equal{#1}{}}{{#2}_{#3}^{#4}}{\tensor*[_{#1}]{#2}{_{#3}^{#4}}}%
      }%
    }%
  }}%
\NewDocumentCommand \vs {O{} O{} O{}} {\vsGen[#1]{\Rrightarrow}[#2][#3]}
\NewDocumentCommand \bvs {O{} O{} O{}} {\vsGen[#1]{\dispdot[0.02ex]{\Rrightarrow}}[#2][#3]}
\NewDocumentCommand \vsL {O{} O{} O{}} {\vsGen[#1]{\Lleftarrow}[#2][#3]}
\NewDocumentCommand \vsE {O{} O{} O{}} %
  {\vsGen[#1]{\Lleftarrow\!\!\!\Rrightarrow}[#2][#3]}
\NewDocumentCommand \pvs {O{} O{} O{}} {\mathord{\vsGen[#1]{{\mid\kern-0.5ex\Rrightarrow\kern-0.25ex}}[#2][#3]\kern0.2ex}}

\newcommand\vsWand{{\displaystyle\equiv\kern-1.6ex-\kern-1.2ex\smash{\raisebox{-0.15ex}{\scalerel*{\vphantom-\ast}{|}}}\kern-0.2ex}}
\NewDocumentCommand \vsW {O{} O{}} {\vsGen[#1]{\vsWand}[#2]}

\newcommand\vsWandInline{{\equiv\kern-1.6ex-\kern-1.28ex\smash{\raisebox{-0.15ex}{\scalerel*{\vphantom-\ast}{\textstyle\sum}}}\kern-0.2ex}}
\NewDocumentCommand \vsWI {O{} O{}} {\vsGen[#1]{\vsWandInline}[#2]}

\newcommand\vsWandStep{{\displaystyle\raisebox{0.106ex}{\scaleobj{0.82}{\later}}\kern-1.65ex\equiv\kern-1.6ex-\kern-1.5ex\smash{\scalerel*{\vphantom-\ast}{\sum}}\kern-0.2ex}}
\NewDocumentCommand \vsWS {O{} O{}} {\vsGen[#1]{\vsWandStep}[#2]}

\NewDocumentCommand \upd {} {\mathop{\dispdot[-0.2ex]{\mid\kern-0.4ex\Rrightarrow\kern-0.25ex}}}

\NewDocumentCommand\Acc{O{} O{} m m}{#3 \mathrel{~\vsGen[#1]{\propto}[#2]~} #4}

\newcommand{\laterCredit}[1]{\text{\textsterling}\hskip 0.1em \ensuremath{#1}}

\newcommand{\curlybracket}[1]{{\left\{#1\right\}}}

\NewDocumentCommand \hoare {m m m O{}}{
	\curlybracket{#1}\spac #2 \spac \curlybracket{#3}_{#4}%
}

\NewDocumentCommand \hoareV {O{c} m m m O{}}{
		{\begin{aligned}[#1]
		&\curlybracket{#2} \\
		&\quad{#3} \\
		&\curlybracket{#4}_{#5}
		\end{aligned}}%
}
\NewDocumentCommand \hoareHV {O{c} m m m O{}}{
	{\begin{aligned}[#1]
	&\curlybracket{#2} \spac {#3} \\
	&\curlybracket{#4}_{#5}
	\end{aligned}}%
}
\NewDocumentCommand \hoareVH {O{c} m m m O{}}{
	{\begin{aligned}[#1]
	&\curlybracket{#2} \\
        & {#3}\spac \curlybracket{#4}_{#5}
	\end{aligned}}%
}

\newcommand{\anglebracket}[1]{{\scaleleftright[2ex]{\langle}{#1}{\rangle}}}

\NewDocumentCommand \ahoare {m m m O{}}{
	\anglebracket{#1}\spac #2 \spac \anglebracket{#3}_{#4}%
}

\NewDocumentCommand \ahoareV {O{c} m m m O{}}{
		{\begin{aligned}[#1]
		&\anglebracket{#2} \\
		&\quad{#3} \\
		&{\anglebracket{#4}}_{#5}
		\end{aligned}}%
}

\NewDocumentCommand \ahoareHV {O{c} m m m O{}}{
	{\begin{aligned}[#1]
	&\anglebracket{#2}\; {#3} \\
	&{\anglebracket{#4}}_{#5}
	\end{aligned}}%
}

\newcommand{\TRUE}{\textlog{True}}
\newcommand{\FALSE}{\textlog{False}}

\newcommand{\expr}{e}
\newcommand{\val}{v}

\newcommand{\step}[1][]{\xrightarrow{{#1}}}
\newcommand{\steps}[1][]{\mathrel{\step[#1]\!\!\vphantom{\step}^*}}
\newcommand{\hstep}[1][]{\xrightarrow{{#1}}_{\textlog{h}}}

\newcommand{\State}{\kern-0.05em\textdom{State}}
\newcommand{\Val}{\kern-0.2em\textdom{Val}}
\newcommand{\Loc}{\kern-0.05em\textdom{Loc}}
\newcommand{\Expr}{\kern-0.05em\textdom{Expr}}
\newcommand{\Var}{\kern-0.2em\textdom{Var}}
\newcommand{\Obs}{\kern-0.1em\textdom{Obs}}
\newcommand{\ThreadPool}{\kern-0.05em\textdom{ThreadPool}}

\newcommand{\atomic}{\textlog{atomic}}

\newcommand{\red}{\textlog{red}}

\def\fillctx#1[#2]{#1 {[}\, #2\,{]} }

\newcommand{\aginj}{\textlog{ag}}

\newcommand{\exinj}{\textlog{ex}}

\newcommand{\authfull}{\mathord{\bullet}}
\newcommand{\authfrag}{\mathord{\circ}}

\tikzstyle{sts_state} = [rectangle, rounded corners, draw, minimum size=1.2cm, align=center]
\tikzstyle{sts_arrows} = [->,arrows={->[scale=1.5]},every node/.style={font=\sffamily\small}]

\newcommand{\mapstoprop}{\mathrel{\kern-0.5ex\tikz[baseline=(m)]{\node at (0,0) (m){}; \draw[line cap=round] (0,0.16) -- (0,-0.004);}\kern-1.5ex\Ra}}

\usepackage{subcaption} %
\usepackage{mathpartir} %
\usepackage{calc,mathtools} %
\usepackage{wrapfig}
\usepackage{tikz}
\usepackage[capitalize,nameinlink,noabbrev]{cleveref} %
\usetikzlibrary{arrows.meta,patterns,patterns.meta}
\usepgflibrary{shadings}
\mprset{sep=1.5em}

\DeclareUnicodeCharacter{22A5}{$\bot$}
\DeclareUnicodeCharacter{03A8}{$\Phi$}

\crefname{section}{\S\!\!}{\S\!\!}
\crefname{subsection}{\S\!\!}{\S\!\!}

\usepackage{caption}
\usepackage[belowskip=-14pt,aboveskip=4pt]{caption}
\makeatletter
\def \mpr@medlineskip {\lineskiplimit=1.2em\lineskip=1.0em plus 0.2em}%
\let \MathparLineskip \mpr@medlineskip
\makeatother

\makeatletter
\def\@parfont{\itshape\bfseries}
\makeatother

\makeatletter
\let\oldhypertarget\hypertarget
\renewcommand{\hypertarget}[2]{%
  \oldhypertarget{#1}{#2}%
    \protected@write\@mainaux{}{%
        \string\expandafter\string\gdef
          \string\csname\detokenize{ hypertarget-#1}\string\endcsname{#2}%
    }%
  }
\makeatother
\newcommand{\refrule}[1]{%
    \textsc{\hyperlink{#1}{\csname hypertarget-#1\endcsname}}%
}

\newcommand{\eg}{\textit{e.g.,}\ }
\newcommand{\ie}{\textit{i.e.,}\ }
\newcommand{\wrt}{\textit{w.r.t.}\ }

\newcommand{\unlog}{\textsc{Ficus}}
\newcommand{\binlog}{\textsc{Rel\unlog}}
\newcommand{\pureunlog}{\textsc{Pure\unlog}}

\colorlet{KeywordColor}{ACMDarkBlue}
\colorlet{EffectTagColor}{ACMOrange!60!black}
\colorlet{CommentColor}{black!65!white}
\colorlet{FunctionColor}{KeywordColor}
\colorlet{GhostColor}{ACMRed}
\colorlet{GhostKeywordColor}{GhostColor!75!KeywordColor}

\newcommand{\keyword}[1]{\text{\textcolor{KeywordColor}{\textbf{\texttt{#1}}}}}
\newcommand{\hole}{\bullet}
\newcommand{\effecttag}[1]{\text{\textcolor{EffectTagColor}{\texttt{#1}}}}
\newcommand{\tags}{\textlog{tags}}
\newcommand{\codecomment}[1]{\text{\textcolor{CommentColor}{\texttt{(*} #1 \texttt{*)}}}}
\newcommand{\function}[1]{\text{\textcolor{FunctionColor}{\texttt{#1}}}}
\newcommand{\ghostcode}[1]{\begingroup\colorlet{KeywordColor}{GhostKeywordColor}\color{GhostColor}#1\endgroup}
\newcommand{\HeapLang}{\textsf{HeapLang}}

\newcommand{\HandlerLang}{\textsf{FicusLang}}
\newcommand{\PureHandlerLang}{\textsf{PureFicusLang}}
\def\Rec #1.{\keyword{rec}~#1.\spac}

\newcounter{codeline}
\newcommand{\codenum}{\stepcounter{codeline}\text{\ttfamily\thecodeline}}
\newenvironment{codeblock}[1]{\small\begin{array}{>{\color{gray}}r@{}#1}\setcounter{codeline}{0}}{\end{array}}

\newcommand\sqsubsetsim{\mathrel{%
  \ooalign{\raise0.2ex\hbox{$\sqsubset$}\cr\hidewidth\raise-0.8ex\hbox{\scalebox{0.9}{$\sim$}}\hidewidth\cr}}}

\newcommand{\natinj}{\textlog{nat}}

\newcommand{\setinj}{\textlog{set}}

\newcommand{\W}{\mathrm{W}}
\newcommand{\E}{\mathcal{E}}
\newcommand{\IsMain}{\mathcal{M}}
\newcommand{\IsChild}{\mathcal{C}}
\newcommand{\efft}{\textit{et}}
\newcommand{\terminated}[1]{#1^{\times}}
\renewcommand{\lbag}{\{\!|}
\renewcommand{\rbag}{|\!\}}
\newcommand{\postcrash}{\vardiamondsuit}
\newcommand{\setup}{\diamondsuit}
\newcommand{\ewpNoArg}{\textlog{ewp}}
\newcommand{\ewpcNoArg}{\textlog{ewpc}}
\NewDocumentCommand\ewp{O{} m O{} m m}%
  {\ewpNoArg^{#1}_{#3}\spac#2\spac{\left\langle#4\right\rangle}\spac{\left\{#5\right\}}}
\NewDocumentCommand\ewpc{O{} m O{} m m}%
  {\ewpcNoArg^{#1}_{#3}\spac#2\spac{\left\langle#4\right\rangle}\spac{\left\{#5\right\}}}

\newcommand{\restrictedstep}[1]{\xrightarrow[{#1}]{}}
\newcommand{\restrictedsteps}[1]{\mathrel{\restrictedstep{#1}\!\!\vphantom{\step}^*}}
\newcommand{\WInvTok}[1]{\textlog{TokI}(#1)}
\newcommand{\WCrashInvTok}[1]{\textlog{CC}(#1)}
\newcommand{\effectatomic}{\textlog{eff-atomic}}
\NewDocumentCommand{\sendrecvprot}{O{} m m O{} m m O{}}{{}!#1 \left(#2\right)\left\{#3\right\}.#7?#4 \left(#5\right)\left\{#6\right\}}

\newcommand{\netmapsto}{\mathrel{\shortmid\kern-0.25ex\leadsto}}
\newcommand{\netmapslb}{\mathrel{\shortmid\kern-0.25ex\leadsto^{\textlog{lb}}}}
\newcommand{\diskmapsto}{\mathrel{\mapsto^{\mathrm{d}}}}
\newcommand{\adiskval}[2]{[#1]#2}
\newcommand{\primproph}[1]{\textlog{primProph}\!\left(#1\right)}
\newcommand{\primprophauth}[1]{\textlog{primProph}_{\!\authfull}\!\left(#1\right)}
\newcommand{\proph}[2]{\textlog{proph}\!\left(#1,#2\right)}
\newcommand{\prophE}[1]{\textlog{prophE}\!\left(#1\right)}
\newcommand{\prophD}[1]{\textlog{prophD}\!\left(#1\right)}
\newcommand{\labelEn}[1]{\textlog{labelEn}\!\left(#1\right)}

\newcommand{\snapshot}[2]{\textlog{snapshot}^{#1}\!\left(#2\right)}
\newcommand{\snapshotlatest}[2]{\textlog{snapshot}_{\!\authfull}^{#1}\!\left(#2\right)}
\newcommand{\maxsteps}[1]{\textlog{maxsteps}\!\left(#1\right)}
\newcommand{\timeReceipt}{\textlog{TR}}
\newcommand{\timeReceiptAuth}{\textlog{TR}_{\!\authfull}}
\newcommand{\Eff}{\kern-0.05em\textdom{Eff}}

\newcommand{\specNoArg}{\textlog{spec}}
\newcommand{\spec}[1]{\specNoArg(#1)}
\newcommand{\specCtxAlt}{\textlog{specCtx}}
\newcommand{\genSpecProtNoArg}{\textlog{genspec}}
\newcommand{\genSpecProt}[3]{\genSpecProtNoArg_{#1}~#2~\langle #3 \rangle}
\newcommand{\specProtNoArg}{\textlog{espec}}
\newcommand{\specprot}[3]{\specProtNoArg_{#1}~#2~\langle #3 \rangle}
\newcommand{\spechandler}[1]{\textlog{handler}_{#1}}
\newcommand{\specthread}[3]{\textlog{spec}^{#1}_{#2}(#3)}
\newcommand{\specthreadprot}[5]{\specProtNoArg_{#3}^{#1;#2}~#4~\langle #5 \rangle}
\newcommand{\specnode}[3]{\textlog{specn}^{#1}_{#2}(#3)}

\newcommand{\specCtx}[3]{\textrm{CTX}^{#1}(#2, #3)}
\newcommand{\isBag}[2]{\textsf{isBag}(#1, #2)}
\def\GFP #1.{\text{gfp}~#1.\spac}
\def\LFP #1.{\text{lfp}~#1.\spac}
\newcommand{\logrefines}{\lesssim}
\newcommand{\ctxrefines}{\lesssim_{\textlog{ctx}}}
\newcommand{\ctxequiv}{\simeq_{\textlog{ctx}}}

\newcommand{\ltrans}{\raisebox{0.3ex}{\scalebox{0.5}{\ensuremath{\ll}}}}
\newcommand{\rtrans}{\raisebox{0.3ex}{\scalebox{0.5}{\ensuremath{\gg}}}}

\newcommand{\tbool}{\keyword{bool}}
\newcommand{\tunit}{\keyword{unit}}

\newcommand{\refappendix}[1]{\Cref{#1}}

\usepackage[disable]{todonotes}

\allowdisplaybreaks

\begin{document}
\renewcommand{\theHtheorem}{\thetheorem}
\def\theHlemma{\thetheorem}
\def\theHcorollary{\thetheorem}
\def\theHproposition{\thetheorem}
\def\theHconjecture{\thetheorem}
\def\theHexample{\thetheorem}
\def\theHdefinition{\thetheorem}
\let\ficusoldcaption\caption
\renewcommand{\caption}[1]{\ficusoldcaption{#1.}}

\title{Building Extensible Program Logics through Effect Handlers\ (Extended Version)}

\author{Zichen Zhang}
\correspondingauthor
\affiliation{%
  \institution{New York University}
  \city{New York}
  \country{USA}
}
\email{zichenzhang@nyu.edu}
\orcid{0009-0004-1151-6149}

\author{Simon Oddershede Gregersen}
\affiliation{%
  \institution{CISPA Helmholtz Center for Information Security}
  \city{Saarbr{\"u}cken}
  \country{Germany}
}
\email{gregersen@cispa.de}
\orcid{0000-0001-6045-5232}

\author{Joseph Tassarotti}
\affiliation{%
  \institution{New York University}
  \city{New York}
  \country{USA}
}
\email{jt4767@nyu.edu}
\orcid{0000-0001-5692-3347}
\authornote{Also affiliated with Amazon Web Services. This paper does not reflect the views of Amazon Web Services.}

\begin{abstract}
    One approach to reasoning about effectful programs is to use program logics that provide proof rules for various effects.
    However, \emph{developing} program logics requires skills that are distinct from those needed for \emph{using} program logics, making the development of new logics less accessible.
    Moreover, when developing new logics, it can be difficult to reuse components from prior logics or combine support for different effects.

    In this paper, we propose an approach for building extensible program logics based on effect handlers.
    Our starting point is an expressive program logic for reasoning about programs written in a pure, sequential language with support for effect handlers.
    Within this language, we implement handlers that model concurrency, distributed execution, and crash-recovery behavior.
    Then, by proving properties of these handlers within the logic, we derive expressive rules for reasoning about these effects.
    In some cases, this approach leads to stronger reasoning rules than those found in prior program logics.

    In addition, we develop a relational logic for proving contextual refinements between programs using effects.
    As with unary reasoning, handlers enable this relational logic to be developed in an extensible way.

\end{abstract}

\ccsdesc[500]{Theory of computation~Separation logic}
\ccsdesc[500]{Theory of computation~Logic and verification}
\ccsdesc[500]{Theory of computation~Program verification}
\ccsdesc[300]{Theory of computation~Control primitives}
\ccsdesc[300]{Theory of computation~Concurrency}
\ccsdesc[300]{Theory of computation~Distributed computing models}
\ccsdesc[100]{Software and its engineering~Software fault tolerance}
\ccsdesc[100]{Software and its engineering~Formal software verification}
\keywords{Iris, Contextual refinement, Prophecy variables, Crash recovery, Perennial}

\maketitle

\section{Introduction}\label{sec:introduction}
Program logics have proven to be powerful tools for program verification. %
As a result, a variety of program logics have been developed for a wide range of program features, including pointers~\citep{reynolds2002-sl}, concurrency~\citep{owicki1976-relyguarantee, ohearn2007-csl, jung2018-iris-groundup}, weak memory~\citep{turon2014-gps, kaiser2017-weakmemory, lahav2015-weak, mevel2021-weak, dang2022-compass, vafeiadis2013-relaxed}, crash recovery~\citep{chen2015-fscq,chajed2019-perennial, raad2020-pog, ntzik2015-fault, chajed2021-gojournal}, distributed execution~\citep{wilcox2017-disel,krogh-jespersen2020-aneris,sharma2023-grove}, and randomness~\citep{tassarotti2019-randomized, barthe2015-prhl, aguirre2024-error-credit, bao2025-bluebell, gregersen2024-clutch, barthe20-psl, aguirre2021-horpl, batz2019-qsl}, among others.

Traditionally, a program logic is developed by proving that a collection of reasoning rules is sound with respect to an operational or denotational semantics for a language.
However, following this traditional approach is challenging for several reasons.
First, the development of program logics requires skills that are distinct from those needed for \emph{using} program logics, making the development of new logics less accessible.
Second, when developing new logics, it can be difficult to reuse components or port a particular feature from a different logic.
In reality, many important computer systems \emph{combine} many of the features described in the previous paragraph, yet developing logics that provide support for such combinations of features requires significant work.

Recently, \citet{vistrup2025-alacarte} proposed an alternative approach to developing program logics that addresses the reusability problem.
Starting from a minimal, pure lambda calculus, they incrementally add effects to this language by giving a denotational semantics in terms of ITrees~\citep{xia2020-itrees}.
On the logic side, one gives rules that logically ``interpret'' or ``handle'' the events in the generated ITree.
The soundness of the logic is established in a modular way by relating these logical handlers for events in the ITree to an interpretation function that ``executes'' the events.
While this approach addresses the problem of reusability, it does not address the accessibility issue: it requires understanding the formalism of ITrees and their denotation, and the adequacy proofs require a form of reasoning that is different from the task of \emph{using} the program logic to reason about programs.

This paper advocates for a third approach to developing program logics by using \emph{effect handlers}~\citep{plotkin2001-effects,plotkin2009-effects} to model all program effects.
Effect handlers are a language feature that enables programmers to define custom effects in a modular and compositional way.
Recent work has shown how to develop program logics for reasoning about effect handlers~\citep{vilhena2021-hazel, vilhena2023-tes, vilhena2025-blaze, vilhena2022-thesis}.
For example, using these logics, it is possible to verify an effect handler that implements a mutable state effect and derive a specification that resembles the usual separation logic rules for reasoning about pointers.
As a result, reasoning about a client program that uses this state effect handler is just like carrying out a standard separation logic proof about a program that uses built-in primitive state.
Because effect handlers are just regular programs, this approach results in a semantics that is executable by construction.
However, prior program logics for effect handlers have considered a setting where effect handlers are added on top of a language that already has various forms of primitive effects built in, such as mutable references and concurrency.
This makes sense for verifying examples that involve a subtle interaction between primitive effects and effect handlers, but the soundness proofs of these logics then combine the complexity of primitive effects and effect handlers.

In this work, we instead use effect handlers to bootstrap an expressive program logic for a range of effects.
Our starting point is a minimal stateless core effect handler language called \HandlerLang{}.
For reasoning about programs written in \HandlerLang{}, we develop \unlog{}, a separation logic for effect handlers that is an adaptation of an earlier logic called Hazel~\citep{vilhena2021-hazel}.
Using \HandlerLang{}, we then implement handlers to model mutable state, shared-memory concurrency, crash recovery with durable state, and distributed execution over unreliable networks.
For each effect, we apply \unlog{} to verify the handler implementations and obtain proof rules that are analogous to the reasoning principles derived in prior specialized program logics for these effects.

The handler-based approach even allows us to derive \emph{stronger} proof rules than prior work.
There are two main reasons.
First, with the handler-based approach, we can build up effects in a hierarchical way, using earlier effects in the definition of handlers for later effects.
Then, when verifying those later effects, we can use the proof rules from the earlier effects.
For example, we show in \Cref{sec:proph} how to derive local \emph{prophecy variables}~\cite{jung2019-pvsl, abadi1988-prophecy} from a simpler global prophecy by implementing local prophecy variables as an effect handler.
Later, when implementing the handlers for crashes and recovery in \Cref{sec:crash}, we attach prophecy variables to non-deterministic choices made during crashes, which allows client proofs to reason about when crashes will occur.

A second source of stronger proof rules arises from using effect handlers to implement non-standard versions of effects that are easier to reason about.
For example, our handler for concurrency generates fewer interleavings than a standard operational semantics for concurrency.
As a result, the ``invariant-opening'' rule that we obtain for this handler is stronger than the standard rule from most concurrent separation logics (CSLs).
To justify the use of these non-standard semantics, we prove that they are equivalent in an appropriate sense to the standard semantics.
To do so, we develop a new relational logic for effect handlers called \binlog.

\paragraph{Contributions} To summarize, our work makes the following contributions:
\begin{itemize}
\item We extend Hazel~\cite{vilhena2021-hazel} to develop \unlog, an extensible unary program logic based on effect handlers (\Cref{sec:overview}).
\unlog{} uses \emph{protocols} to decompose reasoning about handlers and client code using handlers, and adds support for an extensible notion of \emph{worlds} to share resources between client code~(\Cref{sec:conc}).
\item We develop a relational logic, \binlog, for proving contextual refinement in the presence of effect handlers (\Cref{sec:relation}).
\binlog{} adapts \unlog{} protocols to the relational setting, similarly allowing for reuse and extensibility. %
\item As case studies, we show how to reconstruct and extend features from existing program logics using our effect handler approach. This includes:
(1) a derivation of \emph{local} prophecy variables out of global prophecies, along with an approach to \emph{implicitly} make prophecies without annotating a program with prophecy operations (\Cref{sec:proph});
(2) a logic for crash-recovery reasoning with asynchronous durable storage that recovers the features of Perennial~\cite{chajed2019-perennial}, but with a simpler model, and novel support for \emph{crash-aware} prophecy variables (\Cref{sec:crash});
(3) a logic for distributed systems with IronFleet-style~\cite{hawblitzel2015-ironfleet} atomic blocks (\Cref{sec:distributed}).
\end{itemize}
Our work is mechanized in the Iris separation logic framework~\cite{jung2018-iris-groundup} and the Rocq Prover.
\newcommand{\stateHandlerFig}[1]{%
\begin{wrapfigure}{r}{0.585\linewidth}
\vspace{#1}
$\begin{codeblock}{r@{}l@{\,}c@{\,}l}
&\function{run}_{\effecttag{state}}\triangleq&\multicolumn{3}{l}{\Lam \textit{main}\ \textit{init} . \function{go}\ \textit{main}\ ()\ \textit{init}}\\
&\text{where}\ \function{go}\triangleq&\multicolumn{3}{l}{\keyword{rec}\ \function{go}\ k_0\ r\ \sigma.}\\
&&\multicolumn{3}{l}{\quad\keyword{try}\ k_0\ r\ \keyword{with}}\\
&&\qquad\phantom{\mid{}}\ v\ k&\Rightarrow&\keyword{match}\ v\ \keyword{with}\\
&&&&\phantom{\mid{}}\mathmakebox[\widthof{\ensuremath{(\effecttag{write},y)}}][r]{(\effecttag{read},())}\Rightarrow \function{go}\ k\ \sigma\ \sigma\\
&&&&\mid(\effecttag{write},y)\Rightarrow\function{go}\ k\ ()\ y\\
&&&&\mid\mathmakebox[\widthof{\ensuremath{(\effecttag{write},y)}}][r]{(\efft, w)}\Rightarrow\function{go}\ k\ (\keyword{do}\ (\efft, w))\ \sigma\\
&&\qquad\mid \keyword{ret}\ v&\Rightarrow& v
\end{codeblock}$
\caption{Handler for a global state effect}\label{fig:state-handler-code}
\end{wrapfigure}%
}

\section{Program Logics by Effect Handlers}\label{sec:overview}

This section provides an overview of how to build up a program logic using effect handlers.
After introducing \HandlerLang{}, we describe the core features of \unlog, and use them to develop proof rules for reasoning about mutable state.
When writing inference rules, we use two different styles.
When the line dividing the premises from the conclusion is a standard horizontal bar, then this rule should be read as an implication in the meta-logic.
When the horizontal bar is replaced by a $\wand$, we use a rule of the form below as a notation for the separation logic entailment $P_1 \ast \cdots \ast P_n \vdash Q$.
$$\mprset{fraction=--\ast}
\inferrule{P_1\\\cdots\\P_n}
{Q}$$

 \subsection{The \HandlerLang{} Calculus}\label{subsec:lambda_bot}

\begin{figure}[t]
$$\begin{array}{r@{\,}r@{\,}l}
v&::=&()\mid\Rec f\ x. e\mid\cdots\mid\keyword{cont}\ N\\
e&::=&v\mid x\mid e\ e\mid\cdots\mid\keyword{do}\ e\mid\S(N)[v]
\mid(\keyword{try}\ e\ \keyword{with}\ v\ k\Rightarrow\ e\mid\keyword{ret}\ v\Rightarrow e)\mid\keyword{pick}\mid\ghostcode{\keyword{observe}\ e}\\
K&::=&\hole\mid e\ K\mid K\ v\mid\cdots\mid\keyword{do}\ K
\mid(\keyword{try}\ K\ \keyword{with}\ v\ k\Rightarrow\ e\mid\keyword{ret}\ v\Rightarrow e)\\
N&::=&\hole\mid e\ N\mid N\ v\mid\cdots\mid\keyword{do}\ N\\
\end{array}$$
\caption{Syntax of values $v$, expressions $e$, evaluation contexts $K$, and neutral evaluation contexts $N$}\label{fig:syntax}
\end{figure}

\HandlerLang{} is an ML-style lambda calculus with effect handlers.
The syntax of \HandlerLang{} is shown in \Cref{fig:syntax}.
It uses call-by-value and right-to-left evaluation order.
The expression $\keyword{do}\ v$ raises an effect with value $v$, which will later be handled by the closest enclosing effect handler.
Expression $\keyword{try}\ e_0\ \keyword{with}\ v_1\ k\Rightarrow\ e_1\mid\keyword{ret}\ v_2\Rightarrow e_2$ installs a \emph{shallow} effect handler for $e_0$: it evaluates $e_0$ until either $e_0$ raises an effect or becomes a value.
For the first case, the handler will have access to the value $v_1$ raised by the effect and a continuation $k$ that allows the handler to resume $e_0$ from the point the effect was raised.
For the second case, the handler will obtain the result of $e_0$, a value $v_2$.
This type of handler is called a shallow handler because it will disappear in both cases, and it must be reinstalled if the expression $e_0$ may raise effects multiple times.\footnote{\emph{Deep handlers}, which are reinstalled after an effect is raised, can be implemented with shallow handlers and recursion.}
Continuations are used by applying them like functions.
To find the closest handler and capture the continuation when raising an effect, $\keyword{do}\ v$ evaluates to an \emph{active effect} $\S(\hole)[v]$, which then gradually captures its surrounding neutral evaluation context using the rule $N_1[\S(N_2)[v]]\steps \S(N_1\dplus N_2)[v]$, until the immediate surrounding context becomes a \keyword{try} block.
At this point, the $N$ parameter of the active effect construction is the continuation to resume the execution.

\pagebreak\stateHandlerFig{0pt}
\Cref{fig:state-handler-code} shows an example of a handler implementing a global state effect supporting \effecttag{read} and \effecttag{write} operations.
The handler uses a state-passing style. %
The recursive function \function{go} takes a continuation $k_0$, a value $r$ to pass to the continuation, and the current global state $\sigma$.
It runs the continuation under a handler that expects raised effects to be pairs of the form $(\efft, w)$, where $\efft$ is a \emph{tag} indicating whether the operation is a \effecttag{read} or \effecttag{write}.
Based on the tag, it recursively calls \function{go} with the appropriate return value and updated state.
If the tag does not match \effecttag{read} or \effecttag{write}, it re-raises the effect to allow composition with another handler for other effects.

In addition to effect handlers, \HandlerLang{} has two built-in primitive effects.
The first is the \keyword{pick} expression, which non-deterministically evaluates to an arbitrary boolean.
The second is $\ghostcode{\keyword{observe}\ v}$, which performs a labeled transition with the label given by the value $v$.
These $\ghostcode{\keyword{observe}}$ statements are used as a form of ghost code and underlie our support for prophecy variables as discussed in \Cref{sec:proph}.
It might seem that including these two effects as primitives contradicts the claim of being able to bootstrap everything from effect handlers.
In \refappendix{sec:predet} we present an alternative core calculus with
these two effects removed and show that they too can be encoded in terms of effect handlers.
However, in the body of this paper, we include these primitive effects as they are relatively simple.

\subsection{Core \unlog{} Logic}
\label{subsec:unlog-core}

To reason about programs written in \HandlerLang, we make use of \unlog, a separation logic built on top of the Iris framework~\citep{jung2018-iris-groundup}, adapted from the Hazel logic for effect handlers~\citep{vilhena2021-hazel}.
This section first presents the basic core of \unlog, which is essentially a subset of Hazel.
Later sections will describe additional generalizations that go beyond Hazel.

\unlog{} uses a weakest precondition assertion of the form $\ewp[]{e}[]{\Psi}{\Phi}$ for reasoning about programs.
In this assertion, $\expr$ is a program expression, $\Phi$ is a postcondition, and $\Psi$ is a \emph{protocol} that describes the specifications for effect handlers that are active as $e$ executes.
This assertion says that inside an evaluation context that provides handlers satisfying $\Psi$, evaluating $e$ does not get stuck, and if $e$ terminates with value $v$, then the assertion $\Phi(v)$ holds.
More concretely, the protocol $\Psi$ is a predicate of type $\Val \to (\Val \to \iProp) \to \iProp$, where the first argument is the value of the effect, and the second argument is the postcondition at the time the effect is raised.
Throughout this paper, we require all protocols to be \emph{monotonic}~\citep{vilhena2022-thesis}.
A protocol $\Psi$ is monotonic if $(\All w . \Phi(w) \wand \Phi'(w)) \vdash \Psi(v, \Phi) \wand \Psi(v, \Phi')$ for all $v$, $\Phi$, and $\Phi'$.
This is necessary for the monotonicity (\refrule{ewp:mono}) and frame (\refrule{ewp:frame}) rules of $\ewpNoArg$ shown below.
Meanwhile, it enforces a one-shot continuation discipline in the logic, which simplifies our presentation and yet suffices for our purposes.
We will further discuss this design choice in \Cref{sec:related-handlers}.

\Cref{fig:ewp-nofupd} lists a selection of reasoning rules for the \ewpNoArg{}~assertion.
Most of these rules are similar to standard weakest precondition rules in separation logics.
The key rule for reasoning about effects is \refrule{ewp:do}, which says that to raise value $v$, it suffices to show that the protocol holds for the value $v$ and the current postcondition $\Phi$.

\begin{figure}[t]
\begin{mathpar}
\mprset{fraction=--\ast}
\inferrule[\hypertarget{ewp:do}{Ewp-Do}]
{\Psi(v,\Phi)}
{\ewp[]{\keyword{do}\ v}[]{\Psi}{\Phi}}\and
\inferrule[\hypertarget{ewp:mono}{Ewp-Mono}]
{\Psi\sqsubseteq\Psi' \\
\All v .\Phi(v)\wand\Phi'(v)\\\ewp e{\Psi}{\Phi}}
{\ewp e{\Psi'}{\Phi'}}\and
\inferrule[\hypertarget{ewp:frame}{Ewp-Frame}]
{R \\ \ewp[]e[]{\Psi}{\Phi}}
{\ewp[]e[]{\Psi}{\Ret v . R\ast\Phi(v)}}\and
\inferrule[\hypertarget{ewp:value}{Ewp-Value}]
{\Phi(v)}
{\ewp[]v[]{\Psi}{\Phi}}\hspace{\fill}
\inferrule[\hypertarget{ewp:step}{Ewp-Step}]
{\All e'.e\step e'\wand\ewp[]{e'}[]{\Psi}{\Phi}\\\Exists e'.e\step e'}
{\ewp[]{e}[]{\Psi}{\Phi}}\hspace{\fill}
\inferrule[\hypertarget{ewp:bind}{Ewp-Bind}]
{\ewp[]e[]{\Psi}{\Ret v.\smash{\ewp[]{N[v]}[]{\Psi}{\Phi}}}}
{\ewp[]{N[e]}[]{\Psi}{\Phi}}
\end{mathpar}
\caption{Selected reasoning rules about the effect weakest precondition ${\ewp[]{e}[]{\Psi}{\Phi}}$}\label{fig:ewp-nofupd}
\end{figure}

For example, for the state handler in \Cref{fig:state-handler-code}, we use the \textlog{STATE} protocol%
\begin{align*}
\textlog{READ}^\gamma(v,\Phi)&\triangleq\Exists x . v=(\effecttag{read},())\ast S^\gamma(x) \ast(S^\gamma(x)\wand\Phi(x))\\
\textlog{WRITE}^\gamma(v,\Phi)&\triangleq\Exists x,y . v=(\effecttag{write},y)\ast S^\gamma(x)\ast(S^\gamma(y)\wand\Phi(()))\\
\textlog{STATE}^\gamma(v, \Phi)&\triangleq\textlog{READ}^\gamma(v, \Phi)\lor\textlog{WRITE}^\gamma(v, \Phi)
\end{align*}
where $S^\gamma(x)$ is a predicate (defined later) that uses a piece of \emph{ghost state} with the name $\gamma$ to assert that the current value of the global state is $x$.
The first component of the protocol is \textlog{READ}, which says that when the effect tag is \effecttag{read}, then the client must show $S^{\gamma}(x)$ for some $x$, in which case the protocol gives back $S^{\gamma}(x)$ for proving the postcondition $\Phi$ instantiated with the value $x$, indicating that the return value of the effect is $x$.
The second component is the \textlog{WRITE} protocol, which updates the given $S^\gamma$ predicate from value $x$ to the value $y$ being written and returns back the unit value.
Finally, \textlog{STATE} is the disjunction of these two protocols.

By applying \refrule{ewp:do}, we obtain the following derived rules for reasoning with this protocol:
\begin{mathpar}
\mprset{fraction=--\ast}
\inferrule[\hypertarget{ewp:state-read}{Ewp-Read}]
{S^{\gamma}(x)}
{\ewp[]{\keyword{do}\ (\effecttag{read},())}[]{\textlog{STATE}^\gamma}{\Ret v. v = x\ast S^\gamma(x)}}\hfill
\inferrule[\hypertarget{ewp:state-write}{Ewp-Write}]
{S^{\gamma}(x)}
{\ewp[]{\keyword{do}\ (\effecttag{write},y)}[]{\textlog{STATE}^\gamma}{\Ret v. v = () \ast S^{\gamma}(y)}}
\end{mathpar}

\paragraph{Installing Handlers}
So far, we have seen how a client can reason about effects when an appropriate protocol is part of the $\ewpNoArg$ assertion.
Protocols are added to the $\ewpNoArg$ when a handler is installed using the \refrule{ewp:try} rule shown below.
Using the \unlog{} approach, we think of the language and logic as being extended to support new effects by adding handlers, so applying this rule forms the core proof obligation of a developer trying to extend the program logic.
\begin{mathpar}
\mprset{fraction=--\ast}
  \inferrule[\hypertarget{ewp:try}{Ewp-Try}]
  {\raisebox{1.5ex}{\ensuremath{\ewp[]{e}[]{\Psi}{\Phi}}} \\
    {\begin{aligned}[b]
      &\left(\All v_2 . \Phi(v_2)\wand\ewp[]{e_2}[]{\Psi'}{\Phi'}\right) \land{} \\[-0.2em]
      &\left(\All v_1,k_1 .  \Psi(v_1,\Lam w . \ewp[]{k_1\ w}[]{\Psi}{\Phi})\wand\ewp[]{e_1}[]{\Psi'}{\Phi'}\right)
    \end{aligned}}}
{\ewp[]{\keyword{try}\ e\ \keyword{with}\ v_1\ k_1\Rightarrow\ e_1\mid\keyword{ret}\ v_2\Rightarrow e_2}[]{\Psi'}{\Phi'}}
\end{mathpar}

In the \refrule{ewp:try} rule, below the line, we start with a protocol $\Psi'$, and in the first premise above the line,
we reason about the expression $e$, which runs under the new handler, with the protocol $\Psi$.
This premise will be proved by a client who may reason as if $e$ has access to the new effects.

Meanwhile, the second premise makes up the proof obligation that justifies extending the logic with this new protocol $\Psi$.
This premise is a logical conjunction with two parts.
The first conjunct is for the case where $e$ evaluates to a value without raising an effect and requires showing that $e$'s postcondition $\Phi$ implies an $\ewpNoArg$ about the remaining expression $e_2$.
The second conjunct is for the case where $e$ raises an effect.
Recall that when $e$ raises an effect, from the client's perspective it must establish the protocol $\Psi$.
Conversely, that means that here in this premise, we get the protocol $\Psi$ instantiated with the value $v_1$ raised with the effect, and a predicate that captures a specification for the continuation $k_1$.
From this, we must prove an $\ewpNoArg$ about the handler code $e_1$ that will run.
These two conjuncts are joined with $\land$ instead of $*$ because only one of these two outcomes will occur, so the rule does not require separate resources for each conjunct.

To apply this rule for the state handler from \Cref{fig:state-handler-code}, and thereby add the \textlog{STATE} protocol, we first need to more carefully define the $S^\gamma(x)$ assertion.
To do so, we use the underlying Iris logic's support for defining ghost state using \emph{resource algebras} (RAs)~\citep{jung2018-iris-groundup}.
In particular, we define it as the \emph{fragment} copy of an (exclusive) \emph{authoritative} RA: $S^\gamma(x)\triangleq\ownGhost{\gamma}{\authfrag x}$.
This RA comes with two types of ghost resources: an authoritative copy $\ownGhost{\gamma}{\authfull x}$ and a fragment copy $\ownGhost{\gamma}{\authfrag x}$.
When these copies are combined, they are guaranteed to agree, \ie $\ownGhost{\gamma}{\authfull x}\ast\ownGhost{\gamma}{\authfrag x'} \vdash x = x'$, and they can be updated to an arbitrary value $y$, with the rule $\ownGhost{\gamma}{\authfull x}\ast\ownGhost{\gamma}{\authfrag x}\vdash \pvs[][][]\ownGhost{\gamma}{\authfull y}\ast\ownGhost{\gamma}{\authfrag y}$, where $\pvs$ is the \emph{basic update modality}.
The basic update modality is the primitive for manipulating ghost resources in the Iris logic.
The assertion $\pvs P$ says that we can update our ghost resources and obtain $P$.
The modality can be eliminated at any suitable time during program verification.

The handler $\function{run}_{\effecttag{state}}$ allocates ghost states $\smash{\ownGhost{\gamma}{\authfull \textit{init}}}$ and $\smash{\ownGhost{\gamma}{\authfrag \textit{init}}}$ at a fresh ghost name $\gamma$, where $\textit{init}$ is the initial value for the state that is passed in.
It keeps its authoritative copy $\smash{\ownGhost{\gamma}{\authfull \textit{init}}}$, and passes the fragment copy $\smash{\ownGhost{\gamma}{\authfrag \textit{init}}}$, which is $S^{\gamma}(\textit{init})$, to the client.
Whenever the client raises an effect, it must show the value satisfies $\textlog{STATE}^{\gamma}$, which is then passed to the handler code.
The handler proof uses the $S^{\gamma}(x)$ that is included in $\textlog{STATE}^{\gamma}$ and combines it with its corresponding authoritative copy of the ghost state to carry out the read or write.
Note that the handler recursively calls $\function{go}$, thereby reinstalling the handler and running the continuation.
To reason about this recursion, we use L\"{o}b induction from the underlying Iris logic~\citep{jung2018-iris-groundup,Lob_1955}.
Altogether, we obtain the following derived rules for \function{go} and $\function{run}_{\effecttag{state}}$, starting from a base protocol $\bot$ defined as $\bot(v, \Phi) \eqdef\FALSE$.
\begin{mathpar}
\mprset{fraction=--\ast}
\inferrule[\hypertarget{ewp:state-go}{Ewp-StateGo}]
{\ownGhost{\gamma}{\authfull \sigma} \\
\ewp[]{k\ r}{\textlog{STATE}^\gamma}{\Phi}}
{\ewp[]{\function{go}\ k\ r\ \sigma}{\bot}{\Phi}}
\and
\inferrule[\hypertarget{ewp:state-run}{Ewp-StateRun}]
{\All \gamma .  S^\gamma(\textit{init})\wand
\ewp[]{\textit{main}\ ()}{\textlog{STATE}^\gamma}{\Phi}}
{\ewp[]{\function{run}_{\effecttag{state}}\ \textit{main}\ \textit{init}}{\bot}{\Phi}}
\end{mathpar}

\paragraph{Building a Hierarchy of Effects}

The previous example showed how to go from no effects (represented by protocol $\bot$) to the state effect (protocol $\textlog{STATE}^{\gamma}$).
In practice, we want to accumulate effects by nesting additional handlers within the handler for $\textlog{STATE}^{\gamma}$.
To that end, as in Hazel, we define a combinator $\oplus$ on protocols by $(\Psi_1 \oplus \Psi_2)(v, \Phi) \eqdef \Psi_1(v, \Phi) \lor \Psi_2(v, \Phi)$.
In other words, $\Psi_1 \oplus \Psi_2$ represents that a client may choose to use effects from either $\Psi_1$ or $\Psi_2$, or both.
For example, $\textlog{STATE}^{\gamma} = \textlog{READ}^{\gamma} \oplus \textlog{WRITE}^{\gamma}$.
Applying this operation to protocols results in a ``larger'' protocol.
This is formally captured by a preorder relation on protocols $\Psi_1\sqsubseteq \Psi_1 \oplus \Psi_2$.
Intuitively, $\Psi_1 \oplus \Psi_2$ is larger than $\Psi_1$ because the former permits the client to raise more kinds of effects.
Using the \refrule{ewp:mono} rule, we generalize \refrule{ewp:state-read} and \refrule{ewp:state-write} accordingly: rather than requiring exactly the protocol $\textlog{STATE}^{\gamma}$, we require that the protocol is some $\Psi$ such that $\textlog{STATE}^{\gamma} \sqsubseteq \Psi$.

Similarly, the \refrule{ewp:state-go} and \refrule{ewp:state-run} specifications for installing the handler do not need to start from the base protocol $\bot$.
Instead, they can start from an arbitrary protocol $\Psi$, and---so long as $\Psi$ does not already handle the tags for $\effecttag{read}$ and $\effecttag{write}$---the client code would then operate with protocol $\Psi \oplus \textlog{STATE}^{\gamma}$.
This enables the state handler to be composed with an arbitrary context of previously installed handlers, allowing a logic developer to mixin rules for state with other effects.
Formally, a protocol $\Psi$ is said to \emph{only handle} a set of tags $T$ if $\Psi(v, \Phi) \vdash \Exists t, w .
v = (t, w) \land t \in T$.
We write $\tags(\Psi)$ for the smallest set $T$ such that $\Psi$ only handles $T$.
When installing the state protocol, we require $\effecttag{read},\effecttag{write} \notin \tags(\Psi)$
for the protocol $\Psi$ prior to installation.
As another example of effects, we have implemented a handler for a heap effect supporting dynamically allocated higher-order references, with rules for local reads and writes.
This handler uses the \textlog{STATE} protocol to store a global map representing the heap, and provides its own \textlog{HEAP} protocol for clients to use.
\begin{align*}
\textlog{HEAP}&\eqdef\textlog{ALLOC}\oplus\textlog{FREE}\oplus\textlog{LOAD}\oplus\textlog{STORE}\\
\textlog{ALLOC}(v,\Phi)&\eqdef\Exists x.v=(\effecttag{alloc},x)\ast(\All\ell.\ell\mapsto x\wand \Phi(\ell))\\
\textlog{FREE}(v,\Phi)&\eqdef\Exists \ell, x.v=(\effecttag{free},\ell)\ast\ell\mapsto x\ast\Phi(())\\
\textlog{LOAD}(v,\Phi)&\eqdef\Exists \ell,q,x.v=(\effecttag{load},\ell)\ast\ell\stackrel{q}{\mapsto}x\ast(\ell\stackrel{q}{\mapsto}x\wand\Phi(x))\\
\textlog{STORE}(v,\Phi)&\eqdef\Exists \ell,x,y.v=(\effecttag{store},(\ell,y))\ast\ell\mapsto x\ast(\ell \mapsto y\wand\Phi(()))\\
\keyword{ref}\ v\eqdef\keyword{do}\ (\effecttag{alloc},v)&\qquad
!\ell \eqdef\keyword{do}\ (\effecttag{load},\ell)
\qquad \ell\gets v\eqdef \keyword{do}\ (\effecttag{store},(\ell,v))
\end{align*}
The \emph{fractional} points-to connective $\ell\stackrel{q}{\mapsto} v$~\cite{boyland2003-fractional} (with $\ell\mapsto v$ short for $\ell\stackrel{1}{\mapsto} v$) is just a piece of ghost state. It also depends on a ghost name $\gamma_{\textlog{heap}}$, but since $\gamma_{\textlog{heap}}$ is typically globally fixed, we omit this name when the language under consideration only has one heap.
The \textlog{HEAP} protocol provides the standard separation logic interface for the heap, matching the non-atomic heap laws of the \HeapLang{} logic, the example program logic distributed with Iris.\footnote{Here we omit from \textlog{ALLOC} the Iris ``meta token'', which associates ghost state with a location; we also omit the array allocation operation, which allocates several consecutive locations at once.}

In this way, we compositionally build logical support for a collection of effects starting from the $\bot$ protocol, extending the core pure logic with support for these effects.
This approach is grounded in an adequacy theorem, which shows that the logic starting with the $\bot$ protocol is sound.
\begin{theorem}[Adequacy, Core \unlog]\label{thm:unlog-adequacy-core}
  Let $\varphi$ be a first-order predicate (\ie a Rocq \texttt{Prop}).
  If $\vdash \ewp[]e[]{\bot}{\varphi}$, then executing $e$ never gets stuck, and if $e\steps v$ then $\varphi(v)$ holds.
\end{theorem}

As in many other works on program logic, this theorem only considers \emph{partial correctness}, \ie it is a \emph{safety} property and it does not guarantee that the program always terminates.
\section{Concurrency and Extensible Worlds}\label{sec:conc}

In the effects we have seen so far, the handler always immediately returns control to the client that raised the effect.
However, for other kinds of effects, the handler may instead pass control to other client code.
To help reason about these kinds of handlers, \unlog{} uses a new feature called \emph{extensible worlds}.

\begin{wrapfigure}{r}{0.591\linewidth}
\vspace{-5.5mm}
\(\begin{codeblock}{r@{}l@{\,}}
&\function{run}_{\effecttag{conc}}\triangleq{}&\Lam \textit{main} . \function{go}\ \lbag(\textit{main},(),\IsMain)\rbag\\
&\text{where}\ \function{go}\triangleq{}&\keyword{rec}\ \function{go}\ \textit{pool}.\\
\end{codeblock}\)
\(\begin{codeblock}{r@{}l@{\,}c@{\,}l}
&&\multicolumn{3}{l}{\quad\keyword{let}\ ((k_0,r,t),\textit{pool\/}):=\function{choose}\ \textit{pool}\ \keyword{in}}\\
&&\multicolumn{3}{l}{\quad\keyword{try}\ k_0\ r\ \keyword{with}}\\
&&\qquad\phantom{\mid{}} v\ k&\Rightarrow&\keyword{match}\ v\ \keyword{with}\\
&&&&{\phantom{\mid{}}(\effecttag{fork},e)\Rightarrow\function{go}\ (\lbag(e,(),\IsChild),(k,(),t)\rbag\uplus\textit{pool\/})}\\
&&&&{\mid\mathmakebox[\widthof{\ensuremath{(\effecttag{fork},e)}}][r]{(\efft,w)}\Rightarrow\function{go}\ (\lbag(k,\keyword{do}\ (\efft, w),t)\rbag\uplus\textit{pool\/})}\\
&&\qquad\mid \keyword{ret}\ v&\Rightarrow& \keyword{if}\ t=\terminated\IsMain\ \keyword{then}\ v\\
&&&&\keyword{else}\ \function{go}\ (\lbag((\Lam\_ . v),(),\terminated{t})\rbag\uplus\textit{pool\/})\\
\end{codeblock}
  \)
\caption{A handler for concurrency}\label{fig:conc-handler}
\end{wrapfigure}

A key example of an effect where this mechanism is useful is preemptive concurrency.
\Cref{fig:conc-handler} shows an implementation of a concurrency handler.
It represents the collection of threads as a bag (\ie multiset) called \textit{pool}, and uses a function \function{choose} that non-deterministically removes an element from the bag and returns it along with the remaining bag.
Each thread in \textit{pool} is represented by a triple of (continuation, result of last effect, thread type), where the thread type can be either a main thread $\IsMain$, a child thread $\IsChild$, or their terminated variants $\terminated\IsMain$ and $\terminated\IsChild$.

At each iteration, the scheduler non-deterministically chooses one thread from \textit{pool} and lets it run for as many pure steps as it can until it raises an effect or terminates.
If the thread raises a \effecttag{fork} effect, the scheduler will push both the new thread $(e,(),\IsChild)$ and the old thread $(k,(),t)$ to the thread pool.
If the thread raises another type of effect, the scheduler will forward the effect to an outer handler by re-raising it, collect its result, and put the old thread back in the thread pool.
Finally, if the thread terminates, the scheduler will mark the thread as terminated and put it back into \textit{pool}.
If \function{choose} later selects the terminated main thread $\terminated\IsMain$, the scheduler stops and returns the main thread's final value $v$.
Thus, other threads may keep running for a while even after the main thread has terminated, which corresponds to the operational semantics of the \HeapLang~language.
Note that the scheduler itself is agnostic to outer handlers, and in particular, it makes no assumption on the communication model between threads, which is implemented by outer handlers.

Because the handler may pass control to other threads when an effect is raised, we now need to reason about coordination between threads, which is what extensible worlds will enable.

\subsection{Background: Iris Invariants and Fancy Updates}
To motivate extensible worlds, let us first recall how modern concurrent separation logics (CSLs) like Iris handle reasoning about interaction between different threads.
By default, CSL allows for local reasoning about different threads in a concurrent system by dividing up state and resources into separate disjoint parts using separating conjunction, with each thread having ownership of some portion of the state.
However, in some cases, threads need to \emph{share} ownership of state.
To do so, CSLs make use of \emph{invariants}.
In Iris, an invariant assertion $\smash{\knowInv{\mathcal{N}}{P}}$ says that $P$ is an invariant that holds between all program steps.
The $\mathcal{N}$ annotation is a \emph{name} given to this invariant.
These assertions are \emph{persistent}, ensuring that $\smash{\knowInv{\mathcal{N}}{P} \vdash \knowInv{\mathcal{N}}{P} * \knowInv{\mathcal{N}}{P}}$, so each thread can have its own copy of the assertion.
When carrying out a proof about a thread, we access the underlying assertion $P$ by ``opening'' the invariant using the following rule:
\[
\mprset{fraction=--\ast}
\inferrule[\hypertarget{wp:inv-iris}{Wp-InvAcc}]
{\knowInv{\mathcal{N}}{P} \\ \mathcal{N}\subseteq\E \\
\later P\wand \smash{\wpre{e}[\E\setminus\mathcal{N}]{\Ret x. \later P\ast\Phi(x)}} \\
\atomic(e)}
{\smash{\wpre{e}[\E]{\Phi}}}
\]
The third premise of this rule has us prove the weakest precondition under the assumption that $P$ holds (guarded by a later modality $\later$~\citep{nakano2000-later,appel2007-later,birkedal2012-later} for soundness reasons, which intuitively means $P$ holds after the program takes one step), so long as we reestablish $P$ in the postcondition of $e$.
Here, $e$ must be atomic, meaning it reduces to a value in a single step, so that by reestablishing $P$ in the postcondition, we ensure that $P$ will continue to hold before and after each step.
The \emph{mask} $\E$ is a set that tracks invariants that have not yet been opened.
The invariants in $\E$ are said to be \emph{closed} or \emph{enabled}, while all other invariants are \emph{open} or \emph{disabled}.

In fact, in Iris, \refrule{wp:inv-iris} is a derived rule.
Iris uses a more primitive mechanism called a \emph{fancy update modality} of the form $\pvs[\E_1][\E_2]$ that encodes the process of opening and closing invariants.
Informally, the assertion $\pvs[\E_1][\E_2] P$ states that starting with all invariants in $\E_1$ being enabled, and then opening/closing invariants so as to end up with $\E_2$ being enabled, it is possible to prove $P$.
Then the \refrule{wp:inv-iris} rule can be derived from the following two rules.
\begin{mathpar}
\mprset{fraction=--\ast}
\inferrule[\hypertarget{fupd:inv-acc}{Fupd-InvAcc}]
{\knowInv{\mathcal{N}}{P}\\ \mathcal{N}\subseteq\E}
{\pvs[\E][\E\setminus\mathcal{N}]\later P\ast(\later P\wand \pvs[\E\setminus\mathcal{N}][\E]\TRUE)}\and
\inferrule[\hypertarget{wp:atomic}{Wp-Atomic}]
{\pvs[\E_1][\E_2]\wpre{e}[\E_2]{\Ret x. \pvs[\E_2][\E_1]\Phi(x)} \\ \atomic(e)}
{\wpre{e}[\E_1]{\Phi}}
\end{mathpar}
These rules are notationally heavy, but the rule on the left captures the process of opening an invariant with the update modality.
Starting from masks in $\E$, we end up with masks in $\E \setminus \mathcal{N}$, and get $\later P$.
Additionally, we get that by supplying $\later P$ we can close the invariant, as represented by the $\pvs[\E \setminus\mathcal{N}][\E]\TRUE$.
Meanwhile, the rule on the right is what allows us to actually use the fancy update modality to open invariants when reasoning about an atomic expression $e$, so long as the postcondition also includes a modality to close those same invariants.

Under the hood, the semantic model for this $\pvs[\mask_{1}][\mask_{2}]$ modality uses a mechanism called \emph{world satisfaction}.
Essentially, the Iris definition of $\pvs[\mask_{1}][\mask_{2}]$ tracks the set of all of the enabled/disabled invariants, and requires that for each enabled invariant, there are resources ensuring the invariant holds.
This bundle of resources is called a world.

Although the Iris invariant mechanism is very expressive and flexible, it has some limitations.
As a result, some prior projects have found it necessary to modify this notion of invariant.
For example, both Perennial~\citep{chajed2019-perennial} and Nola~\citep{yusuke2025-nola} have considered alternate forms of invariant assertions, the former to encode invariants that govern behavior when a program crashes, and the latter to reason about termination.
One key issue, for our purposes, is that the notion of atomicity and the way invariants can be used in a rule like \refrule{wp:atomic} are closely tied to the built-in preemptive concurrency in Iris.
Instead, we want to allow handler implementers to define a notion of invariant suitable for the kind of effect they are modeling.

\subsection{Extensible Worlds}
To achieve this kind of extensibility, \unlog{} does not fix a single baked-in world in the interpretation of the fancy update.
Instead, \unlog{} parameterizes the $\ewpNoArg$ and fancy update modalities by a customizable notion of world.
The full version of the \unlog{} $\ewpNoArg$ assertion then has the form $\ewp[]{e}[\W_1,\W_2]{\Psi}{\Phi}$,
where $\W_1$ is an arbitrary Iris proposition representing the world at the start of $e$'s execution, and $\W_2$ is the world after $e$ finishes executing.
Meanwhile, the fancy update modality becomes the \emph{world update modality}, defined as $\pvs[\W_1][\W_2] P\eqdef\W_1\wand\pvs(P\ast\W_2)$, which intuitively states that given access to the initial world $\W_1$, we can prove $P$ and the resulting world $\W_2$, potentially performing ghost updates using the basic update modality.
When the starting world $\W$ is the same as the ending world, we simply write $\ewp{e}[\W]{\Psi}{\Phi}$ and $\pvs[\W][][]$.

Worlds are just normal Iris assertions, but it is nevertheless helpful to think of them more abstractly.
The combination of two worlds, written $\W_1 \oplus \W_2$, is defined as $\W_1 * \W_2$.
We impose a preorder $\sqsubseteq$ on worlds defined by $\W_1 \sqsubseteq \W_2 \eqdef \Exists \W'. \left(\W_2 \dashv\vdash \W_1 \oplus \W'\right)$.
Here, larger worlds have more resources, and the minimal element $\bot$ is the proposition $\TRUE$.
Thus, with an update like $\pvs[\W_1][\W_2][] P$, when $\W_2 \sqsubseteq \W_1$, we are shifting to a \emph{smaller} world, and give the difference between $\W_2$ and $\W_1$ to the proof of $P$.
Conversely, shifting to a larger world with $\W_1 \sqsubseteq \W_2$ requires putting in the difference between $\W_1$ and $\W_2$.

The rules we have seen previously for the $\ewpNoArg$ are generalized to account for worlds.
Selected generalized rules about $\ewpNoArg$ and the world update modality are shown in \Cref{fig:ewp-client}.
\refrule{wupd:elim} eliminates a world update modality from an assumption, updating the worlds on the goal accordingly.
\refrule{wupd:intro} introduces a world update $\pvs[\W_1][\W_2] P$ from a larger world into a smaller world and gives a token $\pvs[\W_2][\W_1]\TRUE$ to restore the original world.
\refrule{wupd:frame} allows for ``framing out'' an unnecessary world $\W$ that occurs in both the starting and ending world.

Unlike the Iris \refrule{wp:atomic} rule, which only allows masks to change around an atomic step, the \refrule{ewp:wupd} rule allows us to apply a world update that changes the starting world for any expression, and \refrule{ewp:wupdpost} changes the corresponding ending world.
This is allowed because, unlike standard Iris, where the scheduler could preempt a thread at any point, in the effect handler approach, control can only be transferred when an effect is raised.
This means the starting world does not need to be immediately restored.
Instead, only when an effect is raised must the world be in an appropriate configuration, depending on whether the protocol $\Psi$ requires it or not.
In \refrule{ewp:dowupd}, we start by shifting to the bottom world $\bot$, and then in the continuation passed to the protocol $\Psi$, we must restore $\W_2$.
Finally, %
we can frame out an unused world in $\ewpNoArg$ with \refrule{ewp:world:frame}.

\begin{figure}[t]
\begin{mathpar}
\inferrule[\hypertarget{wupd:elim}{Wupd-Elim}]
{Q\vdash\pvs[\W_2][\W_3] P}
{\left(\smash{\pvs[\W_1][\W_2] Q}\right)\vdash \pvs[\W_1][\W_3] P}
\and
\mprset{fraction=--\ast}
\inferrule[\hypertarget{wupd:intro}{Wupd-Intro}]
{(\pvs[\W_2][\W_1] \TRUE) \wand P\\ \W_2\sqsubseteq\W_1}
{\pvs[\W_1][\W_2] P}
\and
\inferrule[\hypertarget{wupd:frame}{Wupd-Frame}]
{\pvs[\W_1][\W_2] Q}
{\pvs[\W_1 \oplus \W][\W_2 \oplus \W] Q}
\and
\inferrule[\hypertarget{ewp:value-wupd}{Ewp-ValueWupd}]
{\pvs[\W_1][\W_2][] \Phi(v)}
{\ewp[]v[\W_1,\W_2]{\Psi}{\Phi}}\and
\inferrule[\hypertarget{ewp:wupd}{Ewp-WupdPre}]
{\pvs[\W_1][\W_2][]\ \ewp[]e[\W_2,\W_3]{\Psi}{\Phi}}
{\ewp[]e[\W_1,\W_3]{\Psi}{\Phi}}\and
\inferrule[\hypertarget{ewp:wupdpost}{Ewp-WupdPost}]
{\ewp[]e[\W_1,\W_2]{\Psi}{\Ret v . \smash{\pvs[\W_2][\W_3]\Phi(v)}}}
{\ewp[]e[\W_1,\W_3]{\Psi}{\Phi}}\and
\inferrule[\hypertarget{ewp:dowupd}{Ewp-DoWupd}]
{\pvs[\W_1][\bot][]\Psi(v,(\Lam r . \pvs[\bot][\W_2][]\Phi(r)))}
{\ewp[]{\keyword{do}\ v}[\W_1,\W_2]{\Psi}{\Phi}}\and
\inferrule[\hypertarget{ewp:world:frame}{Ewp-WorldFrame}]
{\ewp[]e[\W_1,\W_2]{\Psi}{\Phi}}
{\ewp[]e[\W_1 \oplus \W,\W_2 \oplus \W]{\Psi}{\Phi}}
\and
\end{mathpar}
\caption{Selected reasoning rules for $\pvs[\W_1][\W_2]$ and \ewpNoArg{} with worlds}\label{fig:ewp-client}
\end{figure}

\paragraph{Recovering Iris Invariants}

It is straightforward to recover Iris-style impredicative invariants and the Iris fancy update modality in this more general world setting.
As was described above, the standard Iris definition fixes some particular world in its definition of fancy updates, and
uses ghost state to track the enabled invariants.
Let us write $\WInvTok{\E}$ for the assertion that bundles the world with the ghost state saying that mask $\E$ is enabled.\footnote{Here the letter ``I'' stands for ``invariant'', so the whole notation means a token for invariants.}
Then we recover the following analogue of the \refrule{fupd:inv-acc} rule that we saw earlier.
\begin{mathpar}
\mprset{fraction=--\ast}
\inferrule[\hypertarget{wupd:inv-acc}{Wupd-InvAcc}]
{\knowInv{\mathcal{N}}{P}\\ \mathcal{N}\subseteq\E}
{\pvs[\WInvTok{\E}][\WInvTok{\E\setminus\mathcal{N}}]
  \later P\ast(\later P\wand \pvs[\WInvTok{\E\setminus\mathcal{N}}][\WInvTok{\E}]\TRUE)}
\end{mathpar}
Moreover, by combining this rule with \refrule{ewp:world:frame}, we can support Iris invariants while including other possible components in the world.
As we will see in \Cref{subsec:crash-inv}, this allows us to encode a mechanism similar to Perennial's crash borrows~\citep{tassarotti2022-perennial-crash-borrow} while retaining standard Iris invariants.

\subsection{Protocol for the Concurrency Handler}
Now that we have an extensible mechanism for encoding invariants that hold across threads, we turn to the protocol for the concurrency handler.

One challenge is that the concurrency handler in \Cref{fig:conc-handler} is generic, in the sense that it does not know about the other effects that might be supported by outer handlers.
It simply re-raises those effects to the outer handler and potentially transfers control to another thread when the effect returns.
This means that the outer handlers could, say, implement shared-memory or channel-based message-passing concurrency, or some combination thereof.
Ideally, the protocol we develop should work for different kinds of outer effects.

To achieve this, the first ingredient is a \emph{protocol transformer} $\textlog{ATOM}_{\W}$ that lifts a protocol for these outer effects into a concurrent protocol, where $\W$ is a world that describes shared resources that can be accessed by different threads.
To do this lifting, $\textlog{ATOM}$ transforms $\Psi$ to ensure that as part of raising an effect governed by $\Psi$, the thread must be able to restore the world $\W$.
In addition, when the handler returns control to a thread, it promises that $\W$ will hold.
Formally, this is captured through the following definition:
$$
\textlog{ATOM}_{\W}(\Psi)(v,\Phi)\triangleq\Psi(v,\Lam r . \pvs[\bot][\W][]\pvs[\W][\bot][] \Phi(r))
$$
The first $\pvs[\bot][\W]$ is an obligation for the thread raising the effect to establish $\W$ after the effect completes.
Meanwhile, because the second $\pvs[\W][\bot]$ precedes the continuation $\Phi(r)$, it effectively gives back access to $\W$ before the continuation's $\Phi$ must be proved.

To better understand this definition, let us recall that the intuitive reading of $\pvs[\W_1][\W_2] P$ as a proof obligation is that we are currently in world $\W_1$, and we need to perform an update to the world $\W_2$ and prove $P$. If $\W_1$ is smaller than $\W_2$, we need to prove the difference between $\W_2$ and $\W_1$; conversely, when $\W_1$ is larger than $\W_2$, we obtain the difference between $\W_1$ and $\W_2$.
With this in mind, let us examine what the client needs to prove to raise an effect under the $\textlog{ATOM}_{\W}(\Psi)$ protocol.
Using \refrule{ewp:dowupd}, we derive the following rule:
$$
\mprset{fraction=--\ast}
\inferrule{\pvs[\W_1][\bot]\Psi(v,(\Lam r.\pvs[\bot][\W]\pvs[\W][\bot]\pvs[\bot][\W_2]\Phi(r)))}
{\ewp{\keyword{do}\,v}[\W_1,\W_2]{\textlog{ATOM}_{\W}(\Psi)}{\Phi}}
$$

Here it might be helpful to think of $\W_1$ as the ``local world'' before the effect, $\W$ as the ``shared world'' (for example, containing all the invariants that all threads rely on), and $\W_2$ as the local world after the effect has returned.
In practice, $\W$ is typically larger than or equal to $\W_1$/$\W_2$.
The obligation above the line is equivalent to
$$\pvs[\W_1][\bot]\Psi(v,(\Lam r.\pvs[\bot][\W]\pvs[\W][\W_2]\Phi(r)))$$

The three updates here are used at different places, as depicted in \Cref{fig:atom-explain}.

\begin{itemize}
\item $\pvs[\W_1][\bot]$ is used before raising an effect.
Since it is an update to $\bot$, the client obtains everything in $\W_1$, \eg the ability to open any invariants.
\item $\pvs[\bot][\W]$ is used after the effect has been executed but right before the scheduler switches to another thread.
It requires the client to establish everything in the shared world $\W$, \eg closing all invariants.
Since $\W$ can be larger than $\W_1$, if some invariants were open before raising the effect, the client must also close them now.
\item $\pvs[\W][\W_2]$ is used after the execution of the initial thread resumes.
Again, because $\W$ is larger than $\W_2$, this update allows the client to open more invariants.
\end{itemize}

\begin{figure}[t]
\centering
\begin{tikzpicture}
\fill[color={green!20!white}] (0,0) rectangle (6.5,0.3);
\path[pattern={Lines[angle=-45,distance={4pt},line width=0.2pt]}] (0,0) rectangle (6.5,0.3);
\shade[left color={red!20!white},right color={orange!20!white},middle color={violet!20!white}] (6.5,0) rectangle (9.5,0.3);
\path[pattern={Lines[angle=90,distance={4pt},line width=0.2pt]}] (6.5,0) rectangle (7.5,0.3);
\path[pattern={Dots[radius=0.4pt]}] (7.5,0) rectangle (8.5,0.3);
\path[pattern={Lines[angle=45,distance={4pt},line width=0.2pt]}] (8.5,0) rectangle (9.5,0.3);
\fill[color={green!20!white}] (9.5,0) rectangle (13.5,0.3);
\path[pattern={Lines[angle=-45,distance={4pt},line width=0.2pt]}] (9.5,0) rectangle (13.5,0.3);

\draw (0,0.3) -- +(0,-0.8);
\draw (4,0.3) -- +(0,-0.8);
\draw (6.5,0.3) -- +(0,-0.8);
\draw (9.5,0.3) -- +(0,-0.8);
\draw (13.5,0.3) -- +(0,-0.8);

\node[anchor=base] (cc1) at (2,-0.5) {Client code before};
\node[anchor=north,inner sep=0pt] at (cc1.south) {raising the effect};
\node[anchor=base] (psi) at (5.25,-0.5) {Handling};
\node[anchor=north,inner sep=0pt] at (psi.south) {the effect};
\node[anchor=base] (other) at (8,-0.5) {Other threads};
\node[anchor=base] (cc2) at (11.5,-0.5) {Client code after};
\node[anchor=north,inner sep=0pt] at (cc2.south) {returning from the effect};

\draw[-Stealth] (cc1.west|-,-0.35)--(0,-0.35);
\draw[-Stealth] (cc1.east|-,-0.35)--(4,-0.35);
\draw[-Stealth] (psi.west|-,-0.35)--(4,-0.35);
\draw[-Stealth] (psi.east|-,-0.35)--(6.5,-0.35);
\draw[-Stealth] (other.west|-,-0.35)--(6.5,-0.35);
\draw[-Stealth] (other.east|-,-0.35)--(9.5,-0.35);
\draw[-Stealth] (cc2.west|-,-0.35)--(9.5,-0.35);
\draw[-Stealth] (cc2.east|-,-0.35)--(13.5,-0.35);

\node[anchor=base,inner sep=0pt] (pvs1) at (4,1) {$\pvs[\W_1][\bot]$};
\node[anchor=base,inner sep=0pt] (pvs2) at (6.5,1) {$\pvs[\bot][\W]$};
\node[anchor=base,inner sep=0pt] (pvs3) at (9.5,1) {$\pvs[\W][\W_2]$};

\draw[-Stealth] (4,|-pvs1.south) -- (4,0.3);
\draw[-Stealth] (6.5,|-pvs2.south) -- (6.5,0.3);
\draw[-Stealth] (9.5,|-pvs3.south) -- (9.5,0.3);

\end{tikzpicture}
\caption{Interleaving of threads under the $\textlog{ATOM}_{\W}(\Psi)$ protocol and the use of its world updates}\label{fig:atom-explain}
\end{figure}

As a special case, if we instantiate $\W$ to be $\WInvTok{\top}$, where $\top$ is the full mask saying that all invariants are enabled,
then the $\textlog{ATOM}_{\W}(\Psi)$ protocol requires a thread to close all Iris invariants after an effect completes, just as in the Iris rule \refrule{wp:atomic}.
Under this protocol, we have an analogue of \refrule{wp:inv-iris}, which relaxes the condition $\atomic(e)$ to $\effectatomic(e)$, \ie $e$ consists of several pure steps and one effectful step (\ie a \keyword{do} operation).
\[
\mprset{fraction=--\ast}
\inferrule[\hypertarget{ewp:inv-acc}{Ewp-InvAcc}]
{\knowInv{\mathcal{N}}{P} \\ \mathcal{N}\subseteq\E \\
\later P\wand \smash{\ewp{e}[\WInvTok{\E\setminus\namesp}]{\textlog{ATOM}_{\WInvTok{\top}}(\Psi)}{\Ret x.\later P\ast\Phi(x)}} \\
\effectatomic(e)}
{\smash{\ewp{e}[\WInvTok{\E}]{\textlog{ATOM}_{\WInvTok{\top}}(\Psi)}{\Phi}}}
\]
Although being able to keep invariants open across multiple pure steps might not bring extra expressivity, it results in simpler proofs in some scenarios.
For example, when an invariant contains pure facts needed to perform a pure step, \refrule{ewp:inv-acc} lets us open that invariant once for both the pure step and a subsequent effectful step, whereas the standard Iris rule \refrule{wp:inv-iris} requires opening it twice.
Another benefit is that \refrule{ewp:inv-acc} lets the pure steps of $e$ be used to remove the later modality $\later$ from $P$ after opening the invariant, so that the effectful step afterwards can use the unguarded $P$.
In Iris, this would otherwise require later credits~\cite{spies2022-later}, which not all Iris-based logics support.

To get the final protocol \textlog{CONC} for the concurrency handler, we combine \textlog{ATOM} with a protocol \textlog{FORK} governing the \effecttag{fork} effect.
Because the forked thread will run in the scope of the concurrency handler, \textlog{FORK} must have a recursive dependence on \textlog{CONC}.
\begin{align*}
\textlog{CONC}_\W(\Psi)&\triangleq\textlog{ATOM}_\W(\Psi\oplus\textlog{FORK}_\W(\Psi))\\
\textlog{FORK}_\W(\Psi)(v, \Phi)&\triangleq
\Exists e. v = (\effecttag{fork}, {\Lam\_. e})
\ast
{\later\ewp[]{e}[\W]{\textlog{CONC}_\W(\Psi)}{\Ret \_ . \TRUE} \ast \Phi(())}
\end{align*}
Here, the recursive occurrence of \textlog{CONC} in \textlog{FORK} occurs under the later modality $\later$, so that we can define the result as a \emph{guarded fixed point}~\citep{di-gianantonio2003-contractive-func,jung2018-iris-groundup,ciesielski2007-banach-fixpoint}.
The $\textlog{FORK}$ protocol requires showing an appropriate $\ewpNoArg$ for the forked thread.
Let us introduce a wrapper $\keyword{fork}\ e \eqdef \keyword{do}~(\effecttag{fork}, \Lam\_ . e)$.
We can re-derive the standard $\keyword{fork}$ rule from Iris with this protocol.
\if TT
\begin{mathpar}
  \mprset{fraction=--\ast}
  \inferrule[\hypertarget{ewp:fork}{Ewp-Fork}]
            {{\ewp[]{e}[\W]{\textlog{CONC}_\W(\Psi)}{\Ret\_.\TRUE}}}
            {\ewp[]{\keyword{fork}\ e}[\W]{\textlog{CONC}_\W(\Psi)}{\Ret v. v = ()}}
\end{mathpar}
\else
\begin{mathpar}
  \inferrule[\hypertarget{ewp:fork}{Ewp-Fork}]{}
            {{\ewp[]{e}[\W]{\textlog{CONC}_\W(\Psi)}{\Ret\_.\TRUE}}\proves
            \ewp[]{\keyword{fork}\ e}[\W]{\textlog{CONC}_\W(\Psi)}{\Ret v. v = ()}}
\end{mathpar}
\fi
Finally, we have the specification for $\function{run}_{\effecttag{conc}}$ from \Cref{fig:conc-handler}, which installs the concurrency handler.
\begin{mathpar}
\mprset{fraction=--\ast}
\inferrule[\hypertarget{ewp:conc-run}{Ewp-ConcRun}]
          {\ewp[]{\textit{main}\ ()}[\W]{\textlog{CONC}_\W(\Psi)}{\Phi}\\
          \effecttag{fork} \notin \tags(\Psi)}
          {\ewp[]{\function{run}_\effecttag{conc}\ \textit{main}}[\W]{\Psi}{\Phi}}
\end{mathpar}

In addition to the ability to create threads with \keyword{fork}, we can also model primitive atomic instructions such as compare-and-swap (CAS), atomic exchange (XCHG), or fetch-and-add (FAA).
To do so, we just need to define an additional handler on top of the heap handler and associated protocol \textlog{ATOMHEAP} that models these effects, much like the earlier \textlog{HEAP} protocol did.
\begin{align*}
\textlog{ATOMHEAP}\eqdef{}& \textlog{HEAP}\oplus\textlog{CAS}\oplus\textlog{XCHG}\oplus\textlog{FAA}\\
\textlog{CAS}(v,\Phi)\eqdef{}&\Exists \ell,q,w,x,y.v=(\effecttag{cas},(\ell,x,y))\ast\ell\stackrel{q}{\mapsto} w\ast(q=1\lor w\neq x)\ast{}\\
&(\ell\stackrel{q}{\mapsto}(\text{if }w=x\text{ then } y\text{ else }w)\wand \Phi((w,w=x)))\\
\textlog{XCHG}(v,\Phi)\eqdef{}&\Exists \ell,x,y.v=(\effecttag{xchg},(\ell,y))\ast\ell\mapsto x\ast(\ell\mapsto y\wand\Phi(x))\\
\textlog{FAA}(v,\Phi)\eqdef{}&\Exists\ell,z_1,z_2.v=(\effecttag{faa},(\ell,z_2))\ast\ell \mapsto z_1\ast(\ell\mapsto z_1+z_2\wand \Phi(z_1))
\end{align*}

\subsection{Recovering the \HeapLang{} Logic}
Combining \textlog{ATOMHEAP} and protocols for prophecy variables that will be introduced in \Cref{sec:proph}, and applying the \textlog{CONC} protocol transformer, we obtain a protocol that models all of the operations one finds in \HeapLang{}.
In fact, an $\ewpNoArg$ under these protocols satisfies all reasoning rules that the \HeapLang{} $\textlog{wp}$ satisfies.
Additionally, we have obtained a stronger version of the invariant-opening rule, allowing us to keep invariants open for multiple pure steps.

To facilitate interactive proofs, our Rocq development includes tactics that provide a high-level proof experience similar to that of the \HeapLang{} logic.
Using these tactics, we have ported all programs and specifications in the standard library of \HeapLang{} to \unlog{}, as well as a verification of the Herlihy-Wing queue~\cite{herlihy1990-linear}.
The proofs of these specifications are identical to their \HeapLang{} counterparts except that \HeapLang{} tactics are replaced by corresponding \unlog{} tactics.

\paragraph{Discussion: \unlog{} Concurrency Model}
A careful reader might object that what allows us to derive this stronger invariant-opening rule is the fact that the concurrency handler only transfers control to another thread when an effect is raised.
In contrast, a standard operational semantics for preemptive concurrency typically allows for preemption at \emph{every} step, thereby generating more possible interleavings of thread operations.
Because our handler semantics for concurrency is not generating all of the interleavings that the standard semantics would, one might wonder whether this handler adequately models concurrency.
Fortunately, verifying programs against this handler is still sound because, under partial correctness, the intermediate pure steps between effects are not observable to other threads in the system.
Thus, omitting preemption points between these pure steps cannot hide any safety violations.
In the next section, we introduce a \emph{relational logic} that will allow us to prove this claim rigorously.
\section{Contextual Equivalence of Effectful Programs}\label{sec:relation}

In this section, we develop \binlog{}, a relational logic that allows us to prove correspondences between the behavior of two effectful programs written in \HandlerLang{}.
Just as with unary reasoning, we compositionally derive relational reasoning rules for a number of effects.
Using the resulting logic, we define a logical relations model that can prove contextual equivalences of programs written in typed subsets of \HandlerLang{}.
We apply this logical relations model to prove that inserting additional preemption points in our concurrency handler does not change the set of possible program behaviors, thus justifying the semantics from the previous section  where preemption only occurs when effects are raised.

\subsection{Background: Embedding Relational Logics into Unary Logics}\label{sec:bin:background}

\binlog{} embeds relational reasoning into \unlog{} by encoding a second program as \emph{ghost state}, as in CaReSL~\citep{turon2013-caresl}.
Let us first recall how this ghost state encoding works in modern Iris-based separation logics.
For sequential programs, one first introduces two assertions: $\spec{e}$, which says that the second program (which we call the ``specification'' or ``spec'' program) is currently represented by the expression $e$;
and $\specCtxAlt$, which is an invariant that ensures that the $\spec{e}$ ghost state can only be updated in ways that represent valid transitions in the language's operational semantics.
One needs to further add in assertions to represent the state of this spec program;
for example, a spec points-to assertion $\ell \mapsto_s v$ says that in the spec program, location $\ell$ contains the value $v$, analogous to the standard points-to assertion.
Using these assertions, one derives rules that allow for the spec program to be ``executed'' by applying ghost updates.
To prove a relational property about two programs $e_1$ and $e_2$, it then suffices to derive a judgment of the form
\[ \specCtxAlt * \spec{e_2} \vdash \wpre{e_1}{\Ret v_1. \Exists v_2. \spec{v_2} * \varphi(v_1, v_2)}\]
The soundness theorem of the encoding says that such a derivation implies that, for every execution of $e_1$, if it terminates in a value $v_1$, then there exists a terminating execution of $e_2$ ending in some value $v_2$ such that $\varphi(v_1, v_2)$ holds.
As usual for partial correctness, if an execution of $e_1$ does not terminate, the soundness theorem says nothing about $e_2$.
This basic approach can be generalized to account for concurrency as well by having multiple spec program resources, one for each thread in the concurrent program.

\binlog{} adapts this style of relational reasoning with ghost programs to the setting of effect handlers.
A key challenge is that the usual ghost state encoding requires fixing the primitive effects of the language ahead of time, and requires special treatment in the concurrent case to introduce per-thread spec programs.
In contrast, in \binlog{} these notions are derivable using protocols and worlds, just as with unary reasoning.

\subsection{\binlog{}: A Relational Logic for \HandlerLang}\label{sec:binlog}

To enable extensibility, \binlog{} reasons about spec programs using an \emph{effect specification resource} assertion $\specprot{\W}{e}{\Psi}$ that tracks a spec program $e$ and a protocol $\Psi$ in a world $\W$.
The program $\expr$ can be updated and progressed according to the operational semantics of \HandlerLang{}.
For example, $\specprot{\W}{((\Lam x . e)~v)}{\Psi}$ can be updated to $\specprot{\W}{\subst{e}{x}{v}}{\Psi}$ to reflect the execution of a beta reduction as justified by  \refrule{espec:pure} shown below.
As in \unlog{}, the protocol $\Psi$ describes the effect handlers that are active as $e$ executes.
That is, $\specprot{\W}{N[\keyword{do}~v]}{\Psi}$ can be updated to $\specprot{\W}{N[w]}{\Psi}$ for some $w$ such that $\Phi(w)$ by establishing $\Psi(v, \Phi)$ as enabled by \refrule{espec:do}.
\begin{align*}
  {\specprot{\W}{K[e]}{\Psi} \ast e \steps e'}
  &\vdash
  {\pvs[\W] \specprot{\W}{K[e']}{\Psi}} && \textsc{\hypertarget{espec:pure}{Espec-Pure}} \\
  {\specprot{\W}{N[\keyword{do}~v]}{\Psi}\ast \Psi(v, \Phi)}
  &\vdash
  {\pvs[\W] \Exists w . \specprot{\W}{N[w]}{\Psi} \ast \Phi(w)} && \textsc{\hypertarget{espec:do}{Espec-Do}}
\end{align*}
Using these rules, we obtain derived rules for raising specific effects like global state, much as in the unary case in \Cref{sec:overview}, for example,
\begin{align*}
  {\specprot{\W}{\keyword{do}~(\effecttag{read},())}{\Psi \oplus \textlog{STATE}^\gamma} \ast S^\gamma(v)}
  &\vdash {\pvs[\W] \specprot{\W}{v}{\Psi \oplus \textlog{STATE}^\gamma} \ast S^\gamma(v)} %
  \\
  {\specprot{\W}{\keyword{do}~(\effecttag{write},w)}{\Psi \oplus \textlog{STATE}^\gamma} \ast S^\gamma(v)}
  &\vdash {\pvs[\W] \specprot{\W}{()}{\Psi \oplus \textlog{STATE}^\gamma} \ast S^\gamma(w)} %
\end{align*}
Similarly, we have rules for installing handlers that capture these effects, such as
$$\mprset{fraction=--\ast}
\inferrule{\specprot{\W}{\function{run}_{\effecttag{state}}~\textit{main}~\textit{init}}{\Psi}\\
\effecttag{read},\effecttag{write} \notin \tags(\Psi)}
{\pvs[\W] \Exists \gname . S^\gamma(\textit{init}) \ast \specprot{\W}{\textit{main}~()}{\Psi \oplus \textlog{STATE}^\gamma}}$$

The following adequacy theorem for \binlog{} holds, which requires that both the spec protocol and the $\ewpNoArg$ protocol are $\bot$.
\begin{theorem}[Adequacy, \binlog]\label{thm:binlog-adequacy}
  Let $\varphi$ be a first-order relation. If
  $$ \specprot{\bot}{\expr_{2}}{\bot} \vdash \ewp{\expr_{1}}[\bot]{\bot}{\Ret \val_{1} . \Exists \val_{2} . \specprot{\bot}{\val_{2}}{\bot} \ast \varphi(\val_{1}, \val_{2})}$$
  and $\expr_{1} \steps \val_{1}$ then there exists a value $v_{2}$ such that $\expr_{2} \steps \val_{2}$ and $\varphi(\val_{1}, \val_{2})$.
\end{theorem}
The key to proving this adequacy theorem lies in coming up with a suitable definition of the $\specprot{\W}{e}{\Psi}$ resource that validates the above rules.

\paragraph{Constructing the Effect Specification Resource}
 Much like the $\ewpNoArg$ assertion, the specification resource $\specprot{\W}{e}{\Psi}$ tracks the behavior of the program $\expr$ \emph{under the assumption} that it executes in a program context satisfying the protocol $\Psi$ and a logical world $\W$.
To define $\specProtNoArg$, we first define a more general construction $\genSpecProtNoArg$ which we will specialize to obtain $\specProtNoArg$.
The $\genSpecProtNoArg$ assertion takes some abstract notion of a spec program and transforms it into a version that has protocols and worlds for reasoning about effects.
Specifically, let $\specNoArg\colon \Expr \rightarrow \iProp $ be any predicate that can be angelically updated according to the operational semantics:
$\spec{e} \proves \pvs[\bot] \spec{e'}$ as long as $\expr \steps \expr'$.
Given a choice of a $\specNoArg$ predicate, the $\genSpecProtNoArg$ assertion is defined as
\begin{align*}
  \genSpecProt{\W}{e}{\Psi} &\eqdef{} \Exists K . \spec{K[e]} \ast \spechandler{\W}(\Psi,K)
\end{align*}
where the assertion $\spechandler{\W}(\Psi,K)$ captures that $K$ is an evaluation context that realizes the protocol $\Psi$ indefinitely from the perspective of a program $e$ running inside that context. Formally, this is expressed using a \emph{greatest fixed point} on parameter $K$:\footnote{A guarded fixed point, with the recursive occurrence under a $\later$, would not suffice to derive \refrule{espec:do}: each unfolding would leave a $\later$ that can only be removed when the implementation program takes a step, so the specification program could handle at most one effect per step of the implementation program.}
\begin{align}
 \spechandler{\W}(\Psi,K) \eqdef{} &\All v . \spec{K[v]} \wand \pvs[\W] {\spec{v}}\\
  {}\land{}&
      \All v, N, \Phi .
      \spec{K[\S(N)[v]]} \ast \Psi(v, \Phi) \wand\pvs[\W]\\
  & \phantom{\All v,N,\Phi.}%
    \Exists K', w . \spec{K'[N[w]]} \ast \Phi(w) \ast \spechandler{\W}(\Psi,K')\notag
\end{align}
The two conjuncts of this definition require that
\begin{enumerate}
\item if $\expr$ is a value $v$, then the context terminates with value $v$; and
\item if $\expr$ is an active effect with continuation $N$ and value $v$, \ie $\expr = \S(N)[v]$, and $\Psi(v, \Phi)$ holds, then $N$ is reinstated with some value $w$ in a context $K'$ such that $\Phi(w)$ and $\spechandler{\W}(\Psi,K')$ holds co-recursively.
\end{enumerate}
We can derive generic versions of the \refrule{espec:pure} and \refrule{espec:do} rules for $\genSpecProtNoArg$, as well as examples like the \textlog{STATE} protocol.
Then we obtain $\specProtNoArg$ and specialized versions of these rules by instantiating $\genSpecProtNoArg$ with $\spec{e} \eqdef{} e_{0} \steps e$, where $e_{0}$ is the initial expression that the ghost program starts as.
Later, we will see how instantiating the definition with other choices of the base specification resource allows us to reason about thread-local effects in concurrent execution.

\subsection{Concurrency}\label{sec:relconc}
To reason relationally about effects that do not immediately transfer control back to the raising thread, we need to make use of extensible worlds, just as we did with the unary logic for concurrency.
However, in the relational case, our specification of concurrency has an additional requirement: we want to be able to reason about each thread in the concurrent system individually.
In the encoding of specification programs in CaReSL described above in \Cref{sec:bin:background}, this is achieved by having a specification assertion per thread.
Since each thread can raise effects and use other components of a protocol, we need these per-thread resources to have access to the protocol, just as in $\specProtNoArg$.

To construct this per-thread effect specification resource, we again use the $\genSpecProtNoArg$ construction.
To do so, we first need an underlying per-thread local specification resource $\specthread{\gamma}{t}{e}$ that we will use to instantiate the construction with.
Here, the $\gamma$ parameter is a ghost name and $t$ tracks whether the thread is a main thread $\IsMain$ or a child thread $\IsChild$.
Additionally, we need a \emph{specification context world} $\specCtx{\gamma}{\W}{\Psi}$ that tracks the state of the concurrency handler.
\begin{align*}
  \specthread{\gamma}{t}{e} &\eqdef{} \Exists k, r . \ownGhost{\gamma}{ \authfrag \{ (k, r, t) \}} \ast \left(\All K . \spec{K[k~r]} \wand \pvs[\bot] \spec{K[e]}\right) \\
  \specCtx{\gamma}{\W}{\Psi} &\eqdef{} \Exists B, \textit{pool} . \isBag{B}{\textit{pool}} \ast \ownGhost{\gamma}{\authfull B} \ast \specprot{\W}{\function{run}_{\effecttag{conc}}.\function{go}~\textit{pool}}{\Psi}
\end{align*}

In these definitions, we assert that there is some instance of the concurrency handler from \Cref{fig:conc-handler} installed, and we use ghost state to track the thread pool in the handler.
Specifically, the thread pool is tracked using ghost resources such that $\ownGhost{\gamma}{\authfull B} \ast \ownGhost{\gamma}{ \authfrag B'} \vdash B' \subseteq B$ and $\ownGhost{\gamma}{\authfull B} \vdash \pvs[\W] \ownGhost{\gamma}{\authfull (B \uplus B')} \ast \ownGhost{\gamma}{\authfrag B'}$.
In the definition of $\specthread{\gamma}{t}{e}$ we use this ghost state to assert that a triple of the form $(k, r, t)$ is stored in the pool,  where $k$ is the continuation representing the thread and $r$ is the result of the last effect.
Additionally, we require that there is some way to run $k~r$ so that it will reach the expression $e$.
In other words, the thread currently in the pool may not yet be $e$, but when it is next scheduled to run, it can execute to $e$.
The $\specCtx{\gamma}{\W}{\Psi}$ assertion enforces that in fact there is an underlying $\specProtNoArg$ running the concurrency handler providing the protocol $\Psi$.

This thread-local specification resource fulfills the requirements needed to instantiate $\genSpecProtNoArg$, \ie when $e \steps e'$ then $\specthread{\gamma}{t}{e} \vdash \pvs[\bot] \specthread{\gamma}{t}{e'}$.
When updating $\specthread{\gamma}{t}{e}$ to $\specthread{\gamma}{t}{e'}$ in this derivation, instead of directly updating the underlying base effect specification resource in $\specCtx{\gamma}{\W}{\Psi}$, we accumulate the evidence that there exists a thread in the pool that can be evaluated to the expression $\expr'$ when it is next scheduled.
Here, because a thread takes at most one effectful step before the next scheduling, an update modality is sufficient to carry that evidence; later in \Cref{sec:distributed}, we will see another application of this technique that requires accumulating non-trivial evidence.

Let $\smash{\specthreadprot{\gname}{t}{\W}{e}{\Psi}}$ be the notation for the instantiation of $\genSpecProtNoArg$ with the $\specthread{\gamma}{t}{e}$ assertion.
This thread-local effect specification resource gives us a unified mechanism to reason about both global and thread-local effects.
It supports analogues of the \refrule{espec:pure} and \refrule{espec:do} rules.
In addition, we derive a rule for forking threads,
\begin{mathpar}
  \mprset{fraction=--\ast}
  \inferrule[\hypertarget{spec:conc-fork}{Spec-Fork}]
  {\specthreadprot{\gname}{t}{\W}{N[\keyword{fork}\ e]}{\textlog{CONC}_s^\gamma(\Psi)} \\
  \effecttag{fork}\notin\tags(\Psi')
  }
  {\pvs[\W] \specthreadprot{\gname}{t}{\W}{N[()]}{\textlog{CONC}_{s}^\gamma(\Psi)} \ast
  \specthreadprot{\gname}{\IsChild}{\specCtx{\gname}{\W'}{\Psi'}}{e}{\textlog{CONC}_{s}^\gamma(\Psi')}
  }
\end{mathpar}
where $\textlog{CONC}^{\gamma}_{s}(\Psi) \eqdef{} \textlog{FORK}^{\gamma}_{s} \oplus \Psi$ and $\textlog{FORK}^{\gamma}_{s}(v, \Phi) \eqdef{} \Exists e . v = (\effecttag{fork}, \Lam \_ . e) \ast (\specthread{\gamma}{\IsChild}{e} \wand \Phi(()))$.
The concurrency handler is installed by the following rule.
\begin{mathpar}
  \mprset{fraction=--\ast}
  \inferrule[\hypertarget{spec:conc-run}{Spec-ConcRun}]
  {
    \specprot{\W}{\function{run}_{\effecttag{conc}}~\textit{main}}{\Psi} \\
    \effecttag{fork} \notin \tags(\Psi)
  }
  {
    \pvs[\W] \Exists \gname . \specCtx{\gname}{\W}{\Psi} \ast
    \specthreadprot{\gname}{\IsMain}{\specCtx{\gname}{\W}{\Psi}}{\textit{main}~()}{\textlog{CONC}^\gamma_{s}(\Psi)}
  }
\end{mathpar}

\subsection{Logical Relation for Contextual Equivalence}\label{sec:logrel}

Using \binlog, we next define a \emph{logical relation} in the style of \citet{dreyer2011-llr} for proving contextual equivalence of programs written in typed subsets of \HandlerLang{}.
Two expressions $e_1$ and $e_2$ are contextually equivalent at type $\tau$ if no well-typed context $C$ can distinguish them; that is, replacing $e_1$ with $e_2$ in any client program does not change its observable behavior.
Because programs in \HandlerLang{} may raise effects, we define this relative to a closing handler context $H$ that handles them.
Contextual equivalence $e_1 \ctxequiv^{H} e_2 \colon \tau$ is the symmetric interior of \emph{contextual refinement} $e_1 \ctxrefines^{H} e_2 \colon \tau$, which says that the observable behaviors of $H[C[e_1]]$ are included in those of $H[C[e_2]]$ for every well-typed context $C$:
\begin{align*}
  e_{1} \ctxrefines^{H} e_{2} \colon \tau \quad \eqdef{} \quad  \All b \in \mathbb{B}, C \colon \tau \to \tbool . H[C[e_{1}]] \steps b \Ra H[C[e_{2}]] \steps b.
\end{align*}
Since $H$ encloses both $C$ and $e_i$, the context and the programs may interact through effects.

As an example, we consider a standard System-\textsf{F}-style type system $\Theta \mid \Gamma \vdash e\colon \tau$ with impredicative polymorphism, recursive types, and typing rules for CAS, FAA, and the fork operation (see, \eg \citet{timany2024-semantic-typing} for a complete definition).

The logical relation is entirely standard and follows previous Iris-based models~\citep{timany2024-semantic-typing}, \emph{except} that we define the expression interpretation using \binlog{} instantiated with the atomic heap instructions and concurrency.
The expression interpretation is
\begin{align*}
  \llbracket \tau \rrbracket^E(e_1, e_2)
  & \eqdef{} \All N, t .
    \begin{aligned}[t]
      &\specthreadprot{\gname}{t}{\W_{s}}{N [ e_2 ]}{\Psi_2} \wand \\
      &\ewp{\expr_1}[\W]{\Psi_1}{ \Ret v_1. \Exists v_2 .
        \smash{\specthreadprot{\gname}{t}{\W_{s}}{N [ v_2 ]}{\Psi_2} \ast \llbracket \tau \rrbracket^V(v_1, v_2)}}
    \end{aligned}
\end{align*}
where $\Psi_2 \eqdef{} \textlog{CONC}_{s}^{\gamma}(\Psi')$, $\Psi_1 \eqdef{} \textlog{CONC}_{\W}(\Psi)$, $\W_{s} \eqdef{} \specCtx{\gname}{\bot}{\Psi'}$, $\W \eqdef{} \W_{s} \oplus \WInvTok{\top}$ for $\Psi$ and $\Psi'$ that describe the global stack of effects (state, heap, and atomic heap).
The proof of the fundamental theorem of logical relations is immediate from the existing proofs since all our rules for reasoning about the atomic heap operations and concurrency are identical to the usual separation logic rules.
\begin{theorem}[Fundamental]
  If $\Theta \mid \Gamma \vdash e\colon \tau$ then $\Theta \mid \Gamma \vDash e \logrefines e\colon \tau$.
\end{theorem}
To prove soundness, we consider the closing handler context
\begin{align*}
  H_{\textsf{CONC}} \eqdef{} \function{run}_{\effecttag{state}}~(\Lam \_ . \function{run}_{\effecttag{heap}}~(\Lam \_ . \function{run}_{\effecttag{atomheap}}~(\Lam \_ . \function{run}_{\effecttag{conc}}~(\Lam\_.\hole))))~()
\end{align*}
and use the handler rules, \eg \refrule{ewp:conc-run} and \refrule{spec:conc-run}, and \Cref{thm:binlog-adequacy}.
\begin{theorem}[Soundness]
  If $\cdot \mid \cdot \vDash e_{1} \logrefines e_{2}\colon \tau$ then $e_{1} \ctxrefines^{H_{\normalfont{\textsf{CONC}}}} e_{2}\colon \tau$.
\end{theorem}
\paragraph{Contextual Equivalence of Preemption}\label{sec:yield}

Using our logical relation, we prove that inserting additional preemption points in concurrent programs under the $\function{run}_{\effecttag{conc}}$ scheduler does not change program behaviors.
This justifies the soundness of using a scheduler that only triggers preemption when effects are raised, since it shows that additional preemption would not affect observable behavior.
Formally, we introduce an expression $\keyword{yield}$ that triggers a preemption point by defining $\keyword{yield} \eqdef{} \keyword{fork}~()$, \ie a program that simply forks a thread that terminates immediately.
By forking a thread, $\keyword{yield}$ transfers control to the scheduler, which may choose another thread to continue.

To justify that $\keyword{yield}$ has no effect on the computation, we show that it is contextually equivalent to the unit value, \ie $\keyword{yield} \ctxequiv^{H_{\normalfont{\textsf{CONC}}}} () \colon \tunit$.
The proof is an immediate consequence of the rules \refrule{ewp:fork} and \refrule{spec:conc-fork} for the left-to-right and right-to-left refinements, respectively.
As a corollary, for example, it then follows that $e_1 ;~\keyword{yield};~e_2 \ctxequiv^{H_{\normalfont{\textsf{CONC}}}} e_1;~e_2 \colon \tau$ for any well-typed $e_{1}$ and $e_{2}$.
Similar arguments justify our crash model in \Cref{sec:crash} and stronger atomicity rules for distributed execution in \Cref{sec:distributed}.

Like many results for concurrent relational logics, this equivalence relies on $\function{run}_{\effecttag{conc}}$ being fully non-deterministic: it may pick any thread to run next, and other threads may continue running after the main thread terminates.
We argue that this suffices for safety properties.
An interleaving produced by any other scheduler, including one that preempts between pure steps, can be reproduced under $\function{run}_{\effecttag{conc}}$ by inserting \keyword{yield}s, which do not change the behavior.
Hence, proving an $\ewpNoArg$ under the $\textlog{CONC}$ protocol implies that the program is safe under any scheduler.
\section{Case Study: Prophecy Variables}\label{sec:proph}

When verifying certain concurrent programs in a forward-reasoning style, at some points in the proof it is necessary to know how later operations will be non-deterministically ordered.
Prophecy variables~\cite{abadi1988-prophecy, jung2019-pvsl} are a logical mechanism that allows for ``speculating'' or ``predicting'' these future outcomes during a proof.
In Iris, these prophecy variables are ghost code, and a prover must instrument a program to attach prophecy variables to operations whose values need to be predicted. %
New prophecy variables are allocated using a command $\ghostcode{\keyword{newproph}}$, and then attached to a program value using $\ghostcode{\keyword{resolve}}$.
The proof rules for prophecy variables tell us \emph{at the time of allocation} what the future resolved value will be.
Iris comes with an additional proof showing that these prophecy variable operations can be erased from the program without affecting the outcome.

In this section, we show how to extend \unlog{} with prophecy variables.
Our starting point is a single \emph{global} prophecy variable.
On top of this global prophecy variable, we implement effect handlers that allow for dynamically allocatable local prophecy variables, with an interface similar to that of Iris.
Finally, we observe that by instrumenting the \emph{handlers} for heap operations with prophecy variables, we can automatically extend all heap operations to have prophecies, without requiring the \emph{client} program to be directly instrumented with prophecies.

\paragraph{Global Prophecy}
Recall that \HandlerLang{} has a ghost expression $\ghostcode{\keyword{observe}\ v}$, which records the value $v$ on a global trace.
The entire future value of this trace is predicted in a global prophecy assertion $\primproph{\vv{v}}$, which is used when performing an $\ghostcode{\keyword{observe}}$.
\begin{mathpar}
\mprset{fraction=--\ast}
\inferrule[\hypertarget{ewp:obs}{Ewp-Obs}]
{\primproph{\vv{v}}}
{\ewp{\ghostcode{\keyword{observe}\ w}}[\W]{\Psi}{\Ret \_. \Exists \vv{v}\!'. \vv{v}=w::\vv{v}\!'\ast\primproph{\vv{v}\!'}}}
\end{mathpar}
By making an observation of $w$, we immediately learn that the first element of $\vv{v}$ is indeed $w$, so $\vv{v}$ must equal $w::\vv{v}\!'$ for some unknown $\vv{v}\!'$.\footnote{As usual with prophecy reasoning, if the predicted value at the head of the sequence $\vv{v}$ was \emph{not} $w$, then we derive a contradiction from this rule, and no longer have to reason about this moot execution with a misprediction.}
The full version of the adequacy theorem of \unlog{} provides this prophecy assertion when the program starts.
\begin{theorem}[Adequacy, \unlog]\label{thm:unlog-adequacy}
  Let $\varphi$ be a first-order predicate.
  If\/ $\primproph{\vv{v}}\vdash \ewp[]e[]{\bot}{\varphi}$ for all $\vv v$, then executing $e$ never gets stuck, and if $e\steps v$ then $\varphi(v)$ holds.
\end{theorem}
However, because this prophecy variable is global, it is awkward to use when trying to do local reasoning about data structures that need prophecies.

\paragraph{Encoding Local Prophecy Variables}
We recover Iris-style local prophecy variables by using handlers on top of the global \ghostcode{\keyword{observe}}.
Formally, our local prophecy variables are specified by the following protocols:
\begin{align*}
  \textlog{NEWPROPH}(u,\Phi)\triangleq{}
  &u=(\effecttag{newproph},())\ast (\All p,\vv{v}.\proph{p}{\vv v}\wand \Phi(p))\\
  \textlog{RESOLVE\_PROPH}(u,\Phi)\triangleq{}
  &\Exists v,p,w,\vv{v}.
    \begin{aligned}[t]
      &u=(\effecttag{resolve\_proph},(v,p,w))\ast\proph{p}{\vv{v}}\ast\\
      &(\All\vv{v}\!'.\vv{v}=(v,w)::\vv{v}\!'\ast\proph{p}{\vv{v}\!'}\wand \Phi(v))
    \end{aligned}
\end{align*}
A new prophecy variable is created by raising the effect $\effecttag{newproph}$, which returns a fresh prophecy variable $p$.
Assertion $\proph{p}{\vv{v}}$ is the local version of $\primproph{\vv{v}}$, which says that the trace of resolutions that will occur on prophecy variable $p$ is $\vv{v}$.
The effect \effecttag{resolve\_proph} resolves $p$ to a pair of values $(v,w)$.
The first component $v$ is the primary value that we want to observe, while the second component $w$ is used to tag certain kinds of prophecies with metadata.

Under the hood, these assertions work by slicing the observation trace of the global prophecy into traces of individual prophecy variables.
This is done by making every call to $\ghostcode{\keyword{observe}}$ have the format $(p,v,w)$, where $p$ is the identifier of the prophecy variable that the observation belongs to.
We can now try defining $\proph{p}{\vv{v}}$ as something like $\Exists \vv{v}_0.\primproph{\vv{v}_0}\ast \vv{v}=\textlog{filter}(p,\vv{v}_0)$, where $\textlog{filter}$ is the least fixed point of
\begin{align*}
  \textlog{filter}(p,(p',v,w)::\vv v)
  &\triangleq\text{if }p=p'\text{ then }(v,w)::\textlog{filter}(p,\vv v)\text{ else }\textlog{filter}(p,\vv v) \\
  \textlog{filter}(p,\_ )
  &\triangleq\nil
\end{align*}
The function \textlog{filter} projects out the observations that are associated with the indicated prophecy variable, truncating the trace at the first observation that does not match the expected format.

Recall that resources in CSL are exclusive to one thread in general, so to actually have mutually independent prophecy variables, we need to decompose ownership of the global prophecy $\primproph{\vv v}$ into ownership of individual prophecy variables using an authoritative ghost resource algebra.
\begin{align*}
\proph{p}{\vv{v}}&\triangleq\ownGhost{\gamma_{\textlog{proph}}}{\authfrag\mapsingleton{p}{\vv{v}}} \\
I&\triangleq\Exists \vv{v}_0, M . \primproph{\vv{v}_0}\ast\ownGhost{\gamma_{\textlog{proph}}}{\authfull M}\ast\textstyle\Sep_{p\mapsto\vv{v}\in M} \vv{v}=\textlog{filter}(p,\vv{v}_0)
\end{align*}
Here $\gamma_{\textlog{proph}}$ is an arbitrary but globally fixed ghost name.
Invariant $I$ connects individual prophecy variables to the global prophecy and is maintained by the handler.
The underlying resource algebra guarantees that the entry $p\mapsto\vv{v}$ is always in $M$.

The handler $\function{run}_{\effecttag{proph}}$ providing protocols \textlog{NEWPROPH} and \textlog{RESOLVE\_PROPH} is shown in \refappendix{sec:app:run-proph}.
To create new prophecy variables, it internally maintains a monotonically increasing counter for the next fresh prophecy ID so that it can always ``slice out'' an unused resolving sequence from the global prophecy for a new prophecy variable.
To resolve prophecy variable $p$ to $(v,w)$, it makes an observation of $(p,v,w)$ and updates the ghost resources accordingly.
In practice, $\function{run}_{\effecttag{proph}}$ is installed as the outermost handler so that other handlers can use prophecy variables.

\paragraph{Atomic Prophecies} The resolve effect we have seen so far resolves a value that is returned by evaluating some expression.
In other words, the prophecy is resolved \emph{after} the expression finishes.
However, in some scenarios, it is necessary to atomically execute the expression and prophecy resolution at the same time, particularly for atomic heap operations like CAS and FAA.
Iris provides support for this so-called atomic prophecy resolution, and we can also implement this on top of the local prophecy variables through another effect handler with the following protocol:
\begin{align*}
  \textlog{RESOLVE}(\Psi)(u,\Phi)\triangleq{}
  &\Exists v,p,w,\vv v.
    \begin{aligned}[t]
      &u=(\effecttag{resolve},(v,p,w))\ast\proph{p}{\vv{v}}\ast{}\\
      &\Psi(v,(\Lam r. \All \vv{v}\!'. \vv{v}=(r,w)::\vv{v}\!'\wand\proph{p}{\vv v\!'}\wand\Phi(r)))
    \end{aligned}
\end{align*}
Here, $v$ is a pair of (tag, value) raised by a \keyword{do} expression. For example, to resolve a prophecy variable $p$ to $w$ along with a store expression $\ell\gets x$, one needs to write $\keyword{do}~(\effecttag{resolve},((\effecttag{store},(\ell,x)),p,w))$.
The handler $\function{run}_{\effecttag{atomproph}}$ providing the $\textlog{RESOLVE}$ protocol is shown in \refappendix{sec:app:run-atomproph}.
To handle $\keyword{do}~(\effecttag{resolve},(v,p,w))$, it executes $\keyword{do}~(\effecttag{resolve\_proph},(\keyword{do}~v,p,w))$.
By installing $\function{run}_{\effecttag{atomproph}}$ \emph{inside} $\function{run}_{\effecttag{proph}}$ and the handlers for heap operations, but \emph{outside} $\function{run}_{\effecttag{conc}}$ for concurrency, we ensure that the two \keyword{do}s, \ie $\keyword{do}~v$ and $\keyword{do}~(\effecttag{resolve\_proph},\dots)$, behave as if they executed simultaneously, because the additional \keyword{do} in $\function{run}_{\effecttag{atomproph}}$ does not trigger an additional preemption.

\paragraph{Implicit Prophecies} As in \HeapLang, the above interface for prophecies still requires a proof developer to annotate a program with calls to create new prophecies and to resolve them at relevant points.
However, we can use effect handlers to give \emph{every} heap location an associated prophecy variable that predicts the full trace of operations that will be performed on that heap location.
This ``prophetic heap'' handler interposes on all heap-related effect tags, and adds an extra \effecttag{resolve} operation before re-raising the effect.
With this handler, when a location is allocated, in addition to the standard points-to assertion $\ell \mapsto v$, we also get a $\proph{\ell}{\vv v}$ assertion.
When a heap operation on $\ell$ occurs, we also pass this $\proph{\ell}{\vv v}$ assertion, allowing us to deduce that the value read from or written to the location matches the head value in the trace $\vv{v}$.
This allows for proofs with prophecies \emph{without} having to annotate a client program with explicit prophecy operations.
\refappendix{sec:app:proph-heap} describes the protocols in further detail.
\section{Case Study: Crash-Recovery Reasoning}\label{sec:crash}

\begin{figure}
\begin{minipage}[t]{0.51\linewidth}
  \[
  \begin{codeblock}{l}
&\function{run}_{\effecttag{crash\_trigger}}\triangleq\Lam\textit{main}.\\
&\quad\keyword{if}\ \keyword{pick}\ \keyword{then}\ \keyword{do}\ (\effecttag{crash},())\\
&\quad\keyword{else}\ \keyword{deep-try}\ \textit{main}\ ()\ \keyword{with}\\
&\quad\phantom{\mid{}} \mathmakebox[\widthof{\ensuremath{\keyword{ret}\ v}}][r]{v\ k}\Rightarrow\keyword{let}\ r:=\keyword{do}\ v\ \keyword{in}\\
&\quad\phantom{\mid\keyword{ret}\ v\Rightarrow~}\keyword{if}\ \keyword{pick}\ \keyword{then}\ \keyword{do}\ (\effecttag{crash},())\\
&\quad\phantom{\mid\keyword{ret}\ v\Rightarrow~}\keyword{else}\ k\ r\\
&\quad\mid \keyword{ret}\ v\Rightarrow v
  \end{codeblock}
  \]
\end{minipage}%
\begin{minipage}[t]{0.49\linewidth}
$$\begin{codeblock}{l}
  &\function{run}_{\effecttag{crash}}\triangleq\keyword{rec}\ \textit{run}\ \textit{main}.\\
&\quad\keyword{deep-try}\ \textit{main}\ ()\ \keyword{with}\\
&\quad\phantom{\mid{}} \mathmakebox[\widthof{\ensuremath{\keyword{ret}\ v}}][r]{v\ k}\Rightarrow
  \keyword{match}\ v\ \keyword{with}\\
  &\phantom{\quad\mid \keyword{ret}\ v\Rightarrow}\phantom{\mid{}}(\effecttag{crash},())\Rightarrow\ghostcode{\smash{\keyword{observe}}\ ()}; \textit{run}\  \textit{main}\\
  &\phantom{\quad\mid \keyword{ret}\ v\Rightarrow}\mid(\efft,w)\Rightarrow k\ (\keyword{do}\ (\efft,w))  \\
&\quad\mid \keyword{ret}\ v\Rightarrow v
\end{codeblock}$$
\end{minipage}

\caption{A model of crash-and-recovery execution, where the $\keyword{deep-try}$ expression installs a deep handler that is reinstalled after an effect is raised, which is implemented by using $\keyword{try}$ and recursion}\label{fig:crash-handler}
\end{figure}
In this section, we show how to model the effect of sudden interruption by crashes, like those caused by power loss.
Many software systems that store data on durable media such as disks must be \emph{crash safe}, meaning that the system must be able to recover from such a crash.
When a crash occurs, any data that the system has in volatile memory, such as RAM, will be wiped, but data in durable storage will be preserved.
After the system restarts, it will typically run a recovery procedure that restores system invariants.
A number of program logics and verification frameworks have been developed for reasoning about such systems~\citep{chen2015-fscq,chajed2019-perennial, raad2020-pog, ntzik2015-fault}.
After implementing effect handlers to encode crashes and recovery with durable state, we apply \unlog{} to derive protocols for reasoning about these systems in the style of Perennial~\citep{chajed2019-perennial}, a separation logic for reasoning about the combination of concurrency and crash safety.

The process of crashing and recovering is modeled by the pair of handlers in \Cref{fig:crash-handler}.
The first handler $\function{run}_{\effecttag{crash\_trigger}}$ is installed at the innermost level of a stack of effect handlers for each thread, allowing it to interpose on every \keyword{do}.\footnote{To install $\function{run}_{\effecttag{crash\_trigger}}$ for every child thread, our Rocq mechanization actually integrates $\function{run}_{\effecttag{crash\_trigger}}$ into $\function{run}_{\effecttag{conc}}$.}
When an effect is raised, this handler first re-raises the effect to obtain its result, and then non-deterministically chooses to either trigger a crash by raising \effecttag{crash}, or return the result to the continuation.
Additionally, the handler non-deterministically chooses to crash before $\textit{main}$ starts to execute.
The second handler $\function{run}_{\effecttag{crash}}$ responds to this trigger by
throwing away the captured continuation $k$ and restarting the system by running $\textit{main}$.
Note that this handler does not directly deal with wiping the volatile state of the system.
Instead, this is handled implicitly: because the handlers for volatile state (such as the $\effecttag{heap}$ handler) are installed \emph{after} this crash handler (\ie as part of $\textit{main}$), this volatile state is effectively thrown away as a result of re-running $\textit{main}$ from scratch.
In contrast, durable state can be preserved by installing these handlers \emph{before} installing the crash handler at an outer level.

\subsection{Managing the Crash Invariant}\label{subsec:crash-inv}

In order to establish that a system is crash safe, it is essential to show that when the system restarts, the $\textit{main}$ procedure finds itself in a state that satisfies its precondition.
Perennial maintains a global \emph{crash invariant} $\mathcal{R}$ that must hold before and after each step of execution, and which describes the durable state that the system needs upon restart.

\paragraph{Local Crash Conditions}
Reasoning about a global crash invariant would run counter to the principle of local reasoning in concurrent separation logic.
To recover per-thread reasoning about the crash invariant, Perennial extends the weakest precondition of each thread with an assertion called a \emph{crash condition}.
The crash condition enforces the portion of the global crash invariant that a given thread owns.
In \unlog{}, rather than changing the weakest precondition to add an additional component, we can instead capture this local crash condition through worlds and a protocol transformer called \textlog{DURA}.
We write $\WCrashInvTok{R}$ for a world stating a thread is responsible for ensuring that the local crash condition $R$ holds after each step it takes.\footnote{The notation \textlog{CC} stands for crash condition.}
The \textlog{DURA} protocol forces a thread to show that this $R$ holds after each step of execution.
\begin{equation*}
\textlog{DURA}_{\W}(\Psi)(v,\Phi)\triangleq{}\Exists R. \Psi(v,\Lam r. \pvs[\bot][\W\oplus\WCrashInvTok{R}] (R\land\pvs[\W\oplus\WCrashInvTok{R}][\bot]\Phi(r)))
\end{equation*}
Note that $R$ and the postcondition are connected by a logical conjunction $\land$ because the program can only either crash or continue, so only one of $R$ and the postcondition will be used.

The derived proof rules for the crash handlers then require a specification for the top-level $\textit{main}$ procedure of the following form.
$$\mathcal{R}\vdash\ewp{\textit{main}\ ()}[\W\oplus\WCrashInvTok{\postcrash\mathcal{R}},\bot]{\textlog{DURA}_\W(\Psi)}{\Ret r.\Exists R. \smash{\pvs[\bot][\W\oplus\WCrashInvTok{R}] R\land\Phi(r)}}$$
Here, $\mathcal{R}$ is the global crash invariant that also serves as the precondition for $\textit{main}$, and $\postcrash$ is called the \emph{post-crash modality}~\citep{vindum2025-nextgen,tassarotti2021-perennial-crash-modality}, which captures how crashing modifies volatile and durable resources.
Intuitively, $\postcrash\mathcal{R}$ says that $\mathcal{R}$ holds immediately after a crash.
This top-level specification says that the main thread starts with precondition $\mathcal{R}$, and the initial crash condition requires that $\mathcal{R}$ holds after a crash.
The main thread is free to alter the crash condition (by allocating/returning crash borrows using rules introduced later).
But when raising an effect, as specified by the $\textlog{DURA}$ protocol, the main thread must update to a world containing $\WCrashInvTok{R}$, which is the current crash condition, and prove $R$.
When the main thread terminates, the top-level specification also requires showing what the current crash condition is and proving that crash condition.

\paragraph{Concurrency with Crashes}
To add support for concurrency, we install the concurrency handler inside the outer $\function{run}_{\effecttag{crash}}$ handler.
Because of the crash condition, we need to strengthen the \textlog{CONC} protocol to a protocol called \textlog{CRASHCONC}.
\begin{align*}
\textlog{CRASHCONC}_{\W}(\Psi)\triangleq{}&\textlog{DURA}_{\W} (\Psi\oplus\textlog{FORK}_\W'(\Psi))\\
\textlog{FORK}_\W'(\Psi)(v,\Phi)\triangleq{}&\Exists e. v=(\effecttag{fork},\Lam\_. e)\ast{} \\
&\hspace*{-4.5em}\later\ewp e [\W\oplus\WCrashInvTok{\TRUE},\bot]{\textlog{CRASHCONC}_{\W}(\Psi)}{\Ret\_.\Exists R_e. \smash{\pvs[\bot][\W\oplus\WCrashInvTok{R_e}] R_e}}\ast \Phi(())\notag
\end{align*}
\textlog{CRASHCONC} follows the same structure as \textlog{CONC} and is also a guarded fixed point.
The difference is that here a forked child thread starts with a trivial crash condition and is required to prove its current crash condition when terminating.

When forking a child thread, one would naturally like to move some resources from the parent thread to the child thread.
Similarly, synchronization primitives like locks are logically thought of as transferring ownership of resources in CSL.
However, in order to transfer ownership of durable resources that might be part of the crash condition, we also need a mechanism to transfer the \emph{obligation} to maintain that part of the crash condition.
Crash borrows in Perennial~\citep{tassarotti2022-perennial-crash-borrow} provide a mechanism to ``borrow'' part of the resources governed by the crash condition and transfer them to the child thread.
A crash borrow $\knowInv{}{P\mid R}$ has \emph{content} $P$ that describes the resources currently contained, and an associated \emph{crash obligation} $R$, where $\always(P\wand R)$.

The crash borrow can be understood as a box that packages up a resource $P$ while preserving the obligation to ensure $R$ holds in the event of a crash. Crash borrows are used through the following two key rules.
\begin{mathpar}
\mprset{fraction=--\ast}
\inferrule[\hypertarget{wupd:cbrw-alloc}{Wupd-CBrwAlloc}]
{\later P \\ \always(P\wand R)}
{\pvs[\WCrashInvTok{Q\ast R}][\WCrashInvTok{Q}] \knowInv{}{P\mid R}}
\and
\inferrule[\hypertarget{wupd:cbrw-return}{Wupd-CBrwReturn}]
{\knowInv{}{P\mid R}}
{\pvs[\WCrashInvTok{Q}][\WCrashInvTok{Q\ast R}]\later P}
\end{mathpar}
Rule \refrule{wupd:cbrw-alloc} creates a crash borrow $\knowInv{}{P\mid R}$.
It consumes a resource $P$ that is stronger than $R$ and removes $R$ from the crash condition.
Rule \refrule{wupd:cbrw-return} opens the box $\knowInv{}{P\mid R}$ to extract resource $P$, and in exchange, it adds $R$ to the crash condition.
In Perennial, this crash borrow mechanism is encoded on top of standard Iris invariants in a complex manner that requires extensive use of later credits~\citep{spies2022-later} to avoid inconsistencies from impredicative circularities.
In \unlog{}, the encoding is considerably simpler, because we are able to use a separate world for managing crash conditions and crash borrows.
The complete model can be found in \refappendix{sec:app:crash}.

\subsection{Asynchrony and Crash-Aware Prophecies}\label{subsec:adisk}
\begin{figure}[t]
\begin{align*}
\textlog{ADISK}\triangleq{}&\textlog{ADISK\_LOAD}\oplus\textlog{ADISK\_STORE}\oplus\textlog{ADISK\_BARRIER}\\
\textlog{ADISK\_LOAD}(u,\Phi)\triangleq{}&\Exists \ell, v_c,v.u=(\effecttag{adisk\_load},\ell)\ast \ell\diskmapsto\adiskval{v_c}{v}\ast(\ell\diskmapsto\adiskval{v_c}{v}\wand\Phi(v))\\
\textlog{ADISK\_STORE}(u,\Phi)\triangleq{}&\Exists \ell,v_c,v,w.u=(\effecttag{adisk\_store},(\ell,w))\ast \ell\diskmapsto\adiskval{v_c}{v}\ast{}\\
&(\All v_c'\in\{w,v_c\}. \ell\diskmapsto\adiskval{v_c'}{w}\wand \Phi(()))\notag\\
\textlog{ADISK\_BARRIER}(u,\Phi)\triangleq{}&\Exists M. u=(\effecttag{adisk\_barrier},())\ast  \left(\textstyle\Sep_{\ell\mapsto(v_c,v)\in M}\ell\diskmapsto\adiskval{v_c}{v}\right)\ast{}\\
&\left(\left(\textstyle\Sep_{\ell\mapsto(v_c,v)\in M}v_c=v\ast \ell\diskmapsto\adiskval{v_c}{v}\right)\wand \Phi(())\right)\notag\\
\keyword{adisk\_load}\ \ell\eqdef\keyword{do}\ & (\effecttag{adisk\_load},\ell)\qquad
\keyword{adisk\_store}\ \ell\ v\eqdef\keyword{do}\  (\effecttag{adisk\_store},(\ell,v))\\
&\keyword{adisk\_barrier}\eqdef{}\keyword{do}\ (\effecttag{adisk\_barrier},())
\end{align*}
\caption{The asynchronous disk protocol}\label{fig:adisk-protocol}
\end{figure}

Many durable storage media are \emph{asynchronous}, meaning that when a write is performed, the written value does not immediately become durable.
Instead, the written value is first stored in some volatile buffer.
These buffers are then non-deterministically flushed to disk, making the writes durable.
Alternatively, a program can force a flush by issuing a \emph{barrier}.
If a crash occurs while the value is still in the volatile buffer, then the write is lost.
Reasoning about asynchrony is challenging when trying to prove that a concurrent durable data structure satisfies \emph{durable linearizability}~\citep{DBLP:conf/wdag/IzraelevitzMS16}, because it makes the durability of an operation \emph{future dependent}.

To deal with this challenge, Perennial introduced a \emph{prophetic disk points-to} assertion of the form $\ell\diskmapsto\adiskval{v_c}{v}$ which says that the disk address $\ell$ currently stores the value $v$, and after a crash occurs, the stored value will be $v_c$.
This property is captured in the following entailment that describes how the disk points-to changes under the post-crash modality: $\ell\diskmapsto\adiskval{v_c}{v}\proves\postcrash\ell\diskmapsto\adiskval{v_c}{v_c}$.
Essentially, this assertion bundles a normal points-to with a form of prophecy about the post-crash state.
However, in Perennial, this primitive could not reuse the existing support for prophecies in Iris, and instead has an ad hoc soundness proof.
The issue is that, with standard Iris prophecy variables, there is no way to make a prophecy about whether an event will happen before or after a crash.

In contrast, in \unlog{} it is easy to handle this by incorporating prophecy resolution as part of the implementation of the crash handler.
We use the $\ghostcode{\keyword{observe}\ ()}$ statement in  $\function{run}_{\effecttag{crash}}$ to effectively record that a crash has occurred in the trace of every prophecy variable.
The definition of \textlog{filter} in \Cref{sec:proph} ensures that this truncates the trace of events in the prophecy stream for all variables.
As a result, for every prophecy variable $p$, we have $\proph{p}{\vv{v}}\vdash\postcrash(\vv{v}=\nil)$.
Thus, when inspecting the prophecy stream, it is possible to determine whether a crash will occur before the prophecy is resolved.
We call these resulting prophecy variables \emph{crash-aware}.
Using this mechanism, we implement an asynchronous disk with prophetic points-to assertions by allocating a crash-aware prophecy for each buffered write, which predicts whether the write will be flushed before a crash.
The protocol of the asynchronous disk is found in \Cref{fig:adisk-protocol} and more details can be found in \refappendix{sec:app:crash}.
\textlog{ADISK\_LOAD} returns the current value $v$ of a location.
\textlog{ADISK\_STORE} sets the current value to $w$, and the prophecy allocated for this write determines the post-crash value: the client must handle both $v_c'=w$, where the write is flushed before a crash, and $v_c'=v_c$, where the write is lost and the location keeps its previous post-crash value.
Finally, \textlog{ADISK\_BARRIER} takes ownership of a collection $M$ of disk points-to assertions and returns them together with the fact that $v_c=v$ for each of them.
Because a barrier flushes all pending writes and resolves their prophecies accordingly, the current value of each location is also the value it will have after a crash, so any prophesied future in which a write before the barrier is lost leads to a contradiction.

\subsection{Recovering the Perennial Logic}\label{sec:recovering-perennial}
The effects introduced above provide all of the features Perennial has in order to support crash reasoning, except that
our crash model diverges from the standard crash model considered by prior work like Perennial: just as our concurrency handler only allowed preemption when effects were raised, our crash handler only allows crashing after each effectful step.
In contrast, the typical model allows crashing between every step.
However, just as with concurrency, this crash model is equivalent to the standard model for purposes of partial correctness: the pure steps do not affect the observable state of the system after a crash, so choosing to crash before or after a sequence of pure steps is equivalent to crashing at some point in the middle of them.
To justify the soundness of our model, we again show that \keyword{yield} is contextually equivalent to the unit value under our crash model, similar to what we do in \Cref{sec:logrel}.
This time, because a potential crash could happen after any effectful step, a \keyword{yield} serves as both a possible preemption point and a possible crash point.

Formally, we define a handler context $H_{\textsf{PERENNIAL}}(d)$ that handles the effects for asynchronous disk, crash, heap, and concurrency.
Concretely, it first sets the initial durable state (\ie the state of the disk) to $d$, then installs the $\function{run}_{\effecttag{crash}}$ handler, and finally installs other handlers for volatile state inside.
We then prove $\All d.\keyword{yield}\ctxequiv^{H_{\textsf{PERENNIAL}}(d)}()\colon\tunit$.
The proof is structured by dividing up execution into a sequence of \emph{rounds} separated by crashes.
The proof reuses much of the construction of the spec resource in \Cref{sec:relconc} and the logical relation in \Cref{sec:logrel} to reason about the behavior of two programs within one round of execution.
The major difference is that to show $e_1$ refines $e_2$ within one round, because $e_1$ could crash after any effect, the spec resource of $e_2$ must be able to crash angelically at any step as well.
To derive a multi-round contextual refinement on top of that, we do an induction on the number of rounds and use the logical relation to relate the behavior of two programs in each round.

\subsection{Example: The Durable Pair}\label{sec:durable-pair}
To exercise the features introduced in this section, we verify a durable pair example that uses a crash-safety pattern called a \emph{shadow copy}, which is used in many crash-safe systems.

\begin{figure}[t]
\begin{minipage}{0.6\linewidth}
\[
\begin{codeblock}{l}
&\function{read\_pair}\eqdef\Lam ().\\
&\quad\keyword{let}\ \textit{active} := (\keyword{if}\ \keyword{adisk\_load}\ \ell_d=1\ \keyword{then}\ \ell_1\ \keyword{else}\ \ell_2)\ \keyword{in}\\
&\quad\keyword{let}\ a_0 := \keyword{adisk\_load}\ \textit{active}\ \keyword{in}\\
&\quad\keyword{let}\ a_1 := \keyword{adisk\_load}\ (\textit{active}+1)\ \keyword{in}\\
&\quad(a_0,a_1)\\
&\\
&\\
&\\
\end{codeblock}
\]
\end{minipage}%
\begin{minipage}{0.4\linewidth}
\[
\begin{codeblock}{l}
&\function{write\_pair}\eqdef\Lam (b_0,b_1).\\
\codenum&\quad(\keyword{if}\ \keyword{adisk\_load}\ \ell_d=1\ \keyword{then}\\
\codenum&\qquad\keyword{adisk\_store}\ \ell_2\ b_0;\\
\codenum&\qquad\keyword{adisk\_store}\ (\ell_2+1)\ b_1;\\
\codenum&\qquad\keyword{adisk\_barrier};\\
\codenum&\qquad\keyword{adisk\_store}\ \ell_d\ 2\\
\codenum&\quad\keyword{else} \dots\codecomment{symmetrically});\\
\codenum&\quad\keyword{adisk\_barrier}\\
\end{codeblock}
\]
\end{minipage}

\begin{mathpar}
\mprset{fraction=--\ast}
\inferrule[\hypertarget{ewp:read-pair}{Ewp-ReadPair}]
{\textlog{pair}(a_0,a_1)\\
(\textlog{pair}(a_0,a_1)\wand\Phi ((a_0,a_1)))\land (\postcrash\textlog{pair}(a_0,a_1)\wand R)}
{\ewp{\function{read\_pair}\ ()}
[\WInvTok{\top}\oplus\WCrashInvTok{R}]
{\Psi_{\textit{pair}}}{\Phi}}
\and
\inferrule[\hypertarget{ewp:write-pair}{Ewp-WritePair}]
{\textlog{pair}(a_0,a_1)\\
(\textlog{pair}(b_0,b_1)\wand\Phi (()))\land (\postcrash(\textlog{pair}(a_0,a_1)\lor\textlog{pair}(b_0,b_1))\wand R)}
{\ewp{\function{write\_pair}\ (b_0,b_1)}
[\WInvTok{\top}\oplus\WCrashInvTok{R}]
{\Psi_{\textit{pair}}}{\Phi}}\and
\textlog{pair}(a_0,a_1)\eqdef\ell_d\diskmapsto\adiskval{1}{1}\ast \big(\ell_1\diskmapsto\adiskval{a_0}{a_0}\ast\ell_1+1\diskmapsto\adiskval{a_1}{a_1}\ast\ell_2\diskmapsto\adiskval{\_}{\_}\ast\ell_2+1\diskmapsto\adiskval{\_}{\_}\big)\\
\phantom{\textlog{pair}(a_0,a_1)}\lor\ell_d\diskmapsto\adiskval{2}{2}\ast\big(\ell_2\diskmapsto\adiskval{a_0}{a_0}\ast\ell_2+1\diskmapsto\adiskval{a_1}{a_1}\ast\ell_1\diskmapsto\adiskval{\_}{\_}\ast\ell_1+1\diskmapsto\adiskval{\_}{\_}\big)
\and
\Psi_{\textit{pair}}\eqdef\textlog{DURA}_{\WInvTok{\top}}(\textlog{ATOMHEAP}\oplus\textlog{ADISK})
\end{mathpar}

\caption{Implementation and a simplified specification of the durable pair}\label{fig:durable-pair}
\end{figure}

\paragraph{Implementation}
The implementation and specification of the durable pair are shown in \Cref{fig:durable-pair}.
Logically, each durable pair stores a pair of values that span two consecutive disk addresses.
The goal is to be able to update the pair in a way that is atomic with respect to crashes.
Naively writing one component and then writing the next does not work, because a crash in between the two writes, or a crash after one write's buffer has been flushed but not the other's, can lead to a state after recovery where only one of the components of the pair has been updated.
In order to be able to update the pair in a crash-safe way, the implementation instead maintains two copies of the pair (at $(\ell_1,\ell_1+1)$ and $(\ell_2,\ell_2+1)$, respectively): one that is \emph{active} and one that is \emph{inactive}.
The current active pair is recorded in a selector node $\ell_d$.
To read from the durable pair, $\function{read\_pair}$ reads from the selector node to learn which pair is active and then reads from the active one.
To update the values, $\function{write\_pair}$ first does two writes to update the inactive copy.
It then issues a barrier to ensure both of these writes are durable.
Next, it does one write to the selector node to make this copy the active one, and issues another barrier to make this update durable.
Our Rocq mechanization builds a thread-safe durable pair on top of the durable pair introduced here by wrapping it with a lock.
For simplicity, here we focus on the sequential version, and just reason about crash safety.

\paragraph{Specification}
The assertion $\textlog{pair}(a_0,a_1)$ says that the value of the pair on disk is $(a_0,a_1)$: if the selector node $\ell_d$ equals $1$, then locations $\ell_1$ and $\ell_1 + 1$ hold the pair, while the values for $\ell_2$ and $\ell_2 + 1$ are unspecified, and vice versa.
The specification \refrule{ewp:read-pair} for $\function{read\_pair}$ has two premises.
The first premise is the precondition: we must own $\textlog{pair}(a_0,a_1)$.
The second premise gives the postcondition in continuation-passing style: to prove the postcondition $\Phi$, we can assume ownership of $\textlog{pair}(a_0,a_1)$, and in addition, we must show the crash condition $R$, getting access to $\postcrash\textlog{pair}(a_0,a_1)$, \ie the pair still holds $(a_0,a_1)$ after a crash.
Rule \refrule{ewp:write-pair} follows the same pattern, except that if a crash occurs during the write, we only learn that after the crash the pair holds either the old values $(a_0,a_1)$ or the new values $(b_0,b_1)$.
In other words, the write is atomic with respect to crashes.

\paragraph{Verification}
We sketch the proof of \refrule{ewp:write-pair}; the proof of \refrule{ewp:read-pair} is simpler.
Each disk operation in $\function{write\_pair}$ is an effect, which we step over using \refrule{ewp:bind} and \refrule{ewp:dowupd}.
Because $\Psi_{\textit{pair}}$ is built with \textlog{DURA}, each such step comes with the obligation to show that the crash condition $R$ holds afterwards.
By the second premise of \refrule{ewp:write-pair}, to show $R$, it suffices to show $\postcrash(\textlog{pair}(a_0,a_1)\lor\textlog{pair}(b_0,b_1))$.

We start by case splitting on the disjunction in $\textlog{pair}(a_0,a_1)$.
We will consider the left case here, where the selector $\ell_d$ indicates that the first copy of the pair is active; the other case is symmetric.
Line 1 reads $\ell_d\diskmapsto\adiskval{1}{1}$ and returns $1$.
Lines 2 and 3 write $(b_0,b_1)$ to the inactive copy, and the barrier on line 4 makes these writes durable, giving us $\ell_2\diskmapsto\adiskval{b_0}{b_0}$ and $\ell_2+1\diskmapsto\adiskval{b_1}{b_1}$.
Because $\textlog{pair}(a_0,a_1)$ says nothing about the inactive copy, it still holds after each of these steps, and $\textlog{pair}(a_0,a_1)\proves\postcrash\textlog{pair}(a_0,a_1)$ discharges the crash condition.

The key step is the store to the selector $\ell_d$ on line 5.
By the \textlog{ADISK\_STORE} protocol, we obtain $\ell_d\diskmapsto\adiskval{v_c}{2}$ for some $v_c\in\{1,2\}$, where the prophesied $v_c$ tells us in advance whether this store will survive a crash.
We case split on $v_c$:
if $v_c=2$, the store will be durable, so we now own $\textlog{pair}(b_0,b_1)$, which gives the crash condition through the right disjunct;
if $v_c=1$, a crash would undo the store, so we have $\postcrash\textlog{pair}(a_0,a_1)$, the left disjunct.
Finally, the barrier on line 7 takes $\ell_d\diskmapsto\adiskval{v_c}{2}$ and gives it back with the fact $v_c=2$, since the pending store to $\ell_d$ is now flushed.
In the second case this contradicts $v_c=1$, so only the first case remains, where we own $\textlog{pair}(b_0,b_1)$ and conclude $\Phi(())$ using the second premise.
\section{Case Study: Distributed Systems with IronFleet-Style Atomic Blocks}\label{sec:distributed}

In this section, we consider a distributed system with multiple nodes connected by an unreliable network, in which nodes communicate through messages that may be dropped, delayed, duplicated, or re-ordered.
On top of the network, a global scheduler decides the order of execution of nodes.

\paragraph{Network}

The network provides two operations: $\textlog{NETWORK}\triangleq\textlog{SEND}\oplus\textlog{RECV}$, where
\begin{align*}
\textlog{SEND}(v,\Phi)&\triangleq\Exists s,t,m,M.v=(\effecttag{send},(s,t,m))\ast t\netmapsto M\ast (t\netmapsto\{(s,t,m)\}\cup M\wand\Phi(()))\\
\textlog{RECV}(v,\Phi)&\triangleq\Exists t,M.v=(\effecttag{recv},t)\ast t\netmapsto M\ast
\left(
\begin{gathered}
\All x. t\netmapsto M\ast\big(
x=\keyword{inl}\ ()\lor{}\\(\Exists s,m.x=
\keyword{inr}\ (s,t,m) \land (s,t,m)\in M)\big)\wand \Phi (x)
\end{gathered}
\right)
\end{align*}

Assertion $t\netmapsto M$ says that $M$ is the set of messages that have ever been sent to address $t$.
Since messages can be arbitrarily duplicated, this set is monotonically increasing \wrt the subset relation.
The \textlog{SEND} protocol expects a package of (source address, destination address, message) as input, and adds this package to the message history of the destination address.
The \textlog{RECV} protocol expects the destination address $t$ as input and non-deterministically chooses either not to return a message or to return an arbitrary message that has been sent to $t$.
The dropping of a message is implicitly modeled as just never having it be selected for receipt.
The handler providing protocol $\textlog{NETWORK}$ is called $\function{run}_{\effecttag{network}}$, which implements the network as a soup of messages~\cite{wilcox2015-verdi,krogh-jespersen2020-aneris}.
The code for this handler can be found in \refappendix{sec:app:run-network}.%

\paragraph{Scheduler with IronFleet-Style Atomic Blocks}
Next, we need a scheduler $\function{run}_{\effecttag{distr}}$ (code shown in \refappendix{sec:app:run-distr}) that specifies how nodes run concurrently through a $\textlog{DISTR}_\W^t$ protocol.
\begin{align*}
\textlog{DISTR}_\W^t&\triangleq\textlog{ATOM}_\W(\textlog{SEND}\oplus\textlog{START}_\W^t)\oplus\textlog{RECV}\\
\textlog{START}_\W^t(v, \Phi)&\triangleq
\Exists e. v = (\effecttag{start}, {\Lam\_. e})
\ast
\left(t=\IsChild\lor\later\ewp[]{e}[\W]{\smash{\textlog{DISTR}_\W^{\IsChild}}}{\Ret \_ . \TRUE}\right) \ast \Phi(())
\end{align*}
The $\function{run}_{\effecttag{distr}}$ handler resembles $\function{run}_{\effecttag{conc}}$ and potentially transfers control to different nodes when an effect is raised by a node.
The \textlog{START} protocol is used for initially creating nodes.
In the protocol above, the \textlog{RECV} protocol is \emph{outside} the \textlog{ATOM} protocol transformer.
The reason for this is that the $\function{run}_{\effecttag{distr}}$ scheduler does \emph{not} transfer control to another node when processing a \effecttag{recv} operation.
In other words, a node can receive a series of messages without transferring control to another node.
This modeling choice is inspired by IronFleet~\cite{hawblitzel2015-ironfleet}, which uses a movers-based parallel reduction proof~\citep{DBLP:conf/popl/Lipton75} to treat a sequence of receives followed by a sequence of sends as an atomic step.
Our global scheduler permits the prover to view a series of \effecttag{recv} operations, followed by a series of node-local processing operations, followed by one \effecttag{send} operation as an atomic block.\footnote{We preempt after a single send because a node could diverge after sending. IronFleet avoids this by proving total correctness.}
As a result, because \textlog{RECV} is not included in the \textlog{ATOM} component, we do \emph{not} need to close shared invariants when performing a \effecttag{recv} operation.

Even though this scheduler does not include preemptions at \effecttag{recv}, the absence of these preemptions does not affect the overall set of possible behaviors of the program.
To prove this, we use \binlog{} to prove that an explicit \keyword{yield} preemption is contextually equivalent to the unit value. %
Thus, adding in additional preemptions does not change the program's behavior.
To carry out this proof, we develop a node-local specification resource $\specnode{\gamma}{t}{e}$, similar to the thread-local version described in \Cref{sec:relconc}.
It also uses the evidence accumulation technique,
but additionally accumulates the evidence that a node can delay message receipt without changing behavior.
The idea is that if a message $m$ is received at some time, then if we delay that receipt to some later time, it is still possible to receive $m$.
We capture this evidence using a \emph{monotonic} resource algebra to track the set $M$ of messages.
We then build a logical relation on top of the $\specnode{\gamma}{t}{e}$ resource in the style of \Cref{sec:logrel} to complete the proof.
More details can be found in \refappendix{sec:app:distr}.
\section{Related Work}\label{sec:related}

\subsection{Program Logics for Effect Handlers}\label{sec:related-handlers}
The most closely related work is the Hazel logic for effect handlers~\citep{vilhena2021-hazel}.
As discussed in \Cref{sec:conc}, \unlog{} extends Hazel with support for extensible worlds.
While \unlog{} restricts protocols to be monotonic,
Hazel does not directly impose this requirement, but instead computes an upward closure of the protocol to enforce monotonicity.
Both approaches are identical in practice because protocols defined in a ``standard'' way are already monotonic: protocols directly capturing the input condition and output condition (like the \textlog{READ} and \textlog{WRITE} protocols) are monotonic, and monotonicity is closed under the $\oplus$ combinator and all protocol transformers we have considered.

In both \unlog{} and Hazel, continuations are \emph{one-shot}, meaning that an effect can only invoke the continuation once. Subsequent to Hazel, \citet{vilhena2022-thesis} developed Maze, a separation logic that supports multi-shot continuations.
The lack of multi-shot continuations in \unlog{} means we could not implement backtracking non-determinism. %
However, to support multi-shot continuations using Maze's approach, we would have to discard all non-persistent resources (\eg points-to) when raising a multi-shot effect, and it is not possible to compose this type of reasoning about multi-shot effects with our other effects.
Introducing multi-shot continuations also breaks the frame rule (\refrule{ewp:frame}) because an expression could return  multiple times.

Hazel only handles unary reasoning, whereas \binlog{} supports relational reasoning through an encoding into \unlog{}.
The Blaze logic developed by \citet{vilhena2025-blaze} also supports relational reasoning for effect handlers.
Like \binlog{}, Blaze builds on a unary logic and represents a specification program via ghost state.
However, unlike \binlog, in which the unary logic and the specification program have separate protocols, Blaze instead provides a judgment with a \emph{relational} protocol.
They use this to prove refinements in which the interpretation of effects is different between the two programs.
In contrast, our examples keep effects the same on both sides and prove that client programs under these effects are equivalent.

Among other examples, \citet{vilhena2025-blaze} use Blaze to prove that a handler implementation of concurrency refines a primitive concurrency effect.
This refinement is in some sense the opposite of the direction that motivated our refinement proof in \Cref{sec:relation}: it essentially shows that for every execution of the concurrency handler (which only preempts at effects), there is a corresponding execution using primitive concurrency (which preempts at every step).
In contrast, we show that inserting additional preemption points when using the concurrency handler does not generate new behaviors.
This is morally equivalent to showing that the concurrency handler already covers all possible behaviors that could be generated by a full interleaving semantics.
It would be interesting to apply Blaze's approach to the kinds of applications we have considered here to justify the soundness of alternate implementations of effects that allow for deriving stronger reasoning rules.

Like the language of Hazel, \HandlerLang{} does not natively support effect tags, but instead requires values of effects to have the form (tag, value) and expects the handlers to dispatch and re-raise effects based on the tag, whereas many other works on effect handlers~\citep{plotkin2009-effects,biernacki2017-handle-refine,biernacki2020-lexical-effects,vilhena2025-blaze,vilhena2023-tes,van-rooij2025-effect,tang2025-met,10.1145/2976022.2976033} bake effect tags into the syntax and operational semantics.
The language-level tags make it possible to syntactically decide the set of tags handled by an evaluation context.
Blaze~\citep{vilhena2025-blaze} exploits this for a more general bind rule not limited to (absolutely) neutral contexts, but only requiring that the context does not handle tags of effects that may be raised by the expression being bound.
\unlog{} does not use syntactic tags because restricting to syntactic tags prevents handlers like $\function{run}_{\effecttag{conc}}$ from interposing on all effects that propagate to outer handlers.

Our logical relations consider type systems with a fixed collection of effects and do not provide rules for typing general effect handlers.
Tes~\citep{vilhena2023-tes} and Affect~\citep{van-rooij2025-effect} use logics to construct unary logical relations models for type systems for effect handlers.
\citet{biernacki2017-handle-refine,biernacki2020-lexical-effects} directly construct a binary logical relations model for effects and handlers using biorthogonality and step-indexing.

\subsection{Extensible Program Logics}\label{sec:related-extensible}
As described in the introduction, \citet{vistrup2025-alacarte} develop an approach to extensible program logics using ITrees~\cite{xia2020-itrees}. They use a mechanism called \emph{logical effect handlers} to interpret ITree events for an effect, which has similarities to the way Hazel and \unlog{}'s protocols give a logical interpretation of what a raised effect will do.
Their soundness proofs relate these logical effect handlers to interpretations of the effects.
Similarly, \citet{frumin2024-gitrees} develop \emph{guarded} ITrees as a compositional model of higher-order programming languages with an extensible program logic.
Neither \citet{vistrup2025-alacarte} nor \citet{frumin2024-gitrees} develop a relational logic.
On the other hand, \citeauthor{vistrup2025-alacarte} show how to model other features, such as total correctness and angelic non-determinism, which we do not consider.
Initially, the ITree-based approach of \citet{vistrup2025-alacarte} did not include support for prophecy variables. 
However, in subsequent work on the Mosaic framework~\citep{vistrup2026-mosaic}, \citeauthor{vistrup2026-mosaic} showed how to extend the ITree-based approach to also handle prophecies.
A further benefit of Mosaic is that it allows composing extensions to the \emph{pure} syntax of the language, whereas \unlog{}'s pure calculus is fixed.

Like \unlog{}'s extensible worlds, \citet{yusuke2025-nola} parameterize the Iris update modality and Hoare triples by a notion of a world.
However, they require that the update shifts to the same world before and after, \ie only shifts of the form $\pvs[\W][\W]$.
Hence, they cannot model the use of extensible worlds in \Cref{sec:conc}, in which invariants are kept open across non-preempting steps.

Dijkstra Monads~\cite{swamy2013-dijkstra} offer a framework for deriving pre- and postcondition reasoning about dependently-typed programs with monadic effects,
and, more recently, some aspects of algebraic effect handlers~\citep{DBLP:journals/pacmpl/MaillardAAMHRT19}.
One benefit of Dijkstra Monads is that they work well for representing effectful programs using a monadic, shallow embedding inside of a dependently-typed language like Rocq or Lean.
\citet{loom} have recently used this approach to develop a framework for verifying programs in what they call a \emph{multi-modal} way, which combines a variety of automated and interactive verification techniques in a powerful, common framework.
In particular, they note that their shallow embedding leads to a naturally executable semantics, which they contrast both with standard approaches to building program logics and with the work of \citet{vistrup2025-alacarte}.
While we have not explored this, one advantage of \unlog{} is that it in principle also leads to an executable semantics: one simply needs to write an interpreter for \HandlerLang{}, and then all implementations of effects become executable as well.
It would be interesting as future work to see what other aspects of their multi-modal verification framework can be done using effect handlers.

While \citeauthor{loom} and other prior works demonstrate some powerful benefits of Dijkstra Monads, certain effects and verification tasks are challenging to represent in a shallowly-embedded monadic way as compared to our deep embedding with effect handlers.
For example, existing applications of Dijkstra Monads do not model fine-grained concurrency or crashes, and these may be challenging to encode as monads in a compositional way.
And, some applications of program logics are most naturally carried out with explicit syntax: for example, building logical relations models of type systems, where one needs to refer to the typing rules of the object language.
Finally, existing frameworks based on Dijkstra Monads do not support relational reasoning.
\section{Conclusion}\label{sec:conclusion}
We present \unlog, an extensible program logic that allows users to model new effects by writing programs in the form of effect handlers.
We also present \binlog{}, a relational logic that we have used to justify the soundness of non-standard encodings of various effects.
As case studies, we use \unlog{} to recover Iris's \HeapLang{} logic and crash-safety reasoning from Perennial, and construct a new separation logic for distributed systems in the style of IronFleet.

The key idea underlying \unlog{} is that effect handlers are just regular programs and therefore extending a program logic does not require skills distinct from those needed to use a program logic.
But this does come with a trade-off: the executable nature of effect handlers prevents us from modeling certain language features that cannot be expressed as programs, like angelic choice, non-computable real numbers, and axiomatic weak memory models.

Another limitation is that \unlog{} only covers partial correctness. %
For the core \unlog{} logic, we do not foresee any fundamental obstacle to supporting total correctness.
However, the contextual equivalence between our non-standard scheduler and a fully preemptive scheduler does not hold when termination is observable, so we would need to directly model a handler that can interrupt pure steps, which would require extending the effect handler logic to support asynchronous interruptions.

Finally, \unlog{} is only extensible by adding new effects; its core pure calculus is fixed.
For future work, it would be interesting to find a way to combine the effect handler approach to extensibility with a mechanism for extending
the pure syntax of the language.

\section*{Data Availability Statement}\addcontentsline{toc}{section}{Data Availability Statement}
The Rocq mechanization accompanying this work is available on Zenodo~\cite{artifact} and on GitHub at \url{https://github.com/u8cat/ficus/tree/popl2027-aec-final}.

\begin{acks}
  We thank François Pottier and the anonymous reviewers for their constructive feedback, which helped improve the presentation in the paper.
  This work was supported in part by the \grantsponsor{NSF}{National Science Foundation}{} under Grant No.~\grantnum{NSF}{2319168} and Grant No.~\grantnum{NSF}{2524669}, as well as the \grantsponsor{Carlsberg Foundation}{Carlsberg Foundation}{} under Grant No.~\grantnum{Carlsberg Foundation}{CF23-0791}.
  Any opinions, findings, and conclusions or recommendations expressed in this material are those
  of the authors and do not necessarily reflect the views of these funding agencies.
\end{acks}

\appendix

\captionsetup[figure]{belowskip=0pt,aboveskip=4pt}

\section{Eliminating Primitive Effects via Pre-Determinism}\label{sec:predet}

Although \HandlerLang{} does not come with built-in mutable state, it is not strictly pure because it includes two primitive effects: $\keyword{pick}$ is not deterministic and $\ghostcode{\keyword{observe}}$ adds a special label to the transition relation.
However, it turns out that even these primitives can be removed; we kept them in the main body of the paper for simplicity of presentation.
In this section, we present an alternative core calculus \PureHandlerLang{} that is strictly pure and a logic for it called \pureunlog.
\PureHandlerLang{} is identical to \HandlerLang{} except that the former does not have $\keyword{pick}$ and $\ghostcode{\keyword{observe}}$ primitives.
Instead, it adds a special $\keyword{label}$ primitive.
The $\keyword{label}$ primitive is a trivial, pure operation.
It takes an integer as an argument and simply reduces to unit: $\keyword{label}\ z\hstep ()$.
Its role is simply as a syntactic marker for annotating certain points of the program.

Then, within \PureHandlerLang{}, we model $\keyword{pick}$ and $\ghostcode{\keyword{observe}}$ also as effect handlers.
The non-determinism of $\keyword{pick}$ is simulated by using a handler parameterized by a list of booleans that will be returned to invocations of $\keyword{pick}$.
Once the language is made deterministic, the prophecy behavior needed for $\ghostcode{\keyword{observe}}$ becomes trivial: there is nothing to ``predict'' with prophecies, since the program's future behavior is entirely predetermined.
Below, we first explain in more detail the protocol and implementation of these handlers.
Finally, we justify the soundness of this encoding by connecting it back to the original \HandlerLang{} operational semantics in which these effects were primitives, thereby showing that it is equivalent.

\subsection{Non-Determinism}\label{sec:predet-nondet}
We first define the protocol $\textlog{PICK}(v,\Phi)\eqdef v=(\effecttag{pick},())\ast(\All b\in\mathbb{B}.\Phi(b))$, and set $\function{pick}\eqdef\keyword{do}\ (\effecttag{pick},())$.
It is straightforward to verify that the new $\function{pick}$ in \PureHandlerLang{} satisfies the same client-side reasoning rule as the primitive $\keyword{pick}$ in \HandlerLang.
The handler $\function{run}_{\effecttag{pick}}$ takes as parameter a list of booleans for the future choices to return.
Each time a $\effecttag{pick}$ effect is raised, the handler pops the value from the head of the list and returns it.
In case the list is used up, the handler simply diverges, leading to a trivially safe program under partial correctness.
The implementation of $\function{run}_{\effecttag{pick}}$ is shown in \Cref{sec:app:run-pick}.
The specification \refrule{ewp:pick-run} then guarantees that if a program $\textit{main}$ is verified under the $\textlog{PICK}$ protocol, then an $\ewpNoArg$ holds for the handler when using any list of booleans.
\begin{mathpar}
\mprset{fraction=--\ast}
\inferrule[\hypertarget{ewp:pick-run}{Ewp-PickRun}]
{\effecttag{pick}\notin\tags(\Psi)\\\ewp{\textit{main}\ ()}[\W]{\Psi\oplus\textlog{PICK}}{\Phi}}
{\All \vv b\in\mathbb{B}^*. \ewp{\function{run}_{\effecttag{pick}}\ \textit{main}\ \vv b}[\W]{\Psi}{\Phi}}
\end{mathpar}

Unlike the standard non-determinism semantics where the choice of the value being picked is delayed to the time the \keyword{pick} operation is executed, this handler essentially chooses all non-deterministic values upfront, even before any operation is executed.
Nevertheless, for verifying safety properties, our model is sound \wrt the primitive model of non-determinism used in the main body, because given any prefix of an execution under the primitive non-determinism model, there exists a list of booleans that would yield the same behavior when using the handler.
Thus, because $\vv b$ is universally quantified over all such lists, the conclusion of \refrule{ewp:pick-run} covers all possible prefixes of executions of a program, which is all that is needed for partial correctness and safety properties.

\subsection{Global Prophecy}
Next, to encode the global prophecy rules, we define $\function{observe}\ w\eqdef\keyword{do}\ (\effecttag{observe},w)$ and
\begin{align*}
\textlog{OBSERVE}(v,\Phi)\eqdef{}&\Exists w,\vv v.v=(\effecttag{observe},w)\ast\primproph{\vv v}\\
{}\ast{}&(\All \vv v\!'.\vv v=w::\vv v\!'\ast\primproph{\vv v\!'}\wand \Phi (()))
\end{align*}
Here, $\primproph{\vv v}$ is just a regular piece of authoritative ghost state defined as $\ownGhost{\gamma_p}{\authfrag{\exinj(\vv{v})}}$, where $\gamma_p$ is some globally fixed ghost name allocated at the time the handler is installed.

Meanwhile, the handler for this protocol is effectively just a no-op: to handle an effect of $(\effecttag{observe},v)$, it evaluates $(\keyword{label}\ \effecttag{observe}, v)$ to syntactically ``mark'' the observation effect, discards the result, and calls the continuation with $()$.
As alluded to above, at the logic level, because the language is pure and thus deterministic, the future execution behavior of the program, and thus all of the reductions of $\keyword{label}$ that will occur, are predetermined.
Therefore, we can simply ``pre-compute'' upfront what this trace of reductions will be at the time the handler is installed.\footnote{In our Rocq formalization, this step uses the law of excluded middle.}
Because that pre-computed trace is the only possible trace the program can yield, when an observation effect is raised, the value of that observation must match the value predicted by the global prophecy.

To carry out this argument, we extend the program logic with additional ghost state.
The core new feature is a ghost state mechanism we call \emph{snapshot}, which comes with a $\snapshot{}{e}$ resource witnessing that the program was $e$ at some point in the past.
Additionally, we also need permissions on labels and time receipts~\cite{mevel2019-time-receipt}, and a way to track the surrounding evaluation context that an expression is evaluating under in $\ewpNoArg$.
The details about these features are explained in \Cref{sec:app:pureficus-rules} and a proof based on these features is outlined in \Cref{sec:app:veri-run-observe}.

\subsection{Soundness}
Similar to \Cref{thm:unlog-adequacy}, there is also an adequacy theorem for \pureunlog.
\begin{theorem}[Adequacy, \pureunlog]\label{thm:pureunlog-adequacy}
  Let $\varphi$ be a first-order predicate.
  If\/ $\labelEn{\top}\vdash \ewp e[]{\bot}{\varphi}$, then executing $e$ never gets stuck, and if $e\steps v$ then $\varphi(v)$ holds.
\end{theorem}
Here, $\labelEn{\top}$ is a token saying that all \keyword{label} expressions are permitted to execute.

Further, we reconstruct \HandlerLang{} using \PureHandlerLang, and prove that for every execution in \HandlerLang, there exists a corresponding execution in \PureHandlerLang{} that produces the same result.
To state this result formally, in \HandlerLang, define trivial effect handlers $\function{run}'_{\effecttag{pick}}$ and $\function{run}'_{\effecttag{observe}}$ that simply perform the primitive $\keyword{pick}$ and $\ghostcode{\keyword{observe}}$.
These similarly implement protocols $\textlog{PICK}$ and $\textlog{OBSERVE}$ respectively using the primitive rules for these effects.

Let $H_p$ be a context in \PureHandlerLang{} that handles the non-determinism and global prophecy effects and $H_f$ be a context in \HandlerLang{} with $\function{run}'_{\effecttag{pick}}$ and $\function{run}'_{\effecttag{observe}}$ installed.
\begin{align*}
H_p(\vv b)&\eqdef \function{run}_{\effecttag{observe}}\ (\Lam \_. \function{run}_{\effecttag{pick}}\ (\Lam\_.\hole)\ \vv{b})\\
H_f&\eqdef \function{run}'_{\effecttag{observe}}\ (\Lam \_. \function{run}'_{\effecttag{pick}}\ (\Lam\_.\hole))
\end{align*}
Then, we have the following theorem.

\begin{theorem}[Model Soundness]\label{thm:pureficus-soundness}
Let $e$ be an expression of \HandlerLang{} that does not contain $\keyword{pick}$ and $\ghostcode{\keyword{observe}}$ language primitives, and $\ltrans e\rtrans$ be the equivalent expression of \PureHandlerLang{}. If evaluating $H_p(\vv{b})[\ltrans e\rtrans]$ never gets stuck for all $\vv{b}$, then,
\begin{itemize}
\item evaluating $H_f[e]$ never gets stuck;
\item if $H_f[e]\steps v$, then there exists a boolean list $\vv b$ such that $H_p(\vv b)[\ltrans e\rtrans]\steps \ltrans v\rtrans$.
\end{itemize}
\end{theorem}
This shows that any safety property verified under \PureHandlerLang{} also holds under \HandlerLang{}.
\section{\unlog}

\subsection{Semantics}

\paragraph{Head Reduction Rules} $e\hstep[\ghostcode{\vv{\kappa}}] e'$\\
\begin{minipage}{0.59\linewidth}
$$\begin{array}{l@{\hspace*{0.5em}}r@{}c@{}l}
\textsc{\hypertarget{head:beta}{Hd-Beta}}&(\Rec f\ x. e)\ v&{}\hstep[\ghostcode{\nil}]{} &\subst{e}{f,x}{(\Rec f\ x. e),v}\\
\textsc{\hypertarget{head:eff-ap-r}{Hd-EffApR}}&e_1\ \S(N)[v_2]&{}\hstep[\ghostcode{\nil}]{}&\S(e_1\ N)[v_2]\\
\textsc{\hypertarget{head:eff-ap-l}{Hd-EffApL}}&\S(N)[v_1]\ v_2&{}\hstep[\ghostcode{\nil}]{}&\S(N\ v_2)[v_1]\\
\textsc{\hypertarget{head:eff-do}{Hd-EffDo}}&\keyword{do}\ \S(N)[v]&{}\hstep[\ghostcode{\nil}]{}&\S(\keyword{do}\ N)[v]\\
\end{array}$$
\end{minipage}%
\begin{minipage}{0.41\linewidth}
$$\begin{array}{l@{\hspace*{0.5em}}r@{}c@{}l}
\textsc{\hypertarget{head:cont}{Hd-Cont}}&(\keyword{cont}\ N)\ v&{}\hstep[\ghostcode{\hphantom{[}\nil\hphantom{]}}]{}&N[v]\\
\textsc{\hypertarget{head:do}{Hd-Do}}&\keyword{do}\ v&{}\hstep[\ghostcode{\hphantom{[}\nil\hphantom{]}}]{}&\S(\hole)[v]\\
\textsc{\hypertarget{head:pick}{Hd-Pick}}&\keyword{pick}&{}\hstep[\ghostcode{\hphantom{[}\nil\hphantom{]}}]{}&b\\
\textsc{\hypertarget{head:observe}{Hd-Obs}}&\ghostcode{\keyword{observe}\ v}&{}\hstep[\ghostcode{[v]}]{}&\ghostcode{()}\\
\end{array}$$
\end{minipage}
\vspace*{-0.5ex}
$$\begin{array}{l@{\hspace*{0.5em}}r@{}c@{}l}
\textsc{\hypertarget{head:try-eff}{Hd-TryEff}}&\keyword{try}\ \S(N)[v_0]\ \keyword{with}\ v_1\ k\Rightarrow\ e_1\mid\keyword{ret}\ v_2\Rightarrow e_2&{}\hstep[\ghostcode{\nil}]{}& \; \subst{e_1}{v_1,k}{v_0,\keyword{cont}\ N}\\
\textsc{\hypertarget{head:try-val}{Hd-TryVal}}&\keyword{try}\ v_0\ \keyword{with}\ v_1\ k\Rightarrow\ e_1\mid\keyword{ret}\ v_2\Rightarrow e_2&{}\hstep[\ghostcode{\nil}]{}& \; \subst{e_2}{v_2}{v_0}\\
\end{array}$$

\paragraph{Pure Reduction and Its Reflexive Transitive Closure}
$e\step[\ghostcode{\vv{\kappa}}] e'$ and $e\steps[\ghostcode{\vv{\kappa}}] e'$\\
\begin{align*}
e\step[\ghostcode{\vv{\kappa}}] e'&\triangleq \Exists K,\tilde{e},\tilde{e}' . e=K[\tilde{e}]\land e'=K[\tilde{e}']\land \tilde{e}\hstep[\ghostcode{\vv{\kappa}}] \tilde{e}'\\
e\steps[\ghostcode{\vv{\kappa}}] e'&\triangleq \left(e=e'\ghostcode{\land\vv{\kappa}=\nil}\right) \lor \left(\Exists e''\ghostcode{,\vv{\kappa}_1,\vv{\kappa}_2} . \ghostcode{\vv\kappa=\vv\kappa_1\dplus\vv\kappa_2\land}\smash{e\step[\ghostcode{\vv{\kappa}_1}]e''} \land \smash{e''\steps[\ghostcode{\vv{\kappa}_2}] e'}\right)
\end{align*}

\subsection{Reasoning Rules}

\begin{mathparpagebreakable}
\mprset{fraction=--\ast}
\inferrule[\hypertarget{ewp:value-app}{Ewp-Value}]
{\pvs[\W_1][\W_2] \Phi(v)}
{\ewp v[\W_1,\W_2]{\Psi}{\Phi}}\label{ewp:value}\and
\inferrule[\hypertarget{ewp:do-app}{Ewp-Do}]
{\Psi(v,\Phi)}
{\ewp{\keyword{do}\ v}[\W]{\Psi}{\Phi}}\and
\inferrule[\hypertarget{ewp:dowupd-app}{Ewp-DoWupd}]
{\pvs[\W_1][\bot]\Psi(v,(\Lam r . \pvs[\bot][\W_2]\Phi(r)))}
{\ewp{\keyword{do}\ v}[\W_1,\W_2]{\Psi}{\Phi}}\and
\inferrule[\hypertarget{ewp:mono-app}{Ewp-Mono}]
{\ewp e[\W_1,\W_2]{\Psi}{\Phi}\\ \Psi\sqsubseteq\Psi' \\
\All v .\Phi(v)\wand\pvs[\W_2]\Phi'(v)}
{\ewp e[\W_1,\W_2]{\Psi'}{\Phi'}}\and
\inferrule[\hypertarget{ewp:worldmono-app}{Ewp-WorldMono}]
{\ewp e[\W]{\Psi}{\Phi} \\ \W\sqsubseteq\W'}
{\ewp e[\W']{\Psi}{\Phi}}\and
\inferrule[\hypertarget{ewp:frame-app}{Ewp-Frame}]
{R \\ \ewp e[\W_1,\W_2]{\Psi}{\Phi}}
{\ewp e[\W_1,\W_2]{\Psi}{\Ret v . R\ast\Phi(v)}}\and
\inferrule[\hypertarget{ewp:wupdpre-app}{Ewp-WupdPre}]
{\pvs[\W_1][\W_2]\ \ewp e[\W_2,\W_3]{\Psi}{\Phi}}
{\ewp e[\W_1,\W_3]{\Psi}{\Phi}}\and
\inferrule[\hypertarget{ewp:wupdpost-app}{Ewp-WupdPost}]
{\ewp e[\W_1,\W_2]{\Psi}{\Ret v . \smash{\pvs[\W_2][\W_3] \Phi(v)}}}
{\ewp e[\W_1,\W_3]{\Psi}{\Phi}}\and
\inferrule[\hypertarget{ewp:step-app}{Ewp-Step}]
{\All e'.e\step e'\wand\ewp{e'}[\W_1,\W_2]{\Psi}{\Phi}\\\Exists e'.e\step e'}
{\ewp{e}[\W_1,\W_2]{\Psi}{\Phi}}\and
\inferrule[\hypertarget{ewp:bind-app}{Ewp-Bind}]
{\ewp e[\W_1,\W_2]{\Psi}{\Ret v .\smash{\ewp{N[v]}[\W_2,\W_3]{\Psi}{\Phi}}}}
{\ewp {N[e]}[\W_1,\W_3]{\Psi}{\Phi}}\and
\inferrule[\hypertarget{ewp:pure-bind-app}{Ewp-PureBind}]
{\ewp e[\W_1,\W_2]{\bot}{\Ret v .\smash{\ewp{K[v]}[\W_2,\W_3]{\Psi}{\Phi}}}}
{\ewp {K[e]}[\W_1,\W_3]{\Psi}{\Phi}}\and
\inferrule[\hypertarget{ewp:world:frame-app}{Ewp-WorldFrame}]
{\ewp[]e[\W_1,\W_2]{\Psi}{\Phi}}
{\ewp[]e[\W_1 \oplus \W,\W_2 \oplus \W]{\Psi}{\Phi}}
\and
\inferrule[\hypertarget{ewp:obs-app}{Ewp-Obs}]
{\primproph{\vv{v}}}
{\ewp{\ghostcode{\keyword{observe}\ w}}[\W]{\Psi}{\Ret\_. \Exists \vv{v}\!'. \vv{v}=w::\vv{v}\!'\ast\primproph{\vv{v}\!'}}}
\end{mathparpagebreakable}

\subsection{Model}
\begin{align*}
\ewp v[\W_1,\W_2]{\Psi}{\Phi}\triangleq{}& \pvs[\W_1][\W_2]\Phi(v)\\
\ewp{\S(\hole)[v]}[\W_1,\W_2]{\Psi}{\Phi}\triangleq{}&\pvs[\W_1][\bot]\Psi(v,\Lam w . \pvs[\bot][\W_2]\Phi(w))\\
\ewp{\S(N)[v]}[\W_1,\W_2]{\Psi}{\Phi}\triangleq{}&\pvs[\W_1][\bot]\Psi(v,\Lam w . \pvs[\bot]\later(\ewp{N[w]}[\bot,\W_2]{\Psi}{\Phi}))\\
\ewp e[\W_1,\W_2]{\Psi}{\Phi}\triangleq{}& \ghostcode{\All \vv{\kappa}_1,\vv{\kappa}_2. \primprophauth{\vv{\kappa}_1\dplus\vv{\kappa}_2}\wand} \pvs[\W_1][\bot] \red(e)\ast\notag\\
&\hspace*{-5em}\All e' . e\step[\ghostcode{\vv{\kappa}_1}] e'\wand\laterCredit(1)\wand\pvs[\bot]\later\pvs[\bot]\ghostcode{\primprophauth{\vv{\kappa}_2}\ast}\ewp{e'}[\bot,\W_2]{\Psi}{\Phi}
\end{align*}
$$\primprophauth{\vv{v}}\triangleq\ownGhost{\gamma_p}{\authfull{\exinj(\vv{v})}}\qquad
\primproph{\vv{v}}\triangleq\ownGhost{\gamma_p}{\authfrag{\exinj(\vv{v})}}$$
\section{Protocols for the Prophetic Heap}\label{sec:app:proph-heap}

\begin{align*}
\textlog{PROPH\_ALLOC}(v,\Phi)\triangleq{}&\Exists x.v=(\effecttag{alloc},x)\ast(\All \ell, \vv v.\ell\mapsto x\ast\proph{\ell}{\vv v}\ast\prophE{\ell}\wand\Phi(\ell))\\
\textlog{PROPH\_LOAD}(v,\Phi)\triangleq{}&\Exists \ell, x, \vv{v}. v=(\effecttag{load},\ell)\ast \ell\mapsto x\ast\proph{\ell}{\vv{v}}\ast\prophE{\ell}\ast\\
&(\All \vv{v}\!'.\ell\mapsto x\ast \vv{v}=(x,\textlog{load})::\vv{v}\!'\ast\proph{\ell}{\vv{v}\!'}\ast\prophE{\ell}\wand\Phi(x))\notag\\
\textlog{PROPH\_LOAD}'(v,\Phi)\triangleq{}&\Exists \ell, q, x. v=(\effecttag{load},\ell)\ast \ell\stackrel{q}{\mapsto} x\ast\prophD{\ell}\ast
(\ell\stackrel{q}{\mapsto} x\wand\Phi(x))\\
\textlog{PROPH\_STORE}(v,\Phi)\triangleq{}&\Exists \ell,x,y,\vv{v}.v=(\effecttag{store},(\ell,y))\ast \ell\mapsto x\ast\proph{\ell}{\vv v}\ast\prophE{\ell}\ast\\
&(\All \vv{v}\!'.\ell\mapsto y\ast \vv{v}=((),\textlog{store}(y))::\vv{v}\!'\ast\proph{\ell}{\vv{v}\!'}\ast\prophE{\ell}\wand\Phi(()))\notag\\
\textlog{PROPH\_STORE}'(v,\Phi)\triangleq{}&\Exists \ell,x,y.v=(\effecttag{store},(\ell,y))\ast \ell\mapsto x\ast\prophD{\ell}\ast
(\ell\mapsto y\wand\Phi(()))\\
\textlog{PROPH\_CAS}(v,\Phi)\triangleq{}&\Exists \ell,w,x,y,\vv{v}. v=(\effecttag{cas},(\ell,x,y))\ast \ell \mapsto w\ast\proph{\ell}{\vv{v}}\ast\prophE{\ell}\ast\notag\\
&(\All\vv{v}\!'. \ell\mapsto(\text{if }w=x\text{ then } y\text{ else }w)\ast \vv{v}=((w,w=x),\textlog{cas}(x,y))::\vv{v}\!'\ast\\
& \proph{\ell}{\vv{v}\!'}\ast\prophE{\ell}\wand\Phi((w,w=x)))\notag\\
\textlog{PROPH\_CAS}'(v,\Phi)\triangleq{}&\Exists \ell,q,w,x,y. v=(\effecttag{cas},(\ell,x,y))\ast \ell\stackrel{q}{\mapsto}w\ast (q=1\lor w\neq x)\ast \prophD{\ell}\ast\\
&(\ell\stackrel{q}{\mapsto}(\text{if }w=x\text{ then } y\text{ else }w) \wand\Phi((w,w=x)))
\end{align*}

Each location is associated with a prophecy variable $\proph{\ell}{\vv v}$ and a pair of tokens $\prophE{\ell}$/$\prophD{\ell}$ controlling whether the prophecy variable is enabled or disabled.
The prophecy variable is enabled when the location is allocated, and one can disable it using rule $\prophE{\ell}\proves\pvs\prophD{\ell}$.
Once disabled, a location can no longer be enabled.
In fact, token $\prophD{\ell}$ merely carries the knowledge that one agrees to never use the prophecy variable associated with the location and it is therefore persistent.
To perform an operation on a location, one must either provide the ownership of the associated prophecy variable and the enable token, or provide the disable token.
\section{Details about the Crash Recovery System} \label{sec:app:crash}

This section uses the complete $\WCrashInvTok{\E,R_c}$ token. Compared to $\WCrashInvTok{R_c}$ used in \Cref{sec:crash}, it has one extra parameter for the enabled crash borrows.
The connection between the two tokens is $\WCrashInvTok{R_c}=\WCrashInvTok{\top,R_c}$.

\subsection{Post-Crash Modality}
\begin{align*}
\textlog{crashed}\triangleq{}&\Exists M\colon\textdom{PID}\to\textdom{List}(\textdom{Val}\times\textdom{Val}). \ownGhost{\gamma_{\textlog{proph}}}{\authfull M} \ast \All p\mapsto \vv v \in M. \vv{v}=\nil\\
\postcrash P\triangleq{}&\textlog{crashed}\wand \textlog{crashed}\ast P
\end{align*}

\subsection{Protocol}
\begin{align*}
\textlog{CRASH}(v,\Phi)\triangleq{}&v=(\effecttag{crash},())\ast (\pvs \postcrash \mathcal{R})\\
\textlog{DURA}_{\W}(\Psi)(v,\Phi)\triangleq{}&\Exists R. \Psi(v,(\Lam r. \pvs[\bot][\W\oplus\WCrashInvTok{\top,R}] (R\land\pvs[\W\oplus\WCrashInvTok{\top,R}][\bot]\Phi(r))))\\
\textlog{CRASHCONC}_{\W}(\Psi)\triangleq{}&\textlog{DURA}_{\W} (\Psi\oplus\textlog{FORK}_\W'(\Psi))\\
\textlog{FORK}_\W'(\Psi)(v,\Phi)\triangleq{}&\Exists e. v=(\effecttag{fork},\Lam\_. e)\ast\\
&\hspace*{-8em}\later\ewp e [\W\oplus\WCrashInvTok{\top,\TRUE},\bot]{\textlog{CRASHCONC}_{\W}(\Psi)}{\Ret \_.\Exists R_e. \pvs[\bot][\W\oplus\WCrashInvTok{\top,R_e}] R_e}\ast \Phi(())
\end{align*}
In the \textlog{CRASH} protocol, $\mathcal{R}$ is the global crash invariant, and there is no requirement on $\Phi$ because crash will never return.
\textlog{CRASHCONC} intentionally uses the same tag as the regular \textlog{CONC} protocol to prevent having two schedulers.
The $\textlog{FORK}'$ protocol permits a child thread to change the crash condition during execution, as long as it is consistent with the $\WCrashInvTok{\E,R}$ token.
The crash condition cannot be violated even if a thread terminates.

The relations between these protocols are
\begin{mathparpagebreakable}
\textlog{DURA}_{\W}(\Psi)\sqsubseteq\textlog{ATOM}_{\W}(\Psi)\sqsubsetsim\Psi
\and
\textlog{ATOM}_{\W}(\Psi)\sqsubseteq\textlog{CONC}_{\W}(\Psi)
\and
\textlog{DURA}_{\W}(\Psi)\sqsubseteq\textlog{CRASHCONC}_{\W}(\Psi)
\end{mathparpagebreakable}
\begin{mathparpagebreakable}
\mprset{fraction=--\ast}
\inferrule[\hypertarget{ewp:seq-atom}{Ewp-SeqAtom}]
{\ewp e[\W_l]{\Psi}{\Phi}}
{\ewp e[\W_l\oplus\W]{\textlog{ATOM}_\W(\Psi)}{\Phi}}
\and
\inferrule[\hypertarget{ewp:atom-dura}{Ewp-AtomDura}]
{\always(R'\wand R) \\ R' \\ \ewp e [\W]{\textlog{ATOM}_{\W}(\Psi)}{\Ret v. R'\wand\Phi(v)}}
{\ewp e [\W\oplus\WCrashInvTok{\top,R}]{\textlog{DURA}_{\W}(\Psi)}{\Phi}}
\and
\inferrule[\hypertarget{ewp:conc-crashconc}{Ewp-ConcCrashconc}]
{\ewp e [\W]{\textlog{CONC}_{\W}(\Psi)}{\Phi}}
{\ewp e [\W\oplus\WCrashInvTok{\top,\TRUE}]{\textlog{CRASHCONC}_{\W}(\Psi)}{\Phi}}
\end{mathparpagebreakable}

\subsection{Crash Hoare Logic}
\begin{align*}
  \ewpc[0] e [(\E_1,R_1),(\E_2,R_2)]{\Psi}{\Phi}
  & \eqdef{} \ewp e [\W_{1}, \W_{2}]{\textlog{CRASHCONC}_{\WInvTok{\top}}(\Psi)}{\Phi}\\
  \text{where} \quad \W_{i} &\eqdef{} \WInvTok{\E_i}\oplus\WCrashInvTok{\E_i,R_i}
\end{align*}
The $\ewpcNoArg^0$ assertion does not support the monotonicity rule, but this will not become a restriction in practice because one can always use the upward closure of a non-mask-changing $\ewpcNoArg^0$.
$$\ewpc e [(\E,R)]{\Psi}{\Phi}\triangleq
\All R',\Phi'. (\All v. \Phi(v)\wand\Phi'(v))\land(R\wand R')\wand \ewpc[0] e [(\E,R')]{\Psi}{\Phi'}$$
The connection between $\Phi$ and $R$ is a logical conjunction $\land$ because only one part of this conjunction is needed at a time.
The $\Phi$ part is used during normal execution and the $R$ part is used when the system crashes.

\subsection{Crash Borrow}
\paragraph{Client Rules}\hspace*{0pt}

\begin{mathparpagebreakable}
\mprset{fraction=--\ast}
\inferrule[\hypertarget{wupd:cbrw-alloc-app}{Wupd-CBrwAlloc}]
{\later P \\ \always(\later P\wand\later R)}
{\pvs[\WCrashInvTok{\E,R_c\ast R}][\WCrashInvTok{\E,R_c}] \knowInv{\mathcal{N}}{P\mid R}}
\and
\inferrule[\hypertarget{wupd:cbrw-return-app}{Wupd-CBrwReturn}]
{\knowInv{\mathcal{N}}{P\mid R}\\\mathcal{N}\subseteq\E}
{\pvs[\WCrashInvTok{\E,R_c}][\WCrashInvTok{\E,R_c\ast R}]\later P}
\and
\inferrule[\hypertarget{wupd:cbrw-rename-app}{Wupd-CBrwRename}]
{\knowInv{\mathcal{N}}{P\mid R}\\\mathcal{N}\subseteq \E}
{\pvs[\WCrashInvTok{\E,R_c}]\knowInv{\mathcal{N}'}{P\mid R}}
\and
\inferrule[\hypertarget{wupd:cbrw-acc-update-app}{Wupd-CBrwAccUpdate}]
{\knowInv{\mathcal{N}}{P\mid R}\\\mathcal{N}\subseteq \E}
{\pvs[\WCrashInvTok{\E,R_c}][\WCrashInvTok{\E\setminus\mathcal{N},R_c}] \later P\ast \left(\All Q. \later Q\ast\always(\later Q\wand\later R)\wand\pvs[\WCrashInvTok{\E\setminus\mathcal{N},R_c}][\WCrashInvTok{\E,R_c}]\knowInv{\mathcal{N}}{Q\mid R}\right)}
\and
\inferrule[\hypertarget{wupd:cbrw-mono-app}{Wupd-CBrwMono}]
{\knowInv{\mathcal{N}}{P\mid R} \\ \later\always(P'\wand R')\\
\later\always(P\wand P')\\\later\always(R'\wand R)\\\mathcal{N}\subseteq \E}
{\pvs[\WCrashInvTok{\E,R_c}]\knowInv{\mathcal{N}}{P'\mid R'}}
\and
\inferrule[\hypertarget{wupd:cbrw-split-app}{Wupd-CBrwSplit}]
{\knowInv{\mathcal{N}}{P_1\ast P_2\mid R_1\ast R_2}\\
\always(\later P_1\wand\later R_1)\\\always(\later P_2\wand\later R_2)\\\mathcal{N}\subseteq \E}
{\pvs[\WCrashInvTok{\E,R_c}]\knowInv{\mathcal{N}}{P_1\mid R_1}\ast\knowInv{\mathcal{N}}{P_2\mid R_2}}
\and
\inferrule[\hypertarget{wupd:cbrw-combine-app}{Wupd-CBrwCombine}]
{\knowInv{\mathcal{N}}{P_1\mid R_1}\\ \knowInv{\mathcal{N}}{P_2\mid R_2}\\\mathcal{N}\subseteq \E}
{\pvs[\WCrashInvTok{\E,R_c}]\knowInv{\mathcal{N}}{P_1\ast P_2\mid R_1\ast R_2}}

\end{mathparpagebreakable}

\paragraph{Handler Rules}\hspace*{0pt}

\begin{mathparpagebreakable}
\mprset{fraction=--\ast}
\inferrule[\hypertarget{wsat:cinv-alloc}{Wsat-CinvAlloc}]
{}
{\proves\pvs \Exists \gamma_{\textit{brw}},\gamma_{\textit{cinv}},\gamma_{\textit{cinvset}},\gamma_{\textit{cond}},\gamma_{\textit{active}}. \WCrashInvTok{\top,\mathcal{R}}\ast \ownGhost{\gamma_{\textit{cinvset}}}{\authfrag\exinj(\{\iota\})}\ast\ownGhost{\gamma_{\textit{active}}}{\authfull\exinj(\iota)}\ast\textlog{CInv}(\mathcal{R})}
\and
\inferrule[\hypertarget{wsat:cinv-destruct}{Wsat-CinvDestruct}]
{\raisebox{2ex}{\ensuremath{\ownGhost{\gamma_{\textit{cinvset}}}{\authfrag\exinj(\dom(I))}}}\\
\raisebox{2ex}{\ensuremath{\Sep_{\iota\mapsto R_c\in I}\ownGhost{\gamma_{\textit{cond}}}{\authfrag\mapsingleton{\iota}{\aginj(\latertinj(R_c))}}\ast\later R_c}}\\
\raisebox{2ex}{\ensuremath{\textlog{CInv}(\mathcal{R})}}\\
\raisebox{2ex}{\ensuremath{\textlog{WCBrw}}}\\
\raisebox{2ex}{\ensuremath{\ownGhost{E}{\top}}}}
{\pvs\later\mathcal{R}}
\end{mathparpagebreakable}

\paragraph{Model}\hspace*{0pt}

\begin{mathparpagebreakable}
\knowInv{\mathcal{N}}{P\mid R}\triangleq\Exists i. i\in\mathcal{N}\ast\ownGhost{\gamma_{\textit{brw}}}{\authfrag\mapsingleton{i}{\aginj(\latertinj(P,R))}}
\and
\textlog{CInv}(\mathcal{R})\triangleq\ownGhost{\gamma_{\textit{cinv}}}{\authfrag\exinj(\latertinj(\mathcal{R}))}
\and
\WCrashInvTok{\E,R_c}\triangleq \textlog{WCBrw}\oplus\ownGhost{E}{\E} \oplus\left(\Exists \iota.\ownGhost{\gamma_{\textit{active}}}{\authfrag\exinj( \iota)}\ast\ownGhost{\gamma_{\textit{cond}}}{\authfrag\mapsingleton{\iota}{\aginj(\latertinj(R_c))}}\right)
\and
\textlog{WCBrw}\triangleq{}\Exists B\colon \mathbb{N}\fpfn\iProp\times\iProp,C\colon\mathbb{N}\fpfn\iProp,\mathcal{R}\colon\iProp.\\
\ownGhost{\gamma_{\textit{brw}}}{\authfull\aginj\fmap\latertinj\fmap B}\ast
\ownGhost{\gamma_{\textit{cond}}}{\authfull\aginj\fmap\latertinj\fmap C}\ast\ownGhost{\gamma_{\textit{cinvset}}}{\authfull\exinj(\dom(C))}\ast\ownGhost{\gamma_{\textit{cinv}}}{\authfull\exinj(\latertinj(\mathcal{R}))} \ast\\
\left(\Sep_{i\mapsto(P,R)\in B}\left(\later P \ast\ownGhost{D}{\{i\}}\lor\ownGhost{E}{\{i\}}\right)\ast\always(\later P\wand\later R)\right)\ast\left(\left(\Sep_{(\_,R)\in B}\later R\right)\ast\left(\Sep_{R_c\in C}\later R_c\right)\wand\later\mathcal{R}\right)
\end{mathparpagebreakable}

\subsection{Asynchronous Disk}

Our mechanization actually uses two forms of disk points-to assertions: $\ell\diskmapsto\adiskval{v_c}v$ for the asynchronous disk points-to introduced in \Cref{subsec:adisk}, and $\ell\diskmapsto v$ for a points-to on the physical disk.
When a crash occurs, the asynchronous points-to collapses into a disk points-to: $\ell\diskmapsto\adiskval{v_c}v\vdash\postcrash \ell\diskmapsto v_c$.
When the system recovers from a crash, the recovery procedure will restore disk points-tos to asynchronous points-tos, which is captured by another modality known as the \emph{setup modality}: $\ell\diskmapsto v_c\vdash\setup \ell\diskmapsto\adiskval{v_c}{v_c}$.
The main body of the paper omits the setup modality and directly uses asynchronous points-tos to describe post-crash disk values.

\paragraph{Use of Prophecy Variables in the Effect Handler}
The handler for asynchronous disk is shown in \Cref{sec:app:run-adisk}.
The handler uses a volatile state to store the buffer.
For each buffered location, the handler associates a prophecy variable with it, indicating whether this location will be written back before a crash.
For a buffer item $\ell\mapsto (v,p)$, assertion $\textlog{willWB}$ defined below prophesies that the value $v$ will be written back.
$$\textlog{willWB}(\vv{v})\triangleq\Exists \vv{v}\!'.
\vv{v}=((),\keyword{true})::\vv{v}\!'$$
For an \effecttag{adisk\_load} effect, the handler returns the cached value if location $\ell$ is in the buffer; otherwise, it uses the \keyword{disk\_load} operation to load the value directly from the physical disk.
For an \effecttag{adisk\_store} effect, the handler always writes the result to the buffer, but if the location is already in the buffer, the handler will resolve the associated prophecy variable to \keyword{false}, meaning that the old value in the buffer will never be written back (as it has been overwritten by the new value).
For an \effecttag{adisk\_barrier} effect, the handler writes back the whole buffer and resolves each prophecy variable to \keyword{true}.
In addition to the explicit barrier, after each load and store, $\function{partial\_flush}$ non-deterministically writes back an arbitrary subset of the buffered writes, modeling the disk flushing buffers on its own.
In the event of a crash, all prophecy variables will become $\nil$ and because $\nil\neq ((),\keyword{true})::\_$, we learn that the remaining values in the buffer will never be written back.

Concretely, the asynchronous disk points-to assertion $\ell\diskmapsto\adiskval{v_c}{v}$ is defined as a view of the $\textlog{adp}(B,\ell,v_c,v)$ assertion, where $B$ is the buffer.
\begin{align*}
&\textlog{adp}(B,\ell,v_c,v)\eqdef{} \ell\notin B\ast \ell\diskmapsto v \ast v_c=v\\
&\hspace*{1.5em}\lor\Exists p. B(\ell)=(v,p)\ast\Exists \vv{v}.\proph{p}{\vv v}\ast
(\text{if }\textlog{willWB}(\vv{v})\text{ then }(\Exists x. \ell\diskmapsto x)\ast v_c=v\text{ else }\ell\diskmapsto v_c)
\end{align*}
The assertion consists of two cases.
If $\ell$ is not in the buffer, then $v$ is in the physical disk and $v_c=v$.
If $\ell$ is in the buffer, then $v$ is the buffered value and the value in the physical disk depends on $\textlog{willWB}$.
If $\textlog{willWB}$, then the current value in the physical disk is unknown but also unimportant because it will eventually become $v$; otherwise, the value in the physical disk is $v_c$.
The connection between $\textlog{adp}(B,\ell,v_c,v)$ and $\ell\diskmapsto\adiskval{v_c}{v}$ is enforced by the authoritative resource algebra.
\section{Details about the Node-Local Specification Resource}\label{sec:app:distr}

The spec resource for the IronFleet-style scheduler is defined as
\begin{align*}
\specnode{\gamma}{t}{e}&\triangleq\Exists k,r.\ownGhost{\gamma}{\authfrag\{(k,r,t)\}}\ast \textlog{loop}(k\ r,e)\\
\textlog{loop}(e_0,e)&\triangleq (\All K.\spec{K[e_0]}\wand\pvs[\bot]\spec{K[e]})\\
&\hspace*{-3em}\lor(\All K.\spec{K[e_0]}\vsWI[\bot]\Exists H,t,M.\spec{K[\S(H)[(\effecttag{recv},t)]]}\ast t\netmapslb M\ast\textlog{loop}(H[\keyword{inl}\ ()],e))\\
&\hspace*{-3em}\lor(\All K.\spec{K[e_0]}\vsWI[\bot]\Exists H,s,t,m,M.\spec{K[\S(H)[(\effecttag{recv},t)]]}\ast t\netmapslb M\ast (s,t,m)\in M\\
&\hspace*{23em}\ast\textlog{loop}(H[\keyword{inr}\ (s,t,m)],e))
\end{align*}

It also uses the evidence accumulation technique described in \Cref{sec:relconc},
but additionally accumulates the evidence that a node can delay message receipt without changing behavior.
The idea is that if a message $m$ is received at some time $T$, then if we delay that receipt to some later time $T'$, it is still possible to receive $m$.
This is because the set of messages is monotonically growing.
The evidence that a later \effecttag{recv} can return a given message is accumulated in the \emph{least fixed point} $\textlog{loop}$.
Intuitively, $\textlog{loop}(e_0,e)$ allows $e_0$ to execute to $e$ in three ways:
(1) Taking some pure steps.
(2) First raising a \effecttag{recv} effect that receives nothing and then continuing with the result of \effecttag{recv}.
(3) First raising a \effecttag{recv} effect that receives some message $(s,t,m)$ and then continuing with the result of \effecttag{recv}.
Assertion $t\netmapslb M$ in cases (2) and (3) is a lower-bound resource of $t\netmapsto \cdot$, meaning that $M$ is a subset of messages that have ever been sent to address $t$.
\section{\pureunlog}
\subsection{Reasoning Rules}\label{sec:app:pureficus-rules}
\begin{mathparpagebreakable}
\mprset{fraction=--\ast}
\inferrule[\hypertarget{labelen:disj-union-app}{LabelEn-DisjUnion}]
{}{\labelEn{L_1\uplus L_2}\provesIff\labelEn{L_1}\ast\labelEn{L_2}}
\and
\inferrule[\hypertarget{ewp:label-app}{Ewp-Label}]
{\labelEn{L}\\z\in L}
{\ewp{\keyword{label}\ z}[\W]{\Psi}{\Ret \_.\labelEn{L}}}
\and
\inferrule[\hypertarget{ewp-ectx:bind-app}{EwpEctx-Bind}]
{\ewp[K_o\dplus N] e[\W_1,\W_2]{\Psi}{\Ret v .\smash{\ewp{N[v]}[\W_2,\W_3]{\Psi}{\Phi}}}}
{\ewp [K_o]{N[e]}[\W_1,\W_3]{\Psi}{\Phi}}
\and
\inferrule[\hypertarget{ewp-ectx:pure-bind-app}{EwpEctx-PureBind}]
{\ewp[K_o\dplus K] e[\W_1,\W_2]{\bot}{\Ret v .\smash{\ewp[K_o]{K[v]}[\W_2,\W_3]{\Psi}{\Phi}}}}
{\ewp [K_o]{K[e]}[\W_1,\W_3]{\Psi}{\Phi}}
\and
\inferrule[\hypertarget{ewp:ectx-freeze-app}{Ewp-EctxFreeze}]
{\All K.\ewp[K]{e}[\W_1,\W_2]{\Psi}{\Phi}}
{\ewp{e}[\W_1,\W_2]{\Psi}{\Phi}}
\and
\inferrule[\hypertarget{ewp:ectx-unfreeze-app}{Ewp-EctxUnfreeze}]
{\ewp{e}[\W_1,\W_2]{\Psi}{\Phi}}
{\ewp[K]{e}[\W_1,\W_2]{\Psi}{\Phi}}
\and
\inferrule[\hypertarget{ewp:snapshot-create-app}{Ewp-SnapshotCreate}]
{e\notin \Val\cup\Eff\\\labelEn{\set{l}}\\
\snapshot{l}{K[e]}\wand\ewp[K]{e}[\W_1,\W_2]{\Psi}{\Phi}}
{\ewp[K]{e}[\W_1,\W_2]{\Psi}{\Phi}}
\and
\inferrule[\hypertarget{ewp:snapshot-steps-app}{Ewp-SnapshotSteps}]
{\raisebox{1ex}{\ensuremath{e\notin \Val\cup\Eff}}\\\raisebox{1ex}{\ensuremath{\snapshot{l}{e_0}}}\\
\raisebox{1ex}{\ensuremath{e_0\restrictedsteps{\top\setminus\set{l}} K[e]\ast\snapshot{l}{e_0}\wand\ewp[K]{e}[\W_1,\W_2]{\Psi}{\Phi}}}}
{\ewp[K]{e}[\W_1,\W_2]{\Psi}{\Phi}}
\and
\inferrule[\hypertarget{ewp:snapshot-destroy-app}{Ewp-SnapshotDestroy}]
{e\notin \Val\cup\Eff\\\snapshot{l}{e_0}\\
\labelEn{\set{l}}\wand\ewp[K]{e}[\W_1,\W_2]{\Psi}{\Phi}}
{\ewp[K]{e}[\W_1,\W_2]{\Psi}{\Phi}}
\and
\inferrule[\hypertarget{tr:zero-app}{TR-Zero}]
{}{\proves\pvs\timeReceipt(0)}
\and
\inferrule[\hypertarget{tr:plus-app}{TR-Plus}]
{}{\timeReceipt(m+n)\provesIff\timeReceipt(m)\ast\timeReceipt(n)}
\and
\inferrule[\hypertarget{ewp:maxsteps-app}{Ewp-MaxSteps}]
{e\notin\Val\cup\Eff\\\All m.\maxsteps{m}\wand\ewp{e}[\W_1,\W_2]{\Psi}{\Phi}}
{\ewp{e}[\W_1,\W_2]{\Psi}{\Phi}}
\and
\inferrule[\hypertarget{ewp:timeout-app}{Ewp-Timeout}]
{e\notin\Val\cup\Eff\\\maxsteps{n}\\\timeReceipt(n)}
{\ewp{e}[\W_1,\W_2]{\Psi}{\Phi}}
\end{mathparpagebreakable}

\paragraph{Permission on Labels}
Although $\keyword{label}$ is just a no-op semantically, \pureunlog{} treats it specially: the logic comes with a token $\labelEn{L}$ controlling which labels are permitted to execute, where $L\in\wp(\integer^+)$.
This token can be split into smaller permission tokens using \refrule{labelen:disj-union-app}.
To prove an $\ewpNoArg$ of $\keyword{label}\ z$, one must supply a $\labelEn{L}$ token such that $z\in L$.

\paragraph{Context-Aware $\ewpNoArg$}
When reasoning about a large program, we often use \refrule{ewp:bind-app} to focus on a fragment of this program under a (neutral) evaluation context.
Essentially, this rule reduces a global $\ewpNoArg$ about the whole program $N[e]$ into a local $\ewpNoArg$ about a subprogram $e$, and when verifying $e$, we lose track of the surrounding evaluation context $N$.
\pureunlog{} strengthens $\ewpNoArg$ by adding another parameter $K$ to keep track of the outside evaluation context of the current focused expression: $\ewp[K]{e}[\W_1,\W_2]{\Psi}{\Phi}$.
This strengthened $\ewpNoArg$ is called context-aware $\ewpNoArg$.
Most reasoning rules for $\ewpNoArg$ still hold for the context-aware $\ewpNoArg$, except that the bind rule becomes \refrule{ewp-ectx:bind-app} and \refrule{ewp-ectx:pure-bind-app}.

\refrule{ewp-ectx:bind-app} is the context-aware analogue of \refrule{ewp:bind-app}: because we zoomed into a subprogram, the evaluation context changes from $K_o$ to $K_o\dplus N$.
Critically, after finishing verifying $e$, we have to forget the outside evaluation context and continue with a regular context-unaware $\ewpNoArg$ of $N[v]$.
This is because if $e$ raises an effect, the handler can change the outside evaluation context.
If $e$ never raises an effect, we have a stronger \refrule{ewp-ectx:pure-bind-app} rule that preserves the outside evaluation context after $e$ finishes.
The regular $\ewpNoArg$ is equivalent to a context-aware $\ewpNoArg$ under an arbitrary context.

\paragraph{Snapshots}
A $\snapshot{l}{e}$ is a resource to record the execution history of the program.
In its reasoning rules, $e_1\restrictedstep{L}e_2\eqdef e_1\step e_2 \land (\text{if}\ \textit{first-redex}(e_1)\ \text{is}\ \keyword{label}\ z\ \text{then}\ z\in L)$, and $\restrictedsteps{L}$ is the reflexive transitive closure of $\restrictedstep{L}$.
To create a snapshot, one must supply a $\labelEn{\set{l}}$ token.
Then, the snapshot resource $\snapshot{l}{e_0}$ guarantees that the current expression is reachable from $e_0$, and $\keyword{label}\ l$ has not been executed since $\snapshot{l}{e_0}$ was created.
To regain the permission to execute $\keyword{label}\ l$, one must destroy the snapshot.

\paragraph{Time Receipts}
Because we work in a partial-correctness setting, it is sufficient to consider an arbitrarily long (but finite) execution of a program. While step-indexing and L\"ob induction already \emph{internalize} this reasoning principle into the Iris logic, \pureunlog{} further \emph{externalizes} it using time receipts~\cite{mevel2019-time-receipt}.

When reasoning in \pureunlog, we can assume there is a resource $\maxsteps{m}$.
Each step of execution produces a time receipt $\timeReceipt(1)$, and with $m$ time receipts, we immediately prove any $\ewpNoArg$.

\subsection{Model}
\begin{align*}
\ewp e[\W_1,\W_2]{\Psi}{\Phi}\eqdef{}&\All K.\ewp[K]e[\W_1,\W_2]{\Psi}{\Phi}\\
\ewp [K]v[\W_1,\W_2]{\Psi}{\Phi}\eqdef{}& \pvs[\W_1][\W_2]\Phi(v)\\
\ewp[K]{\S(N)[v]}[\W_1,\W_2]{\Psi}{\Phi}\eqdef{}&\pvs[\W_1][\bot]\Psi(v,\Lam w . \pvs[\bot]\later(\ewp{N[w]}[\bot,\W_2]{\Psi}{\Phi}))\\
\ewp[K]{e}[\W_1,\W_2]{\Psi}{\Phi}\eqdef{}&\All n, m, L.\timeReceiptAuth(n)\ast\maxsteps{m}\ast\snapshotlatest{L}{K[e]}\wand\pvs[\W_1][\bot]\\
n\geq m\lor{}&\red(e)\ast\All e'.e\step e'\wand\laterCredit(1)\wand\pvs[\bot]\later\pvs[\bot]\\
&\timeReceiptAuth(1+n)\ast(\Exists L'.\snapshotlatest{L'}{K[e']})\ast\ewp[K]{e'}[\bot,\W_2]{\Psi}{\Phi}
\end{align*}
\begin{mathparpagebreakable}
\timeReceiptAuth(n)\eqdef\ownGhost{\gamma_{\textit{tr}}}{\authfull\natinj(n)}
\and
\timeReceipt(n)\eqdef\ownGhost{\gamma_{\textit{tr}}}{\authfrag\natinj(n)}
\and
\maxsteps{m}\eqdef\ownGhost{\gamma_{\textit{ms}}}{\aginj(m)}
\and
\snapshotlatest{L}{e}\eqdef\Exists M.\ownGhost{\gamma_{\textit{snapshot}}}{\authfull\exinj\fmap M
}\ast\ownGhost{\gamma_{\textit{labelEn}}}{\authfull\setinj(L)}\ast
\Sep_{l\mapsto e_0\in M}e_0\restrictedsteps{\top\setminus\set{l}} e\ast l\notin L
\and
\snapshot{l}{e}\eqdef\ownGhost{\gamma_{\textit{snapshot}}}{\authfrag\mapsingleton{l}{\exinj(e)}}
\and
\labelEn{L}\eqdef\ownGhost{\gamma_{\textit{labelEn}}}{\authfrag\setinj(L)}
\end{mathparpagebreakable}

\subsection{Verifying the Effect Handler $\function{run}_{\effecttag{observe}}$}\label{sec:app:veri-run-observe}

Define
\begin{align*}
I\eqdef{}&\Exists \vv v, e, m, n.\primprophauth{\vv v}\ast\snapshot{\effecttag{observe}}{e}\ast\timeReceipt(n)\ast\maxsteps{m}\ast\\
&\phantom{\Exists \vv v, e, m, n.} n\leq m\ast \textlog{futureobs}(e,m-n,\vv v)\\
\textlog{futureobs}(\_,0,\vv v)\eqdef{}&\vv{v}=\nil\\
\textlog{futureobs}(e,1+n,\vv v)\eqdef{}&\big(\Exists K, w, \vv v\!'.e\restrictedsteps{\top\setminus\set{\effecttag{observe}}} K[(\keyword{label}\ \effecttag{observe},w)]\\
&\phantom{\Exists K, w, \vv v\!'.}\land \vv{v}=w::\vv{v}\!'\land\textlog{futureobs}(K[((),w)],n,\vv v\!')\big)\\
{}\lor{}&\lnot\Big(\Exists K,w.e\restrictedsteps{\top\setminus\set{\effecttag{observe}}} K[(\keyword{label}\ \effecttag{observe},w)]\Big)\land\vv{v}=\nil
\end{align*}
We are now able to prove this specification by L\"ob induction and maintaining $I$ as an invariant.
\begin{mathpar}
\mprset{fraction=--\ast}
\inferrule[Ewp-ObserveRun]
{\effecttag{observe}\notin\tags(\Psi)\\\effecttag{observe}\in L\\\labelEn{L}\\
\All\vv v.\labelEn{L\setminus\set{\effecttag{observe}}}\wand\primproph{\vv v}\wand\ewp{\textit{main}\ ()}[\W]{\Psi\oplus\textlog{OBSERVE}}{\Phi}}
{\ewp{\function{run}_{\effecttag{observe}}\ \textit{main}}[\W]{\Psi}{\Phi}}
\end{mathpar}
Notably, we use the law of excluded middle to argue that given an expression $e$ and a natural number $n$, there exists a $\vv v$ such that $\textlog{futureobs}(e,n,\vv v)$.
\section{Implementation of Handlers}
\subsection{State}
$$\begin{codeblock}{r@{}l@{\,}c@{\,}lr}
&\function{run}_{\effecttag{state}}\triangleq&\multicolumn{3}{l}{\Lam \textit{main}\ \textit{init}. \function{go}\ \textit{main}\ ()\ \textit{init}}\\
&\text{where}\ \function{go}\triangleq&\multicolumn{3}{l}{\keyword{rec}\ \function{go}\ k\ r\ \sigma.}&\hspace*{-12em}\codecomment{$r$ is the result of the last operation; $\sigma$ is the global state}\\
\codenum&&\multicolumn{3}{l}{\quad\keyword{try}\ k\ r\ \keyword{with}}&\\
\codenum&&\qquad\phantom{\mid{}}\ v\ k&\Rightarrow&(\keyword{match}\ v\ \keyword{with}\\
\codenum&&&&\quad\phantom{\mid{}}\mathmakebox[\widthof{\ensuremath{(\effecttag{write},y)}}][r]{(\effecttag{read},())}\Rightarrow \function{go}\ k\ \sigma\ \sigma&\codecomment{returns the current state}\\
\codenum&&&&\quad\mid(\effecttag{write},y)\Rightarrow\function{go}\ k\ ()\ y&\codecomment{updates the state to $y$}\\
\codenum&&&&\quad\mid\mathmakebox[\widthof{\ensuremath{(\effecttag{write},y)}}][r]{(\efft, v)}\Rightarrow\function{go}\ k\ (\keyword{do}\ (\efft, v))\ \sigma)&\codecomment{re-raises the effect}\\
\codenum&&\qquad\mid \keyword{ret}\ v&\Rightarrow& v&
\end{codeblock}$$

\subsection{Heap}
$$\begin{codeblock}{l}
&\function{run}_{\effecttag{heap}}\triangleq\Lam\textit{main}.\\
\codenum&\quad\keyword{write}\ (0,\varnothing);\\
\codenum&\quad\keyword{deep-try}\ \textit{main}\ ()\ \keyword{with}\ \keyword{ret}\ v\Rightarrow v \mid v\ k\Rightarrow\ \keyword{match}\ v\ \keyword{with}\\
\codenum&\quad\phantom{\mid{}}\mathmakebox[\widthof{\ensuremath{(\effecttag{store},(\ell,w))}}][r]{(\effecttag{alloc},v)}\Rightarrow \keyword{let}\ (\ell,H):=\keyword{read}\ \keyword{in}\ \keyword{write}\ (\ell+1,\mapinsert{\ell}{v}{H});k\ \ell\\
\codenum&\quad \mid\mathmakebox[\widthof{\ensuremath{(\effecttag{store},(\ell,w))}}][r]{(\effecttag{load},\ell)}\Rightarrow \keyword{let}\ (\_,H):=\keyword{read}\ \keyword{in}\ k\ H(\ell)\\
\codenum&\quad \mid(\effecttag{store},(\ell,w))\Rightarrow \keyword{let}\ (n,H):=\keyword{read}\ \keyword{in}\ \keyword{write}\ (n,\mapinsert{\ell}{w}{H}); k\ ()\\
\codenum&\quad \mid\mathmakebox[\widthof{\ensuremath{(\effecttag{store},(\ell,w))}}][r]{(\effecttag{free},\ell)}\Rightarrow \keyword{let}\ (n,H):=\keyword{read}\ \keyword{in}\ \keyword{write}\ (n,\mapdelete{\ell}{H}); k\ ()\\
\codenum&\quad\mid\mathmakebox[\widthof{\ensuremath{(\effecttag{store},(\ell,w))}}][r]{(\efft,v)}\Rightarrow k\ (\keyword{do}\ (\efft,v))
\end{codeblock}$$

\subsection{Atomic Heap}
$$\begin{codeblock}{l}
&\function{run}_{\effecttag{atomheap}}\triangleq\Lam\textit{main}.\\
\codenum&\quad\keyword{deep-try}\ \textit{main}\ ()\ \keyword{with}\ \keyword{ret}\ v\Rightarrow v \mid v\ k\Rightarrow\ \keyword{match}\ v\ \keyword{with}\\
\codenum&\quad\phantom{\mid{}}(\effecttag{cas},(\ell,v,w))\Rightarrow \keyword{let}\ v_0:={!\ell}\ \keyword{in}\\
&\hspace*{10em}\keyword{if}\ v_0=v\ \keyword{then}\ \ell\gets w; k\ (v_0,\keyword{true})\ \keyword{else}\ k\ (v_0,\keyword{false})\\
\codenum&\quad \mid\mathmakebox[\widthof{\ensuremath{(\effecttag{cas},(\ell,v,w))}}][r]{(\effecttag{xchg},(\ell,v))}\Rightarrow \keyword{let}\ v_0:={!\ell}\ \keyword{in}\ \ell\gets v;k\ v_0\\
\codenum&\quad \mid\mathmakebox[\widthof{\ensuremath{(\effecttag{cas},(\ell,v,w))}}][r]{(\effecttag{faa},(\ell,j))}\Rightarrow \keyword{let}\ i:={!\ell}\ \keyword{in}\ \ell\gets i+j; k\ i\\
\codenum&\quad\mid\mathmakebox[\widthof{\ensuremath{(\effecttag{cas},(\ell,v,w))}}][r]{(\efft,v)}\Rightarrow k\ (\keyword{do}\ (\efft,v))
\end{codeblock}$$

\subsection{Concurrency}
$$\begin{codeblock}{r@{}l@{\,}c@{\,}lr}
&\function{run}_{\effecttag{conc}}\triangleq&\multicolumn{3}{l}{\Lam \textit{main}. \function{go}\ \lbag(\textit{main},(),\IsMain)\rbag}&\hspace*{-12em}\codecomment{create a new singleton bag}\\
&\text{where}\ \function{go}\triangleq&\multicolumn{3}{l}{\keyword{rec}\ \function{go}\ \textit{pool}.}&\hspace*{-12em}\codecomment{\textit{pool} is the thread pool}\\
\codenum&&\multicolumn{3}{l}{\quad\keyword{let}\ ((k,r,t),\textit{pool\/}):=\function{choose}\ \textit{pool}\ \keyword{in}}&\hspace{-3em}\codecomment{choose one thread from \textit{pool}}\\
\codenum&&\multicolumn{3}{l}{\quad\keyword{try}\ k\ r\ \keyword{with}}&\\
\codenum&&\qquad\phantom{\mid{}} v\ k&\Rightarrow&(\keyword{match}\ v\ \keyword{with}\\
\codenum&&&&\multicolumn{2}{l}{\phantom{\mid{}}(\effecttag{fork},e)\Rightarrow\function{go}\ (\lbag(e,(),\IsChild),(k,(),t)\rbag\uplus\textit{pool\/})}\\
\codenum&&&&\multicolumn{2}{l}{\mid\mathmakebox[\widthof{\ensuremath{(\effecttag{fork},e)}}][r]{(\efft,v)}\Rightarrow\function{go}\ (\lbag(k,\keyword{do}\ (\efft, v),t)\rbag\uplus\textit{pool\/}))}\\
\codenum&&\qquad\mid \keyword{ret}\ v&\Rightarrow& \keyword{if}\ t=\terminated\IsMain\ \keyword{then}\ v&\\
\codenum&&&&\keyword{else}\ \function{go}\ (\lbag((\Lam\_. v),(),\terminated{t})\rbag\uplus\textit{pool\/})&\\
\end{codeblock}$$

\subsection{Prophecy}\label{sec:app:run-proph}
$$\begin{codeblock}{l}
&\function{run}_{\effecttag{proph}}\triangleq\Lam\textit{main}.\\
\codenum&\quad\keyword{write}\ 0;\\
\codenum&\quad\keyword{deep-try}\ \textit{main}\ ()\ \keyword{with}\ \keyword{ret}\ v\Rightarrow v \mid v\ k\Rightarrow\ \keyword{match}\ v\ \keyword{with}\\
\codenum&\quad\phantom{\mid{}}\mathmakebox[\widthof{\ensuremath{(\effecttag{resolve\_proph},(v,p,w))}}][r]{(\effecttag{newproph},())}\Rightarrow \keyword{let}\ p:=\keyword{read}\ \keyword{in}\ \keyword{write}\ (p+1);k\ p\\
\codenum&\quad \mid(\effecttag{resolve\_proph},(v,p,w))\Rightarrow \ghostcode{\keyword{observe}\ (p,v,w)}; k\ v\\
\codenum&\quad\mid\mathmakebox[\widthof{\ensuremath{(\effecttag{resolve\_proph},(v,p,w))}}][r]{(\efft,v)}\Rightarrow k\ (\keyword{do}\ (\efft,v))
\end{codeblock}$$

\subsection{Atomic Prophecy}\label{sec:app:run-atomproph}
$$\begin{codeblock}{l}
&\function{run}_{\effecttag{atomproph}}\triangleq\Lam\textit{main}.\\
\codenum&\quad\keyword{deep-try}\ \textit{main}\ ()\ \keyword{with}\ \keyword{ret}\ v\Rightarrow v \mid v\ k\Rightarrow\ \keyword{match}\ v\ \keyword{with}\\
\codenum&\quad \phantom{\mid{}}(\effecttag{resolve},(v,p,w))\Rightarrow k\ (\keyword{do}\ (\effecttag{resolve\_proph},(\keyword{do}\ v,p,w)))\\
\codenum&\quad\mid\mathmakebox[\widthof{\ensuremath{(\effecttag{resolve},(v,p,w))}}][r]{(\efft,v)}\Rightarrow k\ (\keyword{do}\ (\efft,v))
\end{codeblock}$$

\subsection{Network}\label{sec:app:run-network}
$$\begin{codeblock}{l}
&\function{run}_{\effecttag{network}}\triangleq\Lam\textit{main}.\\
\codenum&\quad\keyword{write}\ \mapComp{a}{\varnothing}{a\in\vv a};\\
\codenum&\quad\keyword{deep-try}\ \textit{main}\ ()\ \keyword{with}\ \keyword{ret}\ v\Rightarrow v \mid v\ k\Rightarrow\ \keyword{match}\ v\ \keyword{with} \\
\codenum&\quad\phantom{\mid{}}(\effecttag{send},(s,t,m))\Rightarrow\keyword{let}\ \textit{net}:=\keyword{read}\ \keyword{in}\\
\codenum&\hspace*{3em} \keyword{write}\ (\mapinsert{t}{(\{(s,m)\}\cup \textit{net}(t))}{\textit{net}}); k\ ()\\
\codenum&\quad \mid (\effecttag{recv},t)\Rightarrow \keyword{let}\ \textit{net}:=\keyword{read}\ \keyword{in}\\
\codenum&\hspace*{3em} (\keyword{if}\ \textit{net}(t)=\varnothing \mathop{||} \keyword{pick}\ \keyword{then}\ k\ (\keyword{inl}\ ())\\
\codenum&\hspace*{3em} \keyword{else}\ \keyword{let}\ ((s,m),\_) :=\function{choose}\ \textit{net}(t)\ \keyword{in}\ k\ (\keyword{inr}\ (s,t,m)))\\
\codenum&\quad \mid(\efft,v)\Rightarrow k\ (\keyword{do}\ (\efft, v))
\end{codeblock}$$

\subsection{Distributed System}\label{sec:app:run-distr}
$$\begin{codeblock}{r@{}l@{\,}c@{\,}l}
&\function{run}_{\effecttag{distr}}\triangleq&\multicolumn{3}{l}{\Lam \textit{main}. \function{go}\ \lbag(\textit{main},(),\IsMain)\rbag}\\
&\text{where}\ \function{go}\triangleq&\multicolumn{3}{l}{\keyword{rec}\ \function{go}\ \textit{pool}.}\\
\codenum&&\multicolumn{3}{l}{\quad\keyword{let}\ ((k,r,t),\textit{pool\/}):=\function{choose}\ \textit{pool}\ \keyword{in}}\\
\codenum&&\multicolumn{3}{l}{\quad\keyword{let}\ \keyword{rec}\ \function{loop}\ k\ r:=(\keyword{try}\ k\ r\ \keyword{with}}\\
\codenum&&\qquad\phantom{\mid{}} v\ k&\Rightarrow&(\keyword{match}\ v,t\ \keyword{with}\\
\codenum&&&&\phantom{\mid{}}(\effecttag{start},e),\IsMain\Rightarrow\function{go}\ (\lbag(e,(),\IsChild),(k,(),\IsMain)\rbag\uplus\textit{pool\/})\\
\codenum&&&&\mid{}\mathmakebox[\widthof{\ensuremath{(\effecttag{start},e)}}][r]{(\effecttag{start},\_)},\mathmakebox[\widthof{\ensuremath{\IsMain}}][l]{\_}\Rightarrow\function{go}\ (\lbag(k,(),t)\rbag\uplus\textit{pool\/})\ \codecomment{no-op}\\
\codenum&&&&\mid\mathmakebox[\widthof{\ensuremath{(\effecttag{start},e)}}][r]{(\effecttag{recv},v)},\mathmakebox[\widthof{\ensuremath{\IsMain}}][l]{\_}\Rightarrow\function{loop}\ k\ (\keyword{do}\ (\effecttag{recv},v))\\
\codenum&&&&\mid\mathmakebox[\widthof{\ensuremath{(\effecttag{start},e)}}][r]{(\efft,v)},\mathmakebox[\widthof{\ensuremath{\IsMain}}][l]{\_}\Rightarrow\function{go}\ (\lbag(k,\keyword{do}\ (\efft, v),t)\rbag\uplus\textit{pool\/}))\\
\codenum&&\qquad\mid \keyword{ret}\ v&\Rightarrow& \keyword{if}\ t=\terminated\IsMain\ \keyword{then}\ v\\
\codenum&&&&\keyword{else}\ \function{go}\ (\lbag((\Lam\_. v),(),\terminated t)\rbag\uplus\textit{pool\/}))\ \keyword{in}\\
\codenum&&\multicolumn{3}{l}{\quad \function{loop}\ k\ r}
\end{codeblock}$$

\subsection{Crash}
$$\begin{codeblock}{l}
&\function{run}_{\effecttag{crash\_recover}}\triangleq\keyword{rec}\ \textit{run}\ \textit{main}.\\
\codenum&\quad\keyword{write}\ 0;\\
\codenum&\quad\keyword{deep-try}\ \textit{main}\ ()\ \keyword{with}\ \keyword{ret}\ v\Rightarrow v \mid v\ k\Rightarrow\ \keyword{match}\ v\ \keyword{with}\\
\codenum&\quad\phantom{\mid{}}\mathmakebox[\widthof{\ensuremath{(\effecttag{resolve\_proph},(v,p,w))}}][r]{(\effecttag{crash},())}\Rightarrow\ghostcode{\keyword{observe}\ ()}; \textit{run}\ \textit{main}\\
\codenum&\quad\mid\mathmakebox[\widthof{\ensuremath{(\effecttag{resolve\_proph},(v,p,w))}}][r]{(\effecttag{newproph},())}\Rightarrow \keyword{let}\ p:=\keyword{read}\ \keyword{in}\ \keyword{write}\ (p+1);k\ p\\
\codenum&\quad \mid(\effecttag{resolve\_proph},(v,p,w))\Rightarrow \ghostcode{\keyword{observe}\ (p,v,w)}; k\ v\\
\codenum&\quad\mid\mathmakebox[\widthof{\ensuremath{(\effecttag{resolve\_proph},(v,p,w))}}][r]{(\efft,v)}\Rightarrow k\ (\keyword{do}\ (\efft,v))
\end{codeblock}$$

\subsection{Crash Trigger}
$$\begin{codeblock}{l}
&\function{run}_{\effecttag{crash\_trigger}}\triangleq\Lam\textit{main}.\\
\codenum&\quad\keyword{if}\ \keyword{pick}\ \keyword{then}\ \keyword{do}\ (\effecttag{crash},())\\
\codenum&\quad\keyword{else}\ \keyword{deep-try}\ \textit{main}\ ()\ \keyword{with}\\
\codenum&\quad\phantom{\mid{}} \mathmakebox[\widthof{\ensuremath{\keyword{ret}\ v}}][r]{v\ k}\Rightarrow\keyword{let}\ r:=\keyword{do}\ v\ \keyword{in}\\
\codenum&\quad\phantom{\mid\keyword{ret}\ v\Rightarrow}\keyword{if}\ \keyword{pick}\ \keyword{then}\ \keyword{do}\ (\effecttag{crash},())\\
\codenum&\quad\phantom{\mid\keyword{ret}\ v\Rightarrow}\keyword{else}\ k\ r\\
\codenum&\quad\mid \keyword{ret}\ v\Rightarrow v
\end{codeblock}$$

\subsection{Disk}
$$\begin{codeblock}{l}
&\function{run}_{\effecttag{disk}}\triangleq\Lam\textit{main}.\\
\codenum&\quad\keyword{write}\ \mapComp{\ell}{0}{0\leq\ell<\textit{disk\_size}};\\
\codenum&\quad\keyword{deep-try}\ \textit{main}\ ()\ \keyword{with}\ \keyword{ret}\ v\Rightarrow v \mid v\ k\Rightarrow\ \keyword{match}\ v\ \keyword{with}\\
\codenum&\quad \phantom{\mid{}}\mathmakebox[\widthof{\ensuremath{(\effecttag{disk\_store},(\ell,w))}}][r]{(\effecttag{disk\_load},\ell)}\Rightarrow \keyword{let}\ D:=\keyword{read}\ \keyword{in}\ k\ D(\ell)\\
\codenum&\quad \mid(\effecttag{disk\_store},(\ell,w))\Rightarrow \keyword{let}\ D:=\keyword{read}\ \keyword{in}\ \keyword{write}\ (\mapinsert{\ell}{w}{D}); k\ ()\\
\codenum&\quad\mid\mathmakebox[\widthof{\ensuremath{(\effecttag{disk\_store},(\ell,w))}}][r]{(\efft,v)}\Rightarrow k\ (\keyword{do}\ (\efft,v))
\end{codeblock}$$

\subsection{Asynchronous Disk}\label{sec:app:run-adisk}
$$\begin{codeblock}{l}
&\function{run}_{\effecttag{adisk}}\triangleq\Lam\textit{main}.\\
\codenum&\quad\keyword{write}\ \varnothing;\\
\codenum&\quad\keyword{deep-try}\ \textit{main}\ ()\ \keyword{with}\ \keyword{ret}\ v\Rightarrow v \mid v\ k\Rightarrow\ \keyword{match}\ v\ \keyword{with}\\
\codenum&\quad\phantom{\mid{}}(\effecttag{adisk\_load},\ell)\Rightarrow \keyword{let}\ \textit{buf\/}:=\keyword{read}\ \keyword{in}\\
\codenum&\hspace*{3em} \keyword{let}\ r:=(\keyword{if}\ \ell\in\textit{buf\/}\ \keyword{then}\ \keyword{fst}\ \textit{buf\/}(\ell)\ \keyword{else}\ \keyword{disk\_load}\ \ell)\ \keyword{in}\\
\codenum&\hspace*{3em} \function{partial\_flush}\ (); k\ r\\
\codenum&\quad \mid(\effecttag{adisk\_store},(\ell,w))\Rightarrow\keyword{let}\ \textit{buf\/}:=\keyword{read}\ \keyword{in}\\
\codenum&\hspace*{3em} (\keyword{if}\ \ell\in\textit{buf\/}\ \keyword{then}\ \keyword{resolve\_proph}\ (\keyword{snd}\ \textit{buf\/}(\ell))\ \keyword{to}\ \keyword{false});\\
\codenum&\hspace*{3em} \keyword{let}\ p := \keyword{newproph}\ \keyword{in}\ \keyword{write}\ (\mapinsert{\ell}{(w,p)}{\textit{buf\/}}); \function{partial\_flush}\ (); k\ ()\\
\codenum&\quad \mid (\effecttag{adisk\_barrier},())\Rightarrow \keyword{let}\ \textit{buf\/}:=\keyword{read}\ \keyword{in}\\
\codenum&\hspace*{3em} \function{iter}\ \function{write\_back}\ \textit{buf\/}; \keyword{write}\ \varnothing; k\ ()\\
\codenum&\quad\mid(\efft,v)\Rightarrow k\ (\keyword{do}\ (\efft,v))\\
&\function{write\_back}\triangleq\Lam \ell, (v,p).\ \keyword{resolve\_proph}\ p\ \keyword{to}\ \keyword{true};\keyword{disk\_store}\ \ell\ v\\
&\function{partial\_flush}\triangleq\Lam ().\ \keyword{let}\ \textit{buf\/}:=\keyword{read}\ \keyword{in}\\
&\hspace*{3em} \keyword{write}\ (\function{filter}\ (\Lam \ell, (v,p).\ \keyword{if}\ \keyword{pick}\ \keyword{then}\ (\function{write\_back}\ \ell\ (v,p); \keyword{false})\ \keyword{else}\ \keyword{true})\ \textit{buf\/})
\end{codeblock}$$

\subsection{Non-Determinism}\label{sec:app:run-pick}
$$\begin{codeblock}{r@{}l@{\,}c@{\,}l}
&\function{run}_{\effecttag{pick}}\triangleq&\multicolumn{3}{l}{\Lam \textit{main}\ \textit{entropy}. \function{go}\ \textit{main}\ ()\ \textit{entropy}}\\
&\text{where}\ \function{go}\triangleq&\multicolumn{3}{l}{\keyword{rec}\ \function{go}\ k\ r\ \vv b.}\\
\codenum&&\multicolumn{3}{l}{\quad\keyword{try}\ k\ r\ \keyword{with}}\\
\codenum&&\qquad\phantom{\mid{}}\ v\ k&\Rightarrow&(\keyword{match}\ v,\vv b\ \keyword{with}\\
\codenum&&&&\quad\phantom{\mid{}}(\effecttag{pick},()),b_0::\vv b\!'\Rightarrow \function{go}\ k\ b_0\ \vv b\!'\\
\codenum&&&&\quad\mid\mathmakebox[\widthof{\ensuremath{(\effecttag{pick},()),b_0::\vv b\!'}}][r]{(\effecttag{pick},()),\nil}\Rightarrow\function{diverge}\ ()\\
\codenum&&&&\quad\mid\mathmakebox[\widthof{\ensuremath{(\effecttag{pick},()),b_0::\vv b\!'}}][r]{(\efft, v),\vv b}\Rightarrow\function{go}\ k\ (\keyword{do}\ (\efft, v))\ \vv b)\\
\codenum&&\qquad\mid \keyword{ret}\ v&\Rightarrow& v
\end{codeblock}$$

\subsection{Global Prophecy}
$$\begin{codeblock}{l}
&\function{run}_{\effecttag{observe}}\triangleq\Lam\textit{main}.\\
\codenum&\quad\keyword{deep-try}\ \textit{main}\ ()\ \keyword{with}\ \keyword{ret}\ v\Rightarrow v \mid v\ k\Rightarrow\ \keyword{match}\ v\ \keyword{with}\\
\codenum&\quad \phantom{\mid{}}(\effecttag{observe},v)\Rightarrow (\keyword{label}\ \effecttag{observe},v); k\ ()\\
\codenum&\quad\mid\mathmakebox[\widthof{\ensuremath{(\effecttag{observe},v)}}][r]{(\efft,v)}\Rightarrow k\ (\keyword{do}\ (\efft,v))
\end{codeblock}$$

\bibliographystyle{ACM-Reference-Format}
\bibliography{references}
\end{document}